\documentclass[prb,aps,amssymb,twocolumn]{revtex4-1}
\usepackage{amsmath}
\usepackage{amssymb}
\usepackage{amsthm}
\usepackage{amsfonts}
\usepackage{listings}
\usepackage{enumerate}
\usepackage{latexsym}
\usepackage{ulem}

\usepackage{psfrag}

\usepackage{bm}
\usepackage{graphicx}

\usepackage{color}

\newcommand{\beq}{\begin{equation}}
\newcommand{\eneq}{\end{equation}}

\begin{document}

\tolerance 10000

\newcommand{\vk}{{\bf k}}


\title{Symmetry-Protected Topological Interfaces and Entanglement Sequences}

\author{Luiz H. Santos$^1$}
\thanks{
Present address:
Department  of  Physics,  Emory  University,  Atlanta,  GA  30322,  USA.
       }
\author{Jennifer Cano$^2$}
\author{Michael Mulligan$^3$}
\author{Taylor L. Hughes$^1$}
\affiliation{$^1$Department of Physics, and Institute for Condensed Matter Theory, University of Illinois at Urbana-Champaign, Urbana, IL  61801, USA}
\affiliation{$^2$Princeton Center for Theoretical Science, Princeton University, Princeton, NJ 08544, USA}
\affiliation{$^3$Department of Physics and Astronomy, University of California, Riverside, CA 92511, USA}

\date{\today}

\begin{abstract}
Gapped interfaces (and boundaries) of two-dimensional (2D) Abelian topological phases are shown to support a remarkably rich sequence of 1D symmetry-protected topological (SPT) states. We show that such interfaces can provide a physical interpretation for the corrections to the topological entanglement entropy of a 2D state with Abelian topological order found in Ref.~\onlinecite{cano-2015}. The topological entanglement entropy decomposes as $\gamma = \gamma_a + \gamma_s$, where $\gamma_a > 0$ only depends on universal topological properties of the 2D state, while a correction $\gamma_s > 0$ signals the emergence of the 1D SPT state that is produced by interactions along the entanglement cut and provides a direct measure of the stabilizing symmetry of the resulting SPT state. A correspondence is established between the possible values of $\gamma_s$ associated with a given interface - which is named Boundary Topological Entanglement Sequence - and classes of 1D SPT states. We show that symmetry-preserving domain walls along such 1D interfaces (or boundaries) generally host localized parafermion-like excitations that are stable to local symmetry-preserving perturbations.
\end{abstract}

\maketitle
\tableofcontents

\section{Introduction}

Over the past decade, entanglement has played an increasingly important role in characterizing quantum phases of 
matter.\cite{Wen-book,TQC-Review}
Two of the most successful diagnostics are the entanglement entropy\cite{kitaevpreskill,levinwen} and the entanglement spectrum.\cite{lihaldane,pollmann-2010,cirac-2011,regnault-2011}
For a gapped state $|\Psi \rangle$ in two spatial dimensions (2D), the bi-partite entanglement entropy between two spatial regions $A$ and $B$ obeys the scaling law, $S_{\rm ent} \equiv - {\rm Tr}_A (\rho_A \log \rho_A) =\alpha L-\gamma_a$, where the reduced density matrix $\rho_A = {\rm Tr}_B |\Psi \rangle \langle \Psi|.$ 
$S_{\rm ent}$ scales with the linear size $L$ of the boundary of region $A$ (``area law" scaling) and generally contains a sub-leading, constant correction $\gamma_a$ known as the topological entanglement entropy (TEE),
in the limit where $L >> \xi$, where $\xi$ is the correlation length.\cite{kitaevpreskill,levinwen}  
The seminal work of Levin-Wen and Kitaev-Preskill showed that $\gamma_a$ is an important quantum characteristic of a ground state that is determined solely by the topological properties of $|\Psi \rangle$.

When $|\Psi \rangle$ has a vanishingly small correlation length, the entanglement entropy probes the quasi-one-dimensional (1D) gapped state living along the entanglement cut,\cite{qikatsuraludwig, lundgrenentanglement,chen-fradkin-2013,furukawa-kim-2013} i.e., the entanglement is only sensitive to the physics near the interface between regions $A$ and $B$.
Such an interface between (possibly even distinct) 2D gapped states with Abelian topological order (living in regions $A$ and $B$) has been the subject of much recent theoretical work.\cite{kapustin-2011,Levin-2013,Barkeshli-2013-a, Lindner-2012,Clarke-2013,Cheng-2012,Vaezi-2013, Barkeshli-2013-b,Mong-2014,Plamadeala2013, Cano-2014,Cano-2015-a,Khan-2014,Khan-2016, Santos-2017,Lu-2014} 
It was shown in Ref.~\onlinecite{cano-2015} that interactions along the entanglement cut can produce a positive correction to the topological entanglement entropy (even in situations where $\gamma_a = 0$), i.e., $\gamma_a$ increases such that the total entropy decreases. In such cases, these interactions constrain the motion of certain quasiparticles across the interface and, thereby, reduce the total entanglement between regions $A$ and $B.$ 
In Ref.~\onlinecite{cano-2015}, the TEE was extracted from the infinite $L$ limit of a single bipartition calculation, 
in contrast to the procedures used in Refs.~\onlinecite{kitaevpreskill,levinwen} that use linear combinations of entropies.

Nevertheless,
the connection between the 1D state at the interface and the correction to the entanglement entropy in Ref.~\onlinecite{cano-2015} has remained elusive.
The purpose of this work is to provide a physical interpretation for some of these 1D states by relating them to symmetry-protected topological phases (SPTs).\cite{Senthil-SPT-review}
SPTs, the spin-$1$ antiferromagnetic chain\cite{Haldane-1983-a,Haldane-1983-b,Affleck-1987,Affleck-1988} being the earliest known example, possess a gap to bulk excitations, and harbor symmetry-protected boundary modes.
These states have been systematically studied using the tools of 
tensor networks,\cite{pollmann-2010,Schuch-2011} group cohomology,\cite{Chen-2013, Gu-Wen-2014} 
non-linear sigma models,\cite{Bi-2013} 
and effective field theories.\cite{Freedman2004,Fidkowski-2010,Fidkowski-2011, LevinStern2013,Neupert-2011,Santos-2011,Levin2011,LevinStern-2012,Ryu-2012,Levin-2012,Lu-2012,Vishwanath-2013,Metliski-2013-a,Kapustin-2014,C-Wang-2014, J-Wang-2015,Ye-2016}
One of the main results of this work is a general mechanism that generates 1D SPTs protected by Abelian discrete symmetries, that exist at the interface between 2D topological phases of matter. 
Importantly, the discrete symmetry that stabilizes the 1D phase acts solely on the local degrees of freedom along the interface, i.e., it does not affect the (possible) fractionalized quasiparticles of the bulk states. 
As a consequence, the properties of the 1D SPT phase are robust to local perturbations that respect the SPT symmetry. 

In related work, it was shown in Ref.~\onlinecite{Zou-2016} that a non-zero correction to the TEE of topologically trivial bosonic $2$D states can be attributed to 1D SPTs living along the entanglement cut.
In this respect, our work can be viewed as an extension of the relation between TEE corrections and 1D SPTs
to a wide class of 2D Abelian topological phases, which include fermionic and bosonic topologically ordered states. 
Consequently, our results provide new insights into the connection between 1D SPTs and entanglement entropy of 2D gauge theories,\cite{casinihuertarosabal, Ghosh2015, Fliss2017} and may provide useful theoretical guidance to the realization of 1D SPTs from 2D heterostructures.

Although our construction is more general, in this article we primarily focus on explicit examples of gapped interfaces that realize 1D bosonic SPTs protected by $\mathbb{Z}_{n} \times \mathbb{Z}_{n}$ symmetry, where the integer $n > 1$.
In this case, there are $n$ distinct classes of $\mathbb{Z}_{n} \times \mathbb{Z}_{n}$-protected 1D SPTs.
These phases can be viewed as generalizations\cite{Geraedts,Santos-2015-a} of the gapped, Haldane phase of the spin-1 chain\cite{Haldane-1983-a,Haldane-1983-b,Affleck-1987,Affleck-1988} for which $n=2$.
Remarkably, we will show that these 1D SPTs can appear at interfaces between systems whose fundamental constituents are bosonic or fermionic, and that they can arise even when the bulk states are not topologically ordered with fractionalized excitations. 
As we demonstrate in our examples, the existence of these 1D SPT states can be inferred from the presence of symmetry-protected bound states that are localized at any domain wall between distinct classes of the 1D gapped phases.\cite{Santos-2015-a,Geraedts}
Such domain walls can support parafermionic defects\cite{Lindner-2012,Clarke-2013,Cheng-2012,Vaezi-2013,Barkeshli-2013-b,Mong-2014,Khan-2014,Khan-2016,Santos-2017,Fradkin-1980} 
whose properties may be useful for topological quantum computation architectures.\cite{TQC-Review}

From this point of view, it is natural to link the correction to the entanglement entropy found in Ref. \onlinecite{cano-2015} to the emergence of a 1D SPT.
Given an entanglement cut that runs along such a gapped interface, the sub-leading constant in the entanglement entropy can be separated into a sum of two terms,
\begin{equation}
\label{eq: entanglement sequence}
\gamma = \gamma_a + \gamma_{s},
\end{equation}
where $\gamma_a =\log \mathcal{D}$ (with ${\cal D}$ the so-called total quantum dimension\cite{kitaevpreskill,levinwen}) is non-zero whenever the bulk 2D system has deconfined anyonic excitations, i.e., non-trivial topological order,
and we will show below that $\gamma_{s}$ is a non-negative correction that can be associated with the emergent SPT interface. 
For a given set of bulk states that can share a gapped interface, there are many possible discrete values of $\gamma_s$ that can arise; the value of $\gamma_s$ depends on the gapping interactions at the interface. We refer to the set of possible $\gamma_s$ as the {\it boundary topological entanglement sequence} (BTES) corresponding to a given interface. 
It has been shown that $\gamma_a$ depends upon the universal topological data of the 2D state, which for an Abelian topological phase is encoded in its K-matrix (see Sec.~\ref{sec: General Approach}) and results in $\gamma_a = \log{\sqrt{|\textrm{det}(K)|}}.$ 
The correction $\gamma_s$ depends on the properties of the local interactions that give rise to a particular gapped interface, however it still only depends on universal data because the allowed sets of gapping interactions are solely determined by the properties of the K-matrix through the Haldane null-vector criterion.\cite{Haldane-1995}
We will give a simple algorithm for calculating the BTES for Abelian 2D topological phases, and show to what extent $\gamma_s$ can serve to classify the possible SPT states that are supported along 1D interfaces.

For most of the calculations in this article we will imagine a generic setup where two topological phases of matter $L$ and $R$ are connected laterally through a gapped interface. The phases $L$ and $R$ are Abelian topological phases that will be described using the K-matrix Chern-Simons formalism,~\cite{read-1990,wen-1992} and they could be identical or different. The analysis of the possible $1$D gapped interfaces between $L$ and $R$ reveals five different scenarios.
\begin{enumerate}
\item \textit{Fully transmitting Interface}: 
In Sec.~\ref{sec: Fully Transmitting Interface}, we consider interfaces across which every quasiparticle can move. 
Thus, the bulk phases on either side of the interface necessarily have the same topological order.
When a quasiparticle crosses the interface, it either retains its topological properties or changes into another type of quasiparticle.
As we will discuss, a fully transmitting interface does not give rise to a SPT phase.
This is revealed by the absence of an emergent symmetry that is preserved by the interactions that gap the interface,
and has $\gamma_s = 0$. 
\item \textit{Gauging Related Interfaces}: 
In Sec.~\ref{sec: Emergent Interface Symmetry}, we first study heterogeneous interfaces that appear at the boundary between distinct Abelian
topological phases whose total quantum dimensions satisfy 
$
\frac{\mathcal{D}_{R}}{\mathcal{D}_{L}} = k\in \mathbb{Z}
$,
where the $R$ phase is obtained from gauging a discrete $\mathbb{Z}_k$ global symmetry of the $L$ phase.
As such, the local backscattering process that gaps this heterogeneous interface
has an emergent $\mathbb{Z}_{k}$ global symmetry associated with the local degrees of freedom of $L.$ 
In Sec.~\ref{sec:general}, we show that bulk Abelian phases whose K-matrices are either proportional or related by a gauging mechanism share the same BTES and SPT interfaces.
This permits the grouping of different 2D phases into equivalence classes associated to their BTES and SPT interfaces. In the latter half of  Section \ref{sec: Emergent Interface Symmetry}  we study a sandwich structure with two identical Abelian topological phases separated by an intermediate gauged phase. We show that as the intermediate gauged region is thinned down to zero width, there is a clear mechanism for the formation of an emergent symmetry leading to an SPT state and an entanglement correction.  
\item \textit{Homogeneous Self-Gappable SPT Interface}: In Sec.~\ref{sec: Homogeneous Self-Gappable interfaces}, we discuss interfaces separating the same non-chiral 2D bulk phase for which the degrees of freedom living along the left ($L$) and right ($R$) sides of the interface can be gapped independently. 
A 1D SPT can be generated by backscattering processes that mix the local degrees of freedom on the two sides of the interface such that the local interactions are invariant under a symmetry group $\mathcal{G}_{L} \times \mathcal{G}_{R}$, acting on the $L$ and $R$ degrees of freedom with $\mathcal{G}_{L} = \mathcal{G}_{R} = \prod_{i}\,\mathbb{Z}_{d_{i}}$.
This type of interface yields a correction to the entanglement entropy given by $\gamma_{s} = \sum_{i}\,\log d_{i}$. 
We will show that in this case there are (pairs of) symmetry-protected non-Abelian zero modes centered around the positions of the domain walls separating distinct 1D phases.
\item \textit{Heterogeneous Stable Interface}: In Sec.~\ref{sec: Heterogeneous Stable Interfaces}, we consider interfaces between distinct $L$ and $R$ bulk phases that are related by the gauging mechanism studied in Sec.~\ref{sec: Emergent Interface Symmetry}.
We specialize to the case where neither bulk $L$ or $R$ phase has boundary degrees of freedom that can be gapped independently.\cite{Levin-2013}
We refer to the associated interface as ``stable without symmetry," or stable for short.
Although one cannot generate a trivial SPT phase in a heterogeneous stable interface, since one cannot gap the $L$ and $R$ sides independently, such interfaces can meet the topologically trivial 2D vacuum at a junction between gapless edge states that support distinct sets of 
quasiparticle excitations (see Fig.~\ref{fig: heterogeneous-dws}). 
We will show that this type of edge-state junction supports symmetry-protected parafermion-like excitations.
\item \textit{Homogeneous Chiral SPT Interface}: 
In Sec.~\ref{sec: Homogeneous chiral interfaces}, we discuss interfaces formed between the same chiral 2D bulk phases. 
Because the edge theories associated to the $L$ and $R$ phases have non-zero chiral central charge, a gapped interface necessarily requires backscattering processes that mix $L$ and $R$ modes.
When this interaction respects a $\mathcal{G}_L \times \mathcal{G}_R$
symmetry, any endpoint of the gapped interface, terminating at the topologically trivial 2D vacuum, can host a symmetry-protected 
parafermion-like excitation. 
\end{enumerate}

We finish the article with sections on general properties of the boundary topological entanglement sequences (Section \ref{sec:general}) and conclusions (Section \ref{sec:conclusions}).
Several Appendices contain technical details.

\section{General Approach}
\label{sec: General Approach}

In this section, we summarize our conventions for studying
non-trivial gapped interfaces between left ($L$) and right ($R$) 2D Abelian topological phases associated to the $r \times r$ matrices, $K_L$ and $K_R$.
We model the low-energy degrees of freedom living along any interface by an interacting 1D Luttinger liquid.
Defining 
\begin{equation}
\label{eq: interface K-matrix - general}
\mathcal{K} = 
\begin{pmatrix}
K_L & 0
\\
0 & -K_R
\end{pmatrix}
\,,
\end{equation}
the Luttinger liquid Lagrangian is~\cite{Wen-1995}
\begin{equation}
\mathcal{L}_{1D}=\frac{1}{4\pi}\partial_t \Phi_{I}\mathcal{K}_{IJ}\partial_x\Phi^{J}-\frac{1}{4\pi}\partial_x\Phi^{I}V_{IJ}\partial_x\Phi^{J}-H_{int}[\Phi]
\,,
\end{equation}
where $(t,x)$ denote the time and the spatial coordinates along the interface,
$\Phi = (\phi_{L},\phi_{R})$, where $\phi_{L/R}$ are the $r$-component fields on the left and right sides of the interface, $V_{IJ}$ is the forward-scattering/velocity matrix, and repeated indices $I, J \in \{ 1, ..., 2r \}$ are summed over.
In the limit $H_{int}[\Phi] = 0$, the Luttinger liquid Lagrangian has a $U(1)^{r}\times U(1)^{r}$ symmetry, corresponding to the $r$ low-energy modes on each side of the interface.

The bosonic fields satisfy the equal-time commutation relations,
\begin{equation}
\left[\partial_x\Phi_{I}(x), \Phi_{J}(y)\right]=-2\pi i \mathcal{K}^{-1}_{IJ}\delta(x-y).
\end{equation}
We choose the bosonic fields to have unit compactification radii so that 
quasiparticles of the 2D bulk state are created along the edge by the vertex operators,
 \begin{equation}
 V_{\ell}=\exp \left(i\ell^{T}\Phi\right)
 \,,
 \end{equation}\noindent where $\ell$ is a $2r$-component integer vector, i.e., $\ell \in \mathbb{Z}^{2r}.$
Importantly, in the quasiparticle spectrum, there exists a set of local particles $\Psi_{\ell}$ that are identified with the microscopic degrees of freedom of the theory. They are represented by integer vectors $\ell = \mathcal{K}\Lambda$, where $\Lambda \in \mathbb{Z}^{2r}$, such that every quasiparticle in the spectrum braids trivially with respect to the local quasiparticles.~\cite{read-1990,wen-1992}

We require that $K_L$ and $K_R$ permit a gapped interface.
$H_{int}[\Phi]$ contains the possible gap-generating interactions.
Typically, there are many such interactions when $r > 1$.
Heuristically, operators yielding a gapped spectrum pin the $2r$ bosonic (phase) fields to 
$r$ independent
values that minimize $H_{int}[\Phi]$.
The resulting ground state breaks the $U(1)^{r}\times U(1)^{r}$ symmetry down to $U(1)^{r}$. 
In this paper, we are interested in gapped interfaces where this remaining $U(1)^{r}$ symmetry admits
a discrete subgroup $\mathcal{G}_{L} \times \mathcal{G}_{R}$, associated with independent symmetry
transformations acting on the $L$ and $R$ degrees of freedom. 
At the resulting gapped fixed point, the discrete symmetry protects 
a 1D SPT phase, which is characterized by a non-zero value of $\gamma_s$.

Interactions that gap the interface modes describe physical processes coupling local quasiparticles.
A set of $r$ such local interactions takes the form,
\begin{equation}
\label{eq: gapping interaction}
\begin{split}
H_{int}[\Phi] 
&\,
= 
\sum^{r}_{i=1}\,a_{i}\,\cos{\left( \Lambda^{T}_{i}\mathcal{K}\Phi  \right)}
\\
&\,=
\sum^{r}_{i=1}\,a_{i}\,\cos{\left( \Lambda^{T}_{L,i}K_{L}\phi_{L} - \Lambda^{T}_{R,i}K_{R}\phi_{R}  \right),}
\end{split}
\end{equation}
where $a_i \in \mathbb{R},$ and $\Lambda^{T}_{i} \equiv (\Lambda_{L,i}, \Lambda_{R,i})^{T}$ with $i=1,...,r$, is a set a integer vectors satisfying the null condition,\cite{Haldane-1995}:
\begin{equation}
\label{eq: null condition}
\Lambda^{T}_{i}\,\mathcal{K}\,\Lambda_{j} = 0
\,, \quad \forall ~ i, j = 1, ..., r	
\,.
\end{equation}
The null condition \eqref{eq: null condition} guarantees that the $r$ different linear combinations
of fields entering the cosine terms in \eqref{eq: gapping interaction} can be simultaneously ``locked" or pinned.
It is useful to recast Eq.~\eqref{eq: null condition} in the following matrix form,
\begin{equation}
\label{eq: null condition matrix form}
\mathcal{M}_{L}\,K_{L}\,\mathcal{M}^{T}_{L} - \mathcal{M}_{R}\,K_{R}\,\mathcal{M}^{T}_{R} = 0	
\,,
\end{equation}
where 
\begin{equation}
\label{eq: definition of M Left and Right}
\mathcal{M}_{L/R}=
\left(
\begin{array}{c}
\Lambda^{T}_{L/R,1}\\\Lambda^{T}_{L/R,2}\\ \ldots \\ \Lambda^{T}_{L/R,r}
\end{array}
\right)
\,
\end{equation} 
are $r \times r$ integer matrices whose rows are formed by the vectors $\Lambda^{T}_{L/R,i}$.

That the interactions \eqref{eq: gapping interaction} break the $U(1)^{2r}$ symmetry down to $U(1)^{r}$ can
be manifestly seen by considering the operators,
\begin{equation}
\label{eq: generator U(1) symmetry}
\mathcal{S}^{\theta_{j}}_{j}
=
e^
{
i\,\frac{\theta_{j}}{2\pi}\,\int_{\mathcal{I}}\,dx\,\Lambda^{T}_{i}\,\mathcal{K}\,\partial_{x}\Phi	
},	
\quad 
j = 1, ..., r
\,,
\end{equation}
where $\mathcal{I}$ indicates the interface gapped by the interactions Eq.~\eqref{eq: gapping interaction}.
Due to the null condition \eqref{eq: null condition}, the operators in Eq.~\eqref{eq: generator U(1) symmetry} 
commute with each other and with all of the interaction terms in
Eq.~\eqref{eq: gapping interaction} and generate continuous phase rotations on the space of local operators $\Psi_{\ell}$:
\begin{equation}
\begin{split}
&\,
\mathcal{S}^{\theta_{j}}_{j}
\,
\Psi_{\ell}
\,
(\mathcal{S}^{\theta_{j}}_{j})^{-1}
=
e^{i \theta_{j}\,\ell^{T}\,\mathcal{K}\,\Lambda_{j}}
\,
\Psi_{\ell}
\,,
\\
&\,
\Psi_{\ell} = e^{i \ell^{T}\,\mathcal{K}\,\Phi} = e^{i \left( \ell_{L}^{T}\,K_{L}\,\phi_{L} + \ell_{R}^{T}\,K_{R}\,\phi_{R}  \right)}
\,.
	\end{split}	
\end{equation}
Eq.~\eqref{eq: generator U(1) symmetry} then accounts for the generators of $U(1)^{r}$ symmetry 
in the presence of the gapping interactions Eq.~\eqref{eq: gapping interaction}.

The $U(1)$ generators \eqref{eq: generator U(1) symmetry} in general mix
L and R fields.
Yet, in the following subsection we show that whenever $\mathcal{M}_{L}$ and $\mathcal{M}_{R}$ are invertible
matrices, we can identify a subgroup $\mathcal{G}_{L} \times \mathcal{G}_{R}$ of $U(1)^{r}$,
corresponding to independent discrete phase rotations that act separately on the $L$ and $R$ degrees of freedom, 
such that the order of the discrete
groups $\mathcal{G}_{L}$ and $\mathcal{G}_{R}$ is given by $|\textrm{det}(\mathcal{M}_{L})|$ and $|\textrm{det}(\mathcal{M}_{R})|$, respectively.
As we shall see, the existence of the discrete L and R symmetries is directly related
to gapped interfaces with non-zero values of $\gamma_s,$ and the corresponding $1$D SPT state.
As such, we will only focus our attention on independent L and R discrete symmetries, although the $U(1)^{r}$ group may possess other discrete subgroups that mix $L$ and $R$ fields.

\subsection{Left/right discrete symmetries of the interactions in Eq.~$\eqref{eq: gapping interaction}$}
\label{sec: Left/right discrete symmetries}
We now discuss how, quite generally, interactions \eqref{eq: gapping interaction} can possess a discrete global symmetry 
acting on local quasiparticles. 
Consider the following global transformations acting independently on the left and right fields:
\begin{subequations}
\label{eq: global left/right transformation acting on chiral fields}
\begin{equation}
\phi_{L} \rightarrow \phi_{L} + 2\pi\,K^{-1}_{L}\alpha_{L}
\,,
\end{equation}
\begin{equation}
\phi_{R} \rightarrow \phi_{R} + 2\pi\,K^{-1}_{R}\alpha_{R}
\,,
\end{equation}
\end{subequations}
under which local operators $\psi_{L/R} = e^{i\ell_{L/R}^{T}K_{L/R}\phi_{L/R}}$ ($\ell_{L/R} \in \mathbb{Z}^{r}$) transform as
\begin{subequations}
\label{eq: global left/right transformation acting on local fields}
\begin{equation}
\psi_{L} \rightarrow \psi_{L}\,e^{i 2\pi \ell_{L}^{T} \alpha_{L}}	
\,,
\end{equation}
\begin{equation}
\psi_{R} \rightarrow \psi_{R}\,e^{i 2\pi \ell_{R}^{T} \alpha_{R}}	
\,.
\end{equation}
\end{subequations}

The interactions in Eq.~\eqref{eq: gapping interaction} are invariant under the global 
transformations in Eq.~\eqref{eq: global left/right transformation acting on chiral fields} if
\begin{subequations}
\label{eq: condition on M and alpha}
\begin{equation}
\mathcal{M}_{L} \, \alpha_{L} \in \mathbb{Z}^{r}	
\,,
\end{equation}
\begin{equation}
\mathcal{M}_{R} \, \alpha_{R} \in \mathbb{Z}^{r}	
\,.
\end{equation}
\end{subequations}
If $\mathcal{M}_{L/R}$ are invertible,
Eq.~\eqref{eq: condition on M and alpha} implies that $\alpha_{L/R}$ are of the form
$\alpha_{L/R}  = \mathcal{M}^{-1}_{L/R}\,t_{L/R}$, for $t_{L/R} \in$ $\mathbb{Z}^{r}$.
Since values of $\alpha_{L/R}$ that differ by an element of $\mathbb{Z}^r$ yield the same transformations in Eq.~(\ref{eq: global left/right transformation acting on local fields}), we identify $\alpha_{L/R}$ mod $\mathbb{Z}^r$.
Thus, to deduce the symmetry of the gap opening interactions, we must determine which values of $t_{L/R}$ yield integer values of $\alpha_{L/R}$.
This is achieved by computing the Smith Normal Form (SNF) of $\mathcal{M}_{L/R}$:
\begin{equation}
\mathcal{D}_{L/R} = \mathcal{U}_{L/R} \mathcal{M}_{L/R} \mathcal{V}_{L/R},
\label{eq: SNF}
\end{equation}
where $\mathcal{D}_{L/R}$ are diagonal integer matrices with non-zero diagonal entries 
$d_{1,L/R}, d_{2,L/R}, \dots d_{r,L/R},$ and $\mathcal{U}_{L/R}, \mathcal{V}_{L/R}$ are unimodular matrices.
It is useful to define $\alpha'_{L/R} = \mathcal{V}^{-1}_{L/R}\alpha_{L/R}$ and $t'_{L/R} = \mathcal{U}_{L/R}t_{L/R}$ (notice that $\alpha'_{L/R} \in \mathbb{Z}^{r}$ if and only if $\alpha_{L/R} \in \mathbb{Z}^{r}$ and, similarly, 
$t'_{L/R} \in \mathbb{Z}^{r}$ if and only if $t_{L/R} \in \mathbb{Z}^{r}$). 
It follows that 
\begin{equation}
\label{eq: def of alpha prime}
\alpha'_{L/R} = \mathcal{D}^{-1}_{L/R}t'_{L/R}.
\end{equation}
Thus, $\alpha'_{L/R} \in \mathbb{Z}^r$ if and only if $d_{i,L/R}$ divides $( t'_{L/R} )_i$.
We conclude that the interface has $\mathcal{G}_L \times \mathcal{G}_R$ symmetry, where 
\begin{equation}
\mathcal{G}_{L/R} = \mathbb{Z}_{d_{1,L/R}}\times \mathbb{Z}_{d_{2,L/R}} \times \cdots \times\mathbb{Z}_{d_{r,L/R}}.
\label{eq:symmetry}
\end{equation}
(The trivial discrete groups that occur in the decomposition in Eq.~\eqref{eq:symmetry} whenever $d_{i,L/R} = 1$ are denoted by $\mathbb{Z}_1$.)
Using Eq.~(\ref{eq: SNF}), the order, $|\mathcal{G}_{L/R}|$, of the left and right symmetries is equal to the determinant
of the matrices in Eq.~\eqref{eq: definition of M Left and Right} formed by the null vectors,
\begin{equation}
\label{eq: relation between det of M and symmetry order}
|\mathcal{G}_{L/R}| = |\textrm{det}(\mathcal{M}_{L/R})|
\,.
\end{equation}
We stress that Eq.~\eqref{eq: relation between det of M and symmetry order} was derived under the assumption that 
$\mathcal{M}_{L/R}$ are invertible matrices.
However, if the $L$ and $R$ degrees of freedom are not completely chiral, then it is sometimes possible to open a gap without fully mixing them.
In this case, the matrices $\mathcal{M}_{L/R}$ will not be invertible.
We discuss the important case in which the $L$ and $R$ degrees of freedom are ``self-gappable'' in Sec.~\ref{sec: Homogeneous Self-Gappable interfaces}.
In the subsequent formulas, we continue to assume that the backscattering process mix the modes of the L and R sides 
such that $\mathcal{M}_{L/R}$ are invertible matrices and Eq.~\eqref{eq: relation between det of M and symmetry order} holds.

The symmetry generators that implement Eq.~\eqref{eq: global left/right transformation acting on chiral fields} are
\begin{equation}
\label{eq: symmetry generators}
\begin{split}
&\,
S_{L/R}
=
\exp{\left[ i\,\int_{\mathcal{I}}\,dx\, \left( \mathcal{M}^{-1}_{L/R}\,t_{L/R}  \right)^{T} \partial_{x}\phi_{L/R} \right]}
\,,
\end{split}	
\end{equation}
where $\mathcal{I}$ indicates the interface gapped by the local interactions Eq.~\eqref{eq: gapping interaction}.
A faithful representation of $\mathcal{G}_{L/R}$ is formed by choosing the $r$ generators:
\begin{subequations}
\label{eq: faithful rep}
\begin{equation}
\begin{split}
&\,
S_{L/R}^{(p)} = \exp{\left[ i\,\int_{\mathcal{I}}\,dx\, \left( \alpha_{L/R}^{(p)} \right)^{T} \partial_{x}\phi_{L/R} \right]},
\\
&\,
p = 1, ..., r
\,,
\end{split}
\end{equation}
where
\begin{equation}
\label{eq: def alpha}
\alpha_{L/R}^{(p)} = \mathcal{V}_{L/R}\,\mathcal{D}_{L/R}^{-1}\,e_p
\end{equation}
\end{subequations}
and $(e_p)_i = \delta_{ip}$.

\subsection{Symmetry and entanglement}

The symmetry of the interface is detectable via the entanglement entropy.
To show this we now consider the constant, sub-leading term in the entanglement entropy $\gamma = \gamma_a + \gamma_s$ defined in Eq.~(\ref{eq: entanglement sequence}).
We will refer to this term as the topological entanglement entropy (TEE), since both $\gamma_a$ and $\gamma_s$ depend upon the topological data encoded in the K-matrix:
the former is given in terms of the total quantum dimension;~\cite{kitaevpreskill,levinwen} 
the latter is determined by interactions, which are related to the K-matrix through the
Haldane null-vector criterion.\cite{Haldane-1995}
By the complementarity property of entanglement entropy, the TEE can be computed by tracing out either the right or left side of the interface: 
\begin{equation}
\gamma = \gamma_{a,L/R} + \gamma_{s,L/R} ,
\label{eq: TEE complementarity}
\end{equation}
where 
\begin{equation}
\gamma_{a,L/R} = \frac{1}{2}\log{| \det K_{L/R} |}
\label{eq: TEE K}
\end{equation}
and $\gamma_{s,L/R}$ is the ``TEE correction'' found in Ref.~\onlinecite{cano-2015}.
In Appendix~\ref{sec:TEE correction}, we show that for primitive gapping terms, $\gamma_{s,L/R} = \log |\det \mathcal{M}_{L/R} |$ 
(we define primitivity in Sec.~\ref{sec: Primitive Condition of the Null Vectors}).
Thus, Eq.~(\ref{eq: relation between det of M and symmetry order}) yields,
\begin{equation}
\label{eq: TEE correction}
\gamma_{s,L/R} = \log{| \mathcal{G}_{L/R} |}.
\end{equation}
The null conditions \eqref{eq: null condition matrix form} impose the constraint
\begin{equation}
\label{eq: topology and symmetry constraints v1}
\frac
{
|\textrm{det}(K_{L})|	
}	
{
|\textrm{det}(K_{R})|	
}
=
\frac{|\mathcal{G}_R|^{2}}{|\mathcal{G}_L|^{2}}
\,,
\end{equation}
which ensures that Eqs.~(\ref{eq: TEE K}) and (\ref{eq: TEE correction}) satisfy the complementarity property (\ref{eq: TEE complementarity}).
Eq~(\ref{eq: TEE correction}) summarizes one of the main results of this work: interactions can introduce emergent symmetries whose presence is revealed by a correction to the TEE.

Eqs.~\eqref{eq: TEE complementarity} -- \eqref{eq: TEE correction} relate the topological character of the bulk phases to the emergent symmetry respected at their interface. 
An immediate consequence is that the interactions stabilizing a gapped interface between 2D states with distinct fractionalized content, i.e., with $\gamma_{a, R} \neq \gamma_{a, L}$,
\emph{must} have distinct left and right global symmetries. 
For instance, consider the interface between
one-component Abelian theories with $K_L = k\,m^{2}$ and $K_R = k\,n^{2}$ ($k, m, n \in \mathbb{Z}$ and $\textrm{g.c.d}(m,n) = 1$).
The gap instability is driven by a local primitive interaction $H_{\textrm{int}} \sim \left( \psi^{\dagger}_{R} \right)^{m}\,\left( \psi^{\phantom{\dagger}}_{L} \right)^{n} + \textrm{H.c.}$, where $\psi^{\dagger}_{L/R}$ are the local quasiparticle creation operators, 
for which there is a $\mathbb{Z}_{n} \times \mathbb{Z}_{m}$ symmetry associated with left and right degrees of freedom.\cite{Santos-2017}
The left and right local degrees of freedom at the interface transform under different symmetries because the condensate at the interface does not permit every quasiparticle to be fully transmitted from one side to the other.
Interfaces that separate two distinct topological phases behave as anyon Andreev reflectors, as discussed in Ref.~[\onlinecite{Santos-2017}].
We will discuss these types of heterogeneous interfaces in Sec.\ref{sec: Heterogeneous Stable Interfaces}.

If $\gamma_{a, R} = \gamma_{a, L}$, Eqs.~\eqref{eq: TEE complementarity} -- \eqref{eq: TEE correction} imply that 
the interactions allow a global $\mathbb{Z}_{m}\times\mathbb{Z}_m$ symmetry.  When $m > 1$,
we find that this system realizes a 1D bosonic SPT state with symmetry-protected zero modes with $m^2$ degenerate ground states in an open chain.~\cite{Santos-2015-a,Geraedts} 
We will discuss this type of interface in Secs.~\ref{sec: Homogeneous Self-Gappable interfaces} and \ref{sec: Homogeneous chiral interfaces}.

\subsection{Primitivity Condition for the Null Vectors}
\label{sec: Primitive Condition of the Null Vectors}
In searching for solutions of the null condition \eqref{eq: null condition}, we require that 
the null vectors be primitive.~\cite{LevinStern-2012}
An integer vector is primitive if the greatest common divisor of its entries is $1$.
A set of integer vectors, $\Lambda_i$, is primitive, if the linear combination $a_i\Lambda_i$ is a primitive vector for every set of integers $a_i$ whose greatest common divisor is $1$.
As a simple example to illustrate the concept of primitivity, consider the interface between
$\nu=1$ integer quantum Hall (IQH) states for which $K_L = K_R = 1$ in Eq.~\eqref{eq: interface K-matrix - general}. 
This interface can be gapped by a backscattering process, which is represented by the interaction $\cos{\left( \phi_L - \phi_R  \right)}$, where $\phi_{L/R}$ are the left- and right-moving fields. 
The null vector associated with this interaction, $\Lambda^{T} = (1,1)$, is primitive. 
This primitive interaction is not invariant under any non-trivial symmetry.  

Any integer vector of the form $(a,a)$ (with $a \in \mathbb{Z}$) is manifestly null, however, it is non-primitive when $|a| > 1$.
Now, consider the non-primitive interaction $\cos{\left[ a (\phi_L - \phi_R ) \right]}$, for $a > 1$, which is invariant under a 
$\mathbb{Z}_{a} \times \mathbb{Z}_{a}$ symmetry. 
When the non-primitive interaction acquires a non-zero vacuum expectation value (VEV) and gaps the interface, the primitive, local interaction, $\cos{\left( \phi_L - \phi_R  \right)}$, also acquires a non-zero VEV,
which breaks the $\mathbb{Z}_{a} \times \mathbb{Z}_{a}$ symmetry.
Thus, requiring the gapping interactions to be primitive ensures that the symmetry obeyed by such interactions 
cannot be trivially broken by ``more elementary" local null operators.

We will frequently invoke an equivalent condition for primitivity: the null vectors $\Lambda_i$ are primitive if and only if the greatest common divisor of the set of $r\times r$ minors of the $r\times 2r$ integer matrix $\left( \mathcal{M}_{L}, \mathcal{M}_{R} \right)$ is $1$.\cite{LevinStern-2012}
Although the example of the $\nu=1$ IQH interface discussed above suggests that local interactions possessing a global symmetry are always non-primitive, this is only true for one-component edge theories.
In multi-component edge theories, the edge can be gapped with local interactions that are primitive, \emph{and} yet still 
possess an unbroken global discrete symmetry, as we will demonstrate with explicit examples.
When this occurs, the 1D system possesses a discrete global symmetry with a unique (assuming periodic boundary conditions),
gapped symmetric ground state, whose low-energy properties will be identified with that of a 1D SPT state.

\section{Fully Transmitting Interface}
\label{sec: Fully Transmitting Interface}
The conventional interface between two topological phases with the same quasiparticle content is a fully transmitting interface, where
every quasiparticle $\varepsilon$ coming from, say, the left edge is
transformed as it passes across the interface into another quasiparticle $\varepsilon'$ that propagates to the right side.
As such, the types of fully transmitting interfaces are in one-to-one correspondence with anyonic permutation symmetries.\cite{Khan-2014,Lu-Fidkowski-2014}
These permutation symmetries are transformations acting on the quasiparticle spectrum that leave the fusion and braiding properties of the set of quasiparticles invariant. 
Given a 2D Abelian phase described by a K-matrix, $K$, 
an anyonic symmetry can typically be represented by a unimodular matrix $W \in \textrm{SL}(r,\mathbb{Z})$ such that $W\,K\,W^{T} = K$.

Given an anyonic symmetry transformation described by $W$, the following null vectors completely gap the interface between $K$ and itself:\cite{Khan-2014}
\begin{equation}
\label{eq: null vector transmitting interface}
\Lambda^{\textrm{f.t.}}_i =
\begin{pmatrix}	
e_{i}
\\
W^{T}\,e_{i}
\end{pmatrix}
\end{equation}
where $e_i$ with $i = 1, ..., r$ is a set of orthonormal Euclidean vectors and the superscript
f.t. denotes ``fully transmitting."
The null vectors \eqref{eq: null vector transmitting interface} represent a process whereby a quasiparticle $\varepsilon$ on the left side of the interface is transformed to $W^{T}\varepsilon$ as it crosses to the right side of the interface.
From the null vectors in Eq.~\eqref{eq: null vector transmitting interface}, 
we identify $\mathcal{M}_{L} = \textrm{I}_{r\times r}$ and $\mathcal{M}_{R} = W$ in Eq.~\eqref{eq: definition of M Left and Right}.
As a consequence of the unimodular character of $W$, it follows
that $|\mathcal{G}_{L}| = |\textrm{det}(\mathcal{M}_{L})| = 1$ and $|\mathcal{G}_{R}| = |\textrm{det}(\mathcal{M}_{R})| = 1$.
This implies that the local interactions gapping the interface do not have an emergent symmetry, and do not generate an entanglement entropy correction.
The absence of an entanglement entropy correction for fully transmitting interfaces is a consequence of the fact that this type of interface imposes no constraint on which quasiparticles can pass across the interface.

\section{Gauging Related Interfaces and Emergent Interface Symmetry}
\label{sec: Emergent Interface Symmetry}
\subsection{Inhomogeneous gauging related interfaces}
\label{sec: Inhomogeneous gauging related interfaces}
We now present a general mechanism to generate local gapping interactions that are invariant under a discrete symmetry. 
Consider the interface characterized by
\begin{equation}
\label{eq: heterogeneuos KL and KR}
K_{L} = K
\,,\quad
K_{R} = G\,K\,G^{T}
\,,
\end{equation}
where $G \in \textrm{GL}(r,\mathbb{Z})$. 
If $|{\rm det}G|>1$, then the two phases on either side of the interface have different topological order.
The interface can be fully gapped by primitive interactions associated with the null vectors,
\begin{equation}
\label{eq: null vectors heterogeneous interface}
\begin{split}
&\,
\left(\Lambda_{G}\right)^{T}_{i}
=
\left(
e^{T}_{i}\,G,~~ e^{T}_{i}
\right)
\,,\quad
i = 1, ..., r
\,.
	\end{split}	
\end{equation}
Thus, we find that 
\begin{equation}
\mathcal{M}_{L} = G
\,,
\quad
\mathcal{M}_{R} = \textrm{I}_{r \times r}
\,,
\end{equation}
which implies that the interactions associated with the null vectors in Eq.~\eqref{eq: null vectors heterogeneous interface} are invariant 
under a discrete symmetry group $\mathcal{G}_{L}$ acting on the left fields, where $|\mathcal{G}_{L}| = |\textrm{det}(G)|$.

\begin{figure}[h!]
\includegraphics[width=0.4\textwidth]{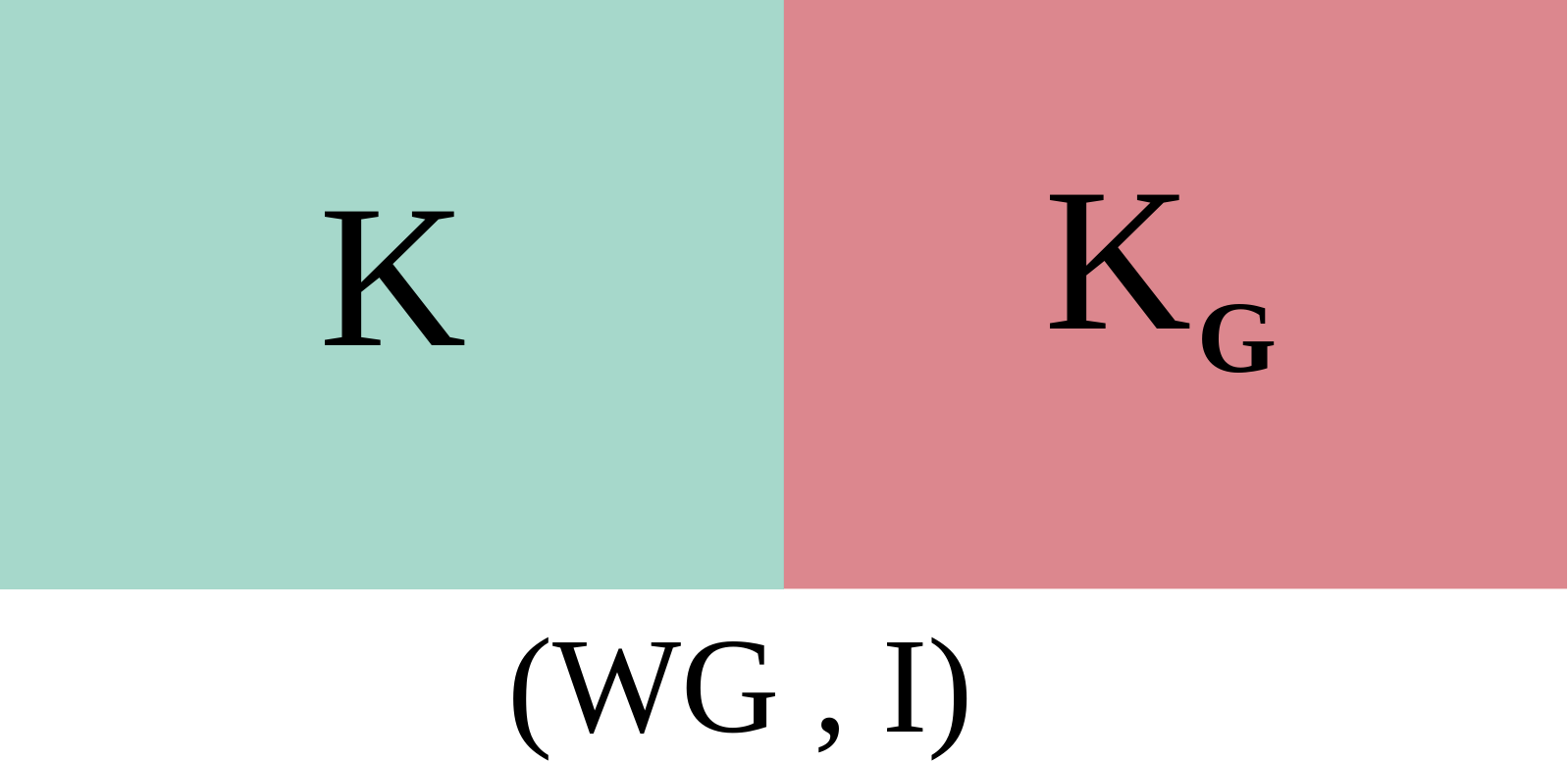}
\caption
{
(Color online)
The interface between the distinct topological phases associated to the K-matrices, $K$ and $K_{G} = G\,K\,G^{T}$, runs along the vertical axis.
The null vectors are parametrized by 
$(\mathcal{M}_{L},\mathcal{M}_{R}) = (W\,G, \textrm{I}_{r \times r})$.
This indicates that the local interactions are invariant under the discrete
symmetry $\mathcal{G}_{L} \times \mathbb{Z}_{1}$, where 
$|\mathcal{G}_{L}| = |\textrm{det}(G)|,$ and $\mathbb{Z}_{1}$ represents the trivial discrete group for the right-hand side of the interface.
}
\label{fig: interface-1}
\end{figure}

For each \emph{unimodular} matrix $W$ that represents an anyonic symmetry of $K_R$, such that $WK_RW^T = K_R$, the set
\begin{equation}
\label{eq: null vectors heterogeneous interface - anyonic symmetry}
\begin{split}
&\,
\left(\Lambda_{(W\,G)}\right)^{T}_{i}
=
\left(
e^{T}_{i}\,(W\,G),~~ e^{T}_{i}
\right)
\,,\quad
i = 1, ..., r
\,,
	\end{split}	
\end{equation}
is also a valid set of null vectors that can gap the heterogeneous interface.
Following Eqs.~\eqref{eq: SNF} -- \eqref{eq:symmetry}, this interface is invariant under the same symmetry, $\mathcal{G}_L$.
This type of heterogeneous interface is depicted in Fig.~\ref{fig: interface-1}.

The gapping interactions of this interface have an emergent symmetry because $K_{L}$ and $K_{R}$ in Eq.~\eqref{eq: heterogeneuos KL and KR} are related by the gauging procedure,
\begin{equation}
\label{eq: gauge mapping K matrices}
K \rightarrow K_{G} \equiv G\,K\,G^{T}	
\,.
\end{equation}
If $G$ is unimodular, Eq.~\eqref{eq: gauge mapping K matrices} merely represents a change of basis. 
However, if $|\textrm{det} {G}| > 1$, then the map \eqref{eq: gauge mapping K matrices} changes the topological structure of the theory: it describes a process whereby a discrete symmetry, $\mathcal{G}$, of the original 2D phase described by $K$ is gauged.
The resulting Abelian state, described by $K_{G}$, contains deconfined fractionalized excitations which, in the Abelian phase described by $K$, represent (confined) fluxes of the global symmetry $\mathcal{G}$ (we will also call these ``twist" 
defects).\cite{Levin-2012,Hung-2013,Lu-2016}
The gapping vectors of the interface between $K$ and its $\mathcal{G}$-gauged state $K_{G}$, parametrized by the null vectors in Eq.~\eqref{eq: null vectors heterogeneous interface}, represent a physical process whereby a set of deconfined excitations of the $K_{G}$ phase is confined along the interface.
The utility of gauging a global symmetry of a gapped phase to characterize the un-gauged, symmetry preserving phase was first pointed out by Levin and Gu in Ref.~[\onlinecite{Levin-2012}].
Here it plays an important role in the identification of symmetry-protected gapped interfaces. 

The deconfinement process in Eq.~\eqref{eq: gauge mapping K matrices} changes 
the total quantum dimension of the Abelian state, and thus the bulk contribution of the TEE, as follows:
\begin{equation}
\label{eq: change in total quantum dimension of K}
\gamma_{a,K} \rightarrow \gamma_{a,K_{G}} = \gamma_{a,K} + \log{|\textrm{det}G|}
\,,
\end{equation}
where $\gamma_{a,K} = \frac{1}{2}\,\log{|\textrm{det}K|}$ and
$\gamma_{a,K_{G}} = \frac{1}{2}\,\log{|\textrm{det}K_{G}|}$. 
If the entanglement cut is exactly at the interface in Fig.~\ref{fig: interface-1}, one also finds the TEE given by Eq.~\eqref{eq: change in total quantum dimension of K}. 
This has two complementary interpretations depending on which side of the cut is traced over. For one side the sub-leading correction to the area law is just the usual TEE of the Abelian topologically ordered phase.
For the other side this value is a combination of the conventional TEE of $K$ plus the correction due to an emergent global symmetry $\mathcal{G}.$ 

\subsection{Homogeneous interface with emergent symmetry}
\label{sec: Homogeneous interface with emergent symmmetry}
\begin{figure}[h!]
\includegraphics[width=0.5\textwidth]{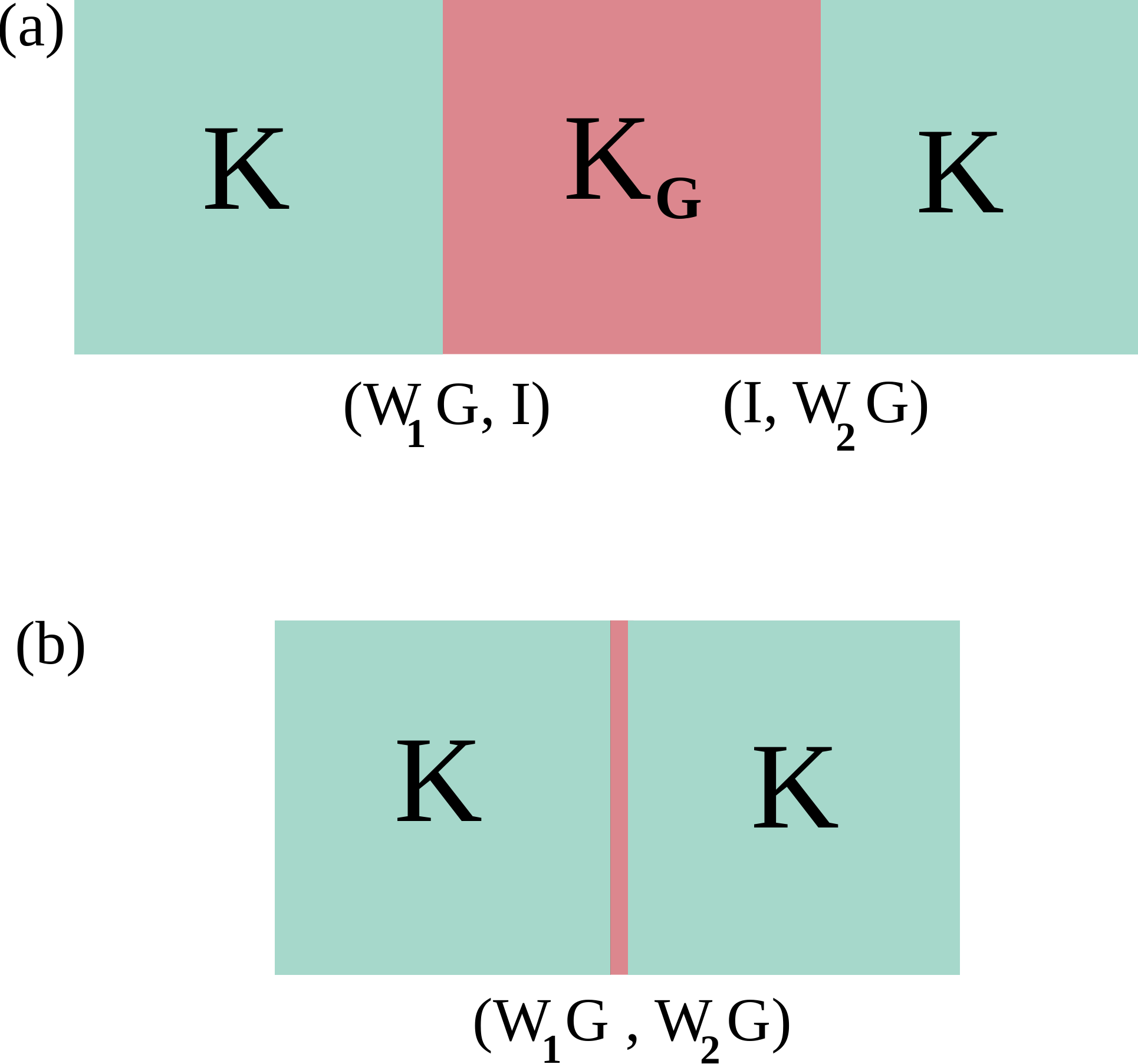}
\caption
{
(Color online)
(a) In the limit where the intermediate slab thickness is large, interfaces 
$1$ and $2$ can be treated as being decoupled from each other. At each interface, the null vectors are parametrized by $(W_{1}\,G, \textrm{I}_{r \times r})$ and
$(\textrm{I}_{r \times r}, W_{2}\,G )$, implying that the local interactions on 
each interface are invariant under the symmetry $\mathcal{G}$, where $|\mathcal{G}| = |\textrm{det}(G)|$.
(b) As the thickness of the intermediate slab goes to zero, one recovers the homogeneous interface,
with null vectors parametrized by $(W_{1}\,G, W_{2}\,G )$, signaling that the local gapping interactions
are invariant under the discrete $\mathcal{G} \times \mathcal{G}$ symmetry. 
The Smith normal form of the matrix $G$ specifies the discrete symmetry,
and the minors of $(W_{1}\,G, W_{2}\,G )$ determine whether these
interactions are primitive.
}
\label{fig: interface-2}
\end{figure}

Consider now the situation depicted in Fig.~\ref{fig: interface-2}(a), where the topological phases associated to $K$ are separated by a slab containing the topological phase associated to $K_{G} = G\,K\,G^{T}$.
We label by $1$ and $2$ the two interfaces, where 
\begin{equation}
\mathcal{K}_{1}
= 
\begin{pmatrix}
K & 0
\\
0 & - G\,K\,G^{T}
\end{pmatrix}
\,,\quad
\mathcal{K}_{2}
= 
\begin{pmatrix}
G\,K\,G^{T} & 0
\\
0 & -K
\end{pmatrix}
\end{equation}
are the effective K-matrices controlling the dynamics of the low-energy modes localized to the two interfaces. 
We assume initially that the thickness of the intermediate slab is large enough so
that we can treat interfaces $1$ and $2$ separately from each other, i.e., we can ignore possible long-range tunneling between the degrees of freedom localized on the two separate interfaces. 
Then, primitive null vectors of interface $1$ read
\begin{equation}
\label{eq: null vectors interfaces 1}
\begin{split}
&\,
\left(\Lambda^{(1)}_{(W_{1}\,G)}\right)^{T}_{i}
=
\left(
e^{T}_{i}\,(W_{1}\,G),~~ e^{T}_{i}
\right)
\,,
\quad i = 1, ..., r
\,,
	\end{split}	
\end{equation}
while those of interface $2$ are
\begin{equation}
\label{eq: null vectors interfaces 2}
\begin{split}
&\,
\left(\Lambda^{(2)}_{(W_{2}\,G)}\right)^{T}_{i}
=
\left(
e^{T}_{i},~~ e^{T}_{i}\,(W_{2}\,G)
\right)
\,,
\quad
i = 1, ..., r
\,,
	\end{split}	
\end{equation}
where $W_{1}, W_{2} \in \textrm{SL}(r,\mathbb{Z})$ are any two anyonic symmetries of $K_{G} = G\,K\,G^{T}$, so as to satisfy the null condition \eqref{eq: null condition}. 

In the limit where the thickness of the intermediate topological slab goes to zero, 
as depicted in Fig.~\ref{fig: interface-2}(b), one recovers the interface between the left and right topological phases described by 
\begin{equation}
\label{eq: K-matrix homogeneous interface -general}
\mathcal{K} = 
\begin{pmatrix}
K & 0
\\
0 & -K
\end{pmatrix}	
\,,
\end{equation}
whose null vectors, obtained by ``fusing" those in Eq.\eqref{eq: null vectors interfaces 1} and Eq.\eqref{eq: null vectors interfaces 2}, are given by\cite{comment-1}
\begin{equation}
\label{eq: null vectors homogeneous interface - general}
\begin{split}
&\,
\left(\Lambda_{(W_{1}G,W_{2}G)}\right)^{T}_{i}
=
\left(
e^{T}_{i}\,(W_{1}\,G),~~ e^{T}_{i}\,(W_{2}\,G)
\right)
\,,
\\
&\,
\quad i = 1, ..., r
\,.
	\end{split}	
\end{equation}
As a result, we have
\begin{equation}
\label{eq: ML and MR homogeneous interface - general}
\mathcal{M}_{L} = W_{1}\,G
\,,
\quad
\mathcal{M}_{R} = W_{2}\,G	
\,.
\end{equation}
Thus, $\mathcal{M}_L$ and $\mathcal{M}_R$ have the same SNF.
Eqs~(\ref{eq: SNF}) -- (\ref{eq:symmetry}) then dictate that the same symmetry is generated on either side of the interface, and that the order of the symmetry is $|\det G|$.
Notice that Eq.~\eqref{eq: ML and MR homogeneous interface - general} admits the more general solution,
$\mathcal{M}_{L} = W_{1}\,G\,\tilde{W}_{L}$ and $\mathcal{M}_{R} = W_{2}\,G\,\tilde{W}_{R}$,
where $\tilde{W}_{L}$ and $\tilde{W}_{R}$ are anyonic symmetries of $K_{L}$ and $K_{R}$, 
and $K_{L} = K_{R} = K$ in this case. However, the multiplication by anyonic symmetries $\tilde{W}_{L}$ and $\tilde{W}_{R}$
can be absorbed into a basis change and, as such, does not change the physical properties of the gapped interface.~\cite{anyonicsymmetry-K}

If the null vectors corresponding to Eq~\eqref{eq: ML and MR homogeneous interface - general}
are primitive, then the TEE obtained by tracing the left (or right) degrees of freedom
of the interface in Fig.~\ref{fig: interface-2}(b)
is given by 
\begin{equation}
\label{eq: modification of TEE by interface interactions}
\gamma = \gamma_{a,K} + \log{|\textrm{det} G|}	
\,,
\end{equation}
which is identical to the right hand side of Eq.~\eqref{eq: change in total quantum dimension of K}.
In other words, the origin of the TEE \textit{correction} for the homogeneous interface can be interpreted as a memory of the fact that the phase was constructed from thin intermediate slabs of $K_G$.

However, the interactions associated with Eq.~\eqref{eq: null vectors homogeneous interface - general} are not necessarily primitive.
For instance, in the trivial case $W_{1} = W_{2} = \textrm{I}_{r \times r}$, the $r\times r$ minors of $(\mathcal{M}_{L},\mathcal{M}_{R})=(G,G)$ are in the set $\{ 0, \pm \textrm{det}G \}$;
since their greatest common divisor is $|{\rm det} G|$, the null vectors are not primitive when $|\textrm{det}G| > 1$ (recall the criteria in Sec.~\ref{sec: Primitive Condition of the Null Vectors} to compute primitivity). If the fused null vectors are not primitive, then the emergent symmetry will generically be spontaneously broken when the intermediate gauged region is thinned, and the TEE correction will disappear, i.e., $\gamma_s = 0$. This is reminiscent of the concept of ``weak symmetry breaking" introduced in Ref. \onlinecite{wanglevin} where it was shown that one could make a ``thick" gapped edge of a 2D SPT phase without breaking a protecting symmetry, but when the thick, gapped edge was narrowed, the system was shown to spontaneously break at least one of the protecting symmetries. In the context of this article we are precisely interested in the case when the weak symmetry breaking mechanism is frustrated so that the symmetry remains intact; the primitivity of the null vectors ensures this outcome.

Now that we have described a general intuitive picture for the emergence of interface symmetries and TEE corrections, in the following sections we will discuss example systems that illustrate that the emergent symmetries can support non-trivial 1D interface SPTs and associated parafermion bound states.

\section{Homogeneous Self-Gappable Interfaces}
\label{sec: Homogeneous Self-Gappable interfaces}

In this section, we consider homogeneous interfaces described by the K-matrix in Eq.~\eqref{eq: K-matrix homogeneous interface -general}, such that each side of the interface is non-chiral
and can be gapped by backscattering processes that do not involve modes from the other side of the interface.
(A necessary condition is that $K$ has zero signature and $\textrm{det}(K) = \pm k^{2}$, where $k \in \mathbb{Z}$; for a complete analysis see Ref.~\onlinecite{Levin-2013}.)
To keep the discussion as simple as possible, we consider the two simplest examples, $K = \sigma_x$ and $K = \sigma_z$, which represent bosonic and fermionic topologically trivial insulators.
(As discussed in Sec~\ref{sec: proportional K matrices share TEE correction}, the discussion can be immediately generalized to cover the cases 
with the bulk K-matrices $K = p\,\sigma_x$ or $K =  p\,\sigma_z$, for any non-zero integer $p$.)

Our approach, in both of these cases, has two steps.
First, we apply the formalism of Sec.~\ref{sec: Emergent Interface Symmetry} to find primitive interactions that have a nontrivial emergent symmetry, and hence a non-zero TEE correction. These gapping interactions, as indicated by the form of the null vectors in Eq.~\eqref{eq: null vectors homogeneous interface - general}, represent correlated backscattering processes between left and right modes (recall that left and right denote degrees of freedom on the sides of the interface, not the chirality of the modes), and preserve a $\mathcal{G} \times \mathcal{G}$ symmetry at the interface.
Second, we identify a different set of interactions that gap the interface without mixing the left and right sides, but which are also invariant under the $\mathcal{G} \times \mathcal{G}$ symmetry.
Since these interactions do not couple the two sides of the interface, they do not yield a TEE correction (or any contribution to the entanglement entropy). As a result, these gapping terms correspond to a trivial SPT phase. 
Remarkably, we will show that the interactions that mix the two sides can be interpreted as a non-trivial SPT and a domain wall between the non-trivial and the trivial SPT interfaces host a pair of zero modes transforming projectively under the action of the $\mathcal{G} \times \mathcal{G}$ symmetry.
An array of domain walls arranged along the interface gives rise to a ground state degeneracy, signaling the ``non-Abelian" nature of these zero modes.

\subsection{$K = \sigma_x$}
\label{sec: sigma x interface}

At an interface with $\mathcal{K}_{x} = \sigma_{x} \oplus (-\sigma_{x})$, 
non-trivial gapped interfaces of the form Eq.~\eqref{eq: ML and MR homogeneous interface - general}
can be parametrized by
\begin{equation}
\label{eq: form of ML and MR - sigma x}
\mathcal{M}^{(n)}_{L} = 
\begin{pmatrix}	
1 & 0
\\
0 & n
\end{pmatrix}
\,,
\quad
\mathcal{M}^{(n)}_{R} = 
W_{2}\,\mathcal{M}^{(n)}_{L} = 
\begin{pmatrix}	
0 & n
\\
1 & 0
\end{pmatrix}
\,,
\end{equation}
where $W_{2} = \sigma_x$, for any non-zero integer $n$.  Eq.\eqref{eq: form of ML and MR - sigma x} encodes the null vectors
\begin{equation}
\label{eq: null vectors Lambda 1 and 2 - sigma x}
\Lambda^{(n)}_1 = (1,0,0,n)
\,,\quad
\Lambda^{(n)}_2 = (0,n,1,0)
\,.
\end{equation}
The $2\times 2$ minors of
\begin{equation}
\left( \mathcal{M}^{(n)}_{L}, \mathcal{M}^{(n)}_{R} \right)
=
\begin{pmatrix}
1 & 0 & 0 & n
\\
0 & n & 1 & 0
\end{pmatrix}
\,	
\end{equation}
are $\{n, 1, 0, 0, -n^{2}, -n \}$; since their greatest common divisor is $1$, the null vectors \eqref{eq: null vectors Lambda 1 and 2 - sigma x} are primitive
for all integer $n$.
Eq.~\eqref{eq: form of ML and MR - sigma x} implies, using 
Eqs.~\eqref{eq: relation between det of M and symmetry order} and~\eqref{eq: TEE correction}, 
that the BTES associated with this interface is
\begin{equation}
\label{eq: BTES sigma x}
\{ \gamma^{(n)}_{s} = \log{|n|}~\,,~ n \in \mathbb{Z}^{*} \}	
\,.
\end{equation}

The null vectors in Eq.~\eqref{eq: null vectors Lambda 1 and 2 - sigma x} correspond to primitive interactions,
\begin{equation}
\label{eq: Zn x Zn interactions - sigma x}
\begin{split}
&\,
\mathcal{H}_{1,x} = \cos{\left( \phi_{2,L} - n\,\phi_{1,R}  \right)}	
\,,
\\
&\,
\mathcal{H}_{2,x} = \cos{\left( \phi_{2,R} - n\,\phi_{1,L} \right)}.	
\,
\end{split}
\end{equation}
The Smith normal form of $\mathcal{M}^{(n)}_{L/R}$ is given by
$
\mathcal{D}^{(n)}_{L/R} = 
\mathcal{U}^{(n)}_{L/R}\,\mathcal{M}^{(n)}_{L/R}\,\mathcal{V}^{(n)}_{L/R}
$, 
where
\begin{equation}
\label{eq: SNF of ML and MR for sigma x}
\begin{split}
&\,
\mathcal{D}^{(n)}_{L}  = \mathcal{D}^{(n)}_{R} 
=
\begin{pmatrix}
n & 0
\\
0 & 1
\end{pmatrix}
\,,
\\
&\,
\mathcal{U}^{(n)}_{L} 
=
I_{2 \times 2}
\,,
\quad
\mathcal{U}^{(n)}_{R} = \mathcal{U}^{(n)}_{L}\,\sigma_{x} = \sigma_{x}
\,,
\\
&\,
\mathcal{V}^{(n)}_{L}  = \mathcal{V}^{(n)}_{R}
=
I_{2 \times 2}
\,,
	\end{split}	
\end{equation}
which shows that
the gapping interactions in Eq.~\eqref{eq: Zn x Zn interactions - sigma x}
are invariant under a $\mathbb{Z}_{n}\times\,\mathbb{Z}_{n}$ symmetry acting separately on the left $(\phi_{1,L},\phi_{2,L})$ and right $(\phi_{1,R},\phi_{2,R})$ degrees of freedom. They are generated by (Eq.\eqref{eq: faithful rep})
\begin{equation}
\label{eq: left symmetry generator sigma_x}
\begin{split}
S^{(n)}_{x,L}
&\,=
\exp{\left(\frac{i}{n}\,\int_{\mathcal{I}_{x}}\,dx\, \partial_{x}\phi_{2L} \right)}	
\,,
\end{split}
\end{equation}
\begin{equation}
\label{eq: right symmetry generator sigma_x}
\begin{split}
S^{(n)}_{x,R}
&\,=
\exp{\left(\frac{i}{n}\,\int_{\mathcal{I}_{x}}\,dx\, \partial_{x}\phi_{2R} \right)}	
\,,
\end{split}
\end{equation}
where the domain of integration $\mathcal{I}_{x}$ contains the region of the interface that is gapped by the local interactions Eq.~\eqref{eq: Zn x Zn interactions - sigma x}. 
The $\mathbb{Z}_{n}\times\,\mathbb{Z}_{n}$ symmetry, whose generators are given by 
the expressions above, form a subgroup of the $U(1)\times U(1)$ symmetry with generators given in Eq.~\eqref{eq: generator U(1) symmetry}.

Following Sec.~\ref{sec: Emergent Interface Symmetry}, 
we give an interpretation for the null vectors associated with
Eq.~\eqref{eq: form of ML and MR - sigma x} in terms of a gauging process. Notice that 

\begin{equation}
\label{eq: gauged K-matrix - sigma x}
\begin{split}
K_{n,x} 
&\,\equiv 
\mathcal{M}^{(n)}_{L/R}\,\sigma_{x}\,\left( \mathcal{M}^{(n)}_{L/R} \right)^{T}
= 
n\,\sigma_{x}
\,,
\end{split}
\end{equation}
where $K_{n,x}$ is identified as the K-matrix of a deconfined $\mathbb{Z}_n$ gauge theory.
The charge and flux quasiparticle excitations of this gauge theory are represented, respectively,
by $\mathsf{e} = (1,0)$ and $\mathsf{m} = (0,1)$.
The $\mathbb{Z}_{n}$ gauge theory 
describes the deconfined phase of the symmetry fluxes associated with the global $\mathbb{Z}_n$ 
symmetries given by Eq.\eqref{eq: left symmetry generator sigma_x} and~\eqref{eq: right symmetry generator sigma_x},
where the flux associated to the $\mathbb{Z}_{n}$ transformation is 
given by the vertex operator $e^{\frac{i}{n}\phi_{2,L/R}}$. 

To explain the origin of the gapping interactions in Eq.~\eqref{eq: Zn x Zn interactions - sigma x} from the gauged theory, 
consider, as in Fig.~\ref{fig: interface-2}(a), a slab of the $\mathbb{Z}_{n}$ gauge theory in between two regions of the 
$K=\sigma_x$ theory. At the interfaces $1$ and $2$, the effective K matrices are given, respectively, 
by $\mathcal{K}_{1} = \sigma_{x} \oplus (-n\,\sigma_{x})$ 
[with low energy modes described by the fields $(\phi_{1L},\phi_{2L},\theta_{1},\theta_{2})$] 
and $\mathcal{K}_{2} = n\,\sigma_{x} \oplus (-\sigma_{x})$
[with low energy modes described by the fields $(\theta_{1}^{'},\theta_{2}^{'},\phi_{1R},\phi_{2R})$].
Consider gapping out the modes of $\mathcal{K}_{1}$ with the primitive null vectors
$
\tilde{\Lambda}_{a,1} = (1,0,\underline{1},\underline{0})
$ 
and 
$
\tilde{\Lambda}_{b,1} = (0,n,\underline{0},\underline{1})
$,
where the underlined entries are associated with the $\mathbb{Z}_{n}$ gauge theory. 
The interactions associated with these null vectors read 
$V_{a,1} = \cos{(\phi_{2L} - n\,\theta_{2})}$
and 
$V_{b,1} = \cos{(n\,\phi_{1L} - n\,\theta_{1})}$. 
The interaction $V_{a,1}$ enforces the formation of a bound state 
between the symmetry flux $e^{\frac{i}{n}\phi_{2L}}$ and $\bar{\mathsf{m}} = e^{-i\theta_{2}}$,
and the interaction $V_{b,1}$ enforces the formation of a bound state 
between the local boson $e^{i\phi_{1L}}$ and $\bar{\mathsf{e}} = e^{-i\theta_{1}}$.

Likewise, consider gapping out the modes of $\mathcal{K}_{2}$ with the primitive null vectors 
$
\tilde{\Lambda}_{a,2} = (\underline{1},\underline{0},0,n)
$ 
and 
$
\tilde{\Lambda}_{b,2} = (\underline{0},\underline{1},1,0)
$.
The interactions associated with these null vectors read 
$V_{a,2} = \cos{(n\,\theta_{2}^{'} - n\,\phi_{1R})}$
and 
$V_{b,2} = \cos{(n\,\theta_{1}^{'}  - \phi_{2R})}$. 
The interaction $V_{a,2}$ enforces the formation of a bound state 
between the local boson $e^{-i\phi_{1R}}$ and $\mathsf{m} = e^{i\theta_{2}^{'}}$,
and interaction $V_{b,2}$ enforces the formation of a bound state 
between the symmetry flux $e^{-\frac{i}{n}\phi_{2R}}$ and $\mathsf{e} = e^{i\theta_{1}^{'}}$.
Now, imagine a process whereby the intermediate slab of the topological phase is thinned out, as described
in Fig.~\ref{fig: interface-2}(b), so as to give rise to the original interface we proposed to study described by 
$\mathcal{K} = \sigma_x \oplus (-\sigma_x)$. 
This process then naturally yields the null vectors Eq.~\eqref{eq: null vectors Lambda 1 and 2 - sigma x}
by ``fusing" $\tilde{\Lambda}_{a,1}$ with $\tilde{\Lambda}_{a,2}$ 
(which involves the formation a new bound state between $e^{\frac{i}{n}\phi_{2L}}$ and $e^{-i\phi_{1R}}$
after the process $\bar{\mathsf{m}} \times \mathsf{m} \sim 1$) 
and by ``fusing" $\tilde{\Lambda}_{b,1}$ with $\tilde{\Lambda}_{b,2}$
(which involves the formation of a new bound state between $e^{-\frac{i}{n}\phi_{2R}}$ and $e^{i\phi_{1L}}$
after the process $\bar{\mathsf{e}} \times \mathsf{e} \sim 1$). 
Note however that when the interface $2$ is gapped by the pair of null vectors
$
\tilde{\Lambda}^{'}_{a,2} = (\underline{1},\underline{0},1,0)
$ 
and 
$
\tilde{\Lambda}^{'}_{b,2} = (\underline{0},\underline{1},0,n)
$,
the previous analysis gives that the homogeneous interface with $\mathcal{K} = \sigma_x \oplus \sigma_x$ is gapped by the interactions
$\cos{(\phi_{2L}-\phi_{2R})}$ and $\cos{(n\,\phi_{1L}-n\,\phi_{1R})}$, which, despite
$\mathbb{Z}_{n} \times \mathbb{Z}_n$ symmetric, are non-primitive. This means that the ground
state spontaneously breaks the symmetry with $n$ symmetry broken vacua characterized by the expectation
value of $\phi_{1L} - \phi_{1R}$. Furthermore, the nature of the symmetry
broken phase can be probed with the local operator $\cos{(\phi_{1L}-\phi_{1R})}$.

We can also find trivial gapping terms that do not mix the left and right sides of the interface, such as
\begin{equation}
\label{eq: trivial null vectors - sigma x}
\Lambda^{(0)}_{1} = (1,0,0,0)
\,
\quad
\Lambda^{(0)}_{2} = (0,0,1,0)
\,,	
\end{equation}
whose associated gapping interactions,
\begin{equation}
\label{eq: trivial gapping interactions - sigma x}
\begin{split}
&\,
\mathcal{H}^{(0)}_{1,x} = \cos{\left(  \phi_{2 L} \right)}
\,,
\\
&\,
\mathcal{H}^{(0)}_{2,x} = \cos{\left(  \phi_{2 R} \right)}
\,,		
	\end{split}	
\end{equation}
are manifestly invariant under the action of the left and right symmetry generators in Eqs.~\eqref{eq: left symmetry generator sigma_x} and~\eqref{eq: right symmetry generator sigma_x}.

Let us now consider a series of domain walls separating segments gapped by the
interactions in Eq.~\eqref{eq: trivial gapping interactions - sigma x} from segments gapped by 
the interactions in Eq.~\eqref{eq: Zn x Zn interactions - sigma x}. Let 
$
R^{x}_{0} = \cup_{i}\,(  x_{2i}+\varepsilon, x_{2i+1}-\varepsilon)
$
be the region gapped solely by the interactions Eq.~\eqref{eq: trivial gapping interactions - sigma x},
and
$
R^{x}_{n} = \cup_{i}\,(  x_{2i-1}+\varepsilon, x_{2i}-\varepsilon)
$
be the region gapped solely by the interactions Eq.~\eqref{eq: Zn x Zn interactions - sigma x}, where $\varepsilon = 0^{+}$ is an infinitesimally small distance regulator.
We assume also that the length of each of these segments, $|x_{k+1} - x_{k}| \gg \xi$, where $\xi$ is the correlation length
set by the energy gap due to the interactions.
We are going to show that there is a symmetry-protected ground state degeneracy associated
with this configuration, which is a consequence of the existence of a pair of $\mathbb{Z}_{n}$
parafermions localized at each domain wall, each possessing the quantum dimension $\sqrt{n}$.
In order to show the existence of these zero mode excitations,
consider the set of \textit{mutually commuting} operators:
\begin{subequations}
\label{eq: def An and Bn operators sigma x}
\begin{equation}
A^{(n)}_{2i-1,2i} = \exp{\Big( \frac{i}{n}\,\int^{x_{2i}+\varepsilon}_{x_{2i-1}-\varepsilon}\,dx\,\partial_{x}\phi_{2L}  \Big)},
\end{equation}
\begin{equation}
B^{(n)}_{2i-1,2i} = \exp{\Big( \frac{i}{n}\,\int^{x_{2i}+\varepsilon}_{x_{2i-1}-\varepsilon}\,dx\,\partial_{x}\phi_{2R}  \Big)}.
\end{equation}
\end{subequations}
Moreover, consider another set of \textit{mutually commuting} operators:
\begin{subequations}
\label{eq: def A0 and B0 operators sigma x}
\begin{equation}
A^{(0)}_{2i,2i+1} = \exp{\Big[ \frac{i}{n}\,\int^{x_{2i+1}+\varepsilon}_{x_{2i}-\varepsilon}\,dx\,
\partial_{x} ( \phi_{2R}  - n\,\phi_{1L}) \Big]},
\end{equation}
\begin{equation}
B^{(0)}_{2i,2i+1} = \exp{\Big[ \frac{i}{n}\,\int^{x_{2i+1}+\varepsilon}_{x_{2i}-\varepsilon}\,dx\,
\partial_{x} ( \phi_{2L}  - n\,\phi_{1R}) \Big]}.
\end{equation}
\end{subequations}
The two sets of non-local operators defined in Eqs.~\eqref{eq: def An and Bn operators sigma x} and~\eqref{eq: def A0 and B0 operators sigma x} commute with all the the gapping interactions in
Eqs.~\eqref{eq: trivial gapping interactions - sigma x} and \eqref{eq: Zn x Zn interactions - sigma x}. 
Hence, the commutation relations between these operators offer useful information about the ground state manifold, 
for the states belonging to this manifold form a representation of the operator algebra,
whose dimension is the ground state degeneracy.

Making use of the result derived in Appendix~\ref{sec: Commutation relation between vertex operators},
we find that the only non-trivial commutation relations read:
\begin{equation}
\label{eq: commutation relations non-local ops - sigma x}
\begin{split}
&\,
A^{(n)}_{2i-1,2i}\,A^{(0)}_{2j,2j+1}
=
e^{i\,\frac{2\pi}{n}\left( \delta_{j,i} - \delta_{j,i-1}  \right)}
\,
A^{(0)}_{2j,2j+1}\,A^{(n)}_{2i-1,2i}	
\,,
\\
&\,
B^{(n)}_{2i-1,2i}\,B^{(0)}_{2j,2j+1}
=
e^{-i\,\frac{2\pi}{n}\left( \delta_{j,i} - \delta_{j,i-1}  \right)}
\,
B^{(0)}_{2j,2j+1}\,B^{(n)}_{2i-1,2i}	
\,.
\end{split}
\end{equation}
Eq.~\eqref{eq: commutation relations non-local ops - sigma x} implies that the commutation relations
between these non-local operators break into two independent sets. In each of these sets, say the first one,
the operators $A^{(n)}_{2i-1,2i}$ act as raising or lowering operators with respect to the 
neighboring operators $A^{(0)}_{2i-2,2i-1}$ and $A^{(n)}_{2i,2i+1}$. Since $\Big(A^{(n)}_{2i-1,2i}\Big)^{n}$
and $\Big(A^{(0)}_{2i-1,2i}\Big)^{n}$ act as the identity operator, each of the algebras appearing in 
Eq.~\eqref{eq: commutation relations non-local ops - sigma x} can be represented by conventional ``clock" operators,
where the minimum dimension of the representation of each of these algebras gives the ground degeneracy. 
For a configuration with $2g$ domain walls, Eq.~\eqref{eq: commutation relations non-local ops - sigma x}
implies ground state degeneracy of $n^{g-1} \times n^{g-1}$.

\begin{figure}[h!]
\includegraphics[width=0.4\textwidth]{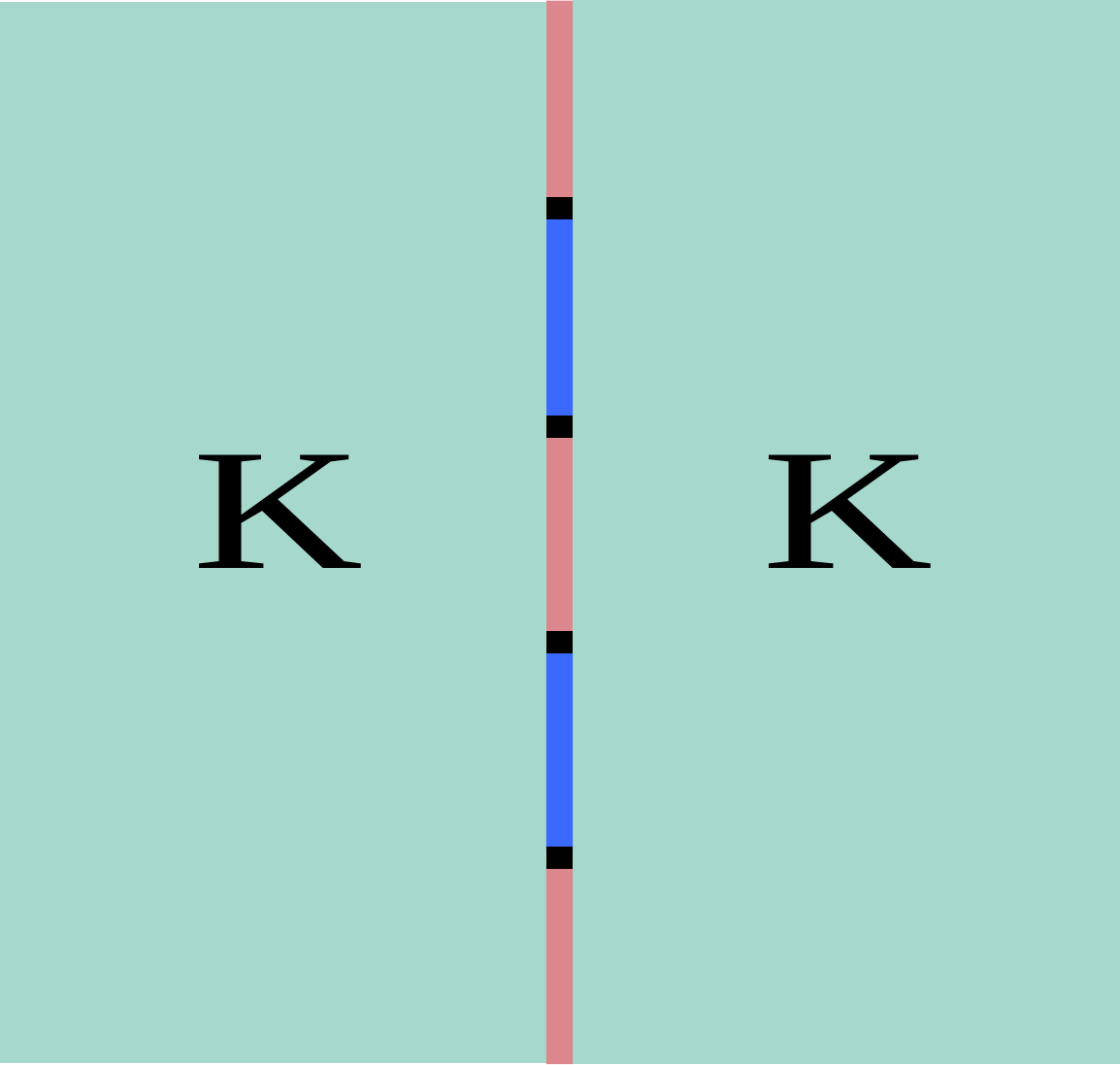}
\caption
{
(Color online)
Zero-dimensional domain walls (black dots) separating $1$D $\mathbb{Z}_{n}\times \mathbb{Z}_{n}$
SPT states (red) formed by the interactions in Eq.~\eqref{eq: Zn x Zn interactions - sigma x}  
from the trivial SPT states (blue) formed by the interactions in Eq.~\eqref{eq: trivial gapping interactions - sigma x}. 
Each domain wall supports a pair of parafermions defined in Eqs.~\eqref{eq: alpha parafermions - sigma x interface}
and \eqref{eq: beta parafermions - sigma x interface}.
}
\label{fig: interface-domainwalls}
\end{figure}

This ground state degeneracy is due to the presence of a pair of zero modes, $(\alpha_{i},\beta_{i})$, at the domain walls:
\begin{equation}
\label{eq: alpha parafermions - sigma x interface}
\begin{split}
&\,
\alpha_{2i} 
=
e^
{
\frac{i}{n}\,\left( \phi_{2R} - n\,\phi_{1L} \right)(x_{2i} - \varepsilon)
+
\frac{i}{n}\,\phi_{2L}(x_{2i} + \varepsilon)
}	
\,,
\\
&\,
\alpha_{2i+1} 
=
e^
{
\frac{i}{n}\,\phi_{2L}(x_{2i+1} - \varepsilon)
+
\frac{i}{n}\,\left( \phi_{2R} - n\,\phi_{1L} \right)(x_{2i+1} + \varepsilon)
}	
\,,	
	\end{split}	
\end{equation}
and
\begin{equation}
\label{eq: beta parafermions - sigma x interface}
\begin{split}
&\,
\beta_{2i} 
=
e^
{
\frac{i}{n}\,\left( \phi_{2L} - n\,\phi_{1R} \right)(x_{2i} - \varepsilon)
+
\frac{i}{n}\,\phi_{2R}(x_{2i} + \varepsilon)
}	
\,,
\\
&\,
\beta_{2i+1} 
=
e^
{
\frac{i}{n}\,\phi_{2R}(x_{2i+1} - \varepsilon)
+
\frac{i}{n}\,\left( \phi_{2L} - n\,\phi_{1R} \right)(x_{2i+1} + \varepsilon)
}	
\,.	
\end{split}	
\end{equation}
These zero modes satisfy the algebra $\alpha_{2i+1}\,\alpha_{2i} = e^{-i\frac{2\pi}{n}}\,\alpha_{2i}\,\alpha_{2i+1}$,
$\beta_{2i+1}\,\beta_{2i} = e^{i\frac{2\pi}{n}}\,\beta_{2i}\,\beta_{2i+1}$ and $\alpha_{i}\,\beta_{j} = \beta_{j}\,\alpha_{i}$,
which we identify as two independent sets of parafermions\cite{Fradkin-1980} with quantum dimension $\sqrt{n}$.
The non-local operators in Eqs.~\eqref{eq: def An and Bn operators sigma x} and \eqref{eq: def A0 and B0 operators sigma x} can be expressed as parafermion bilinears such as
$\alpha^{\dagger}_{2i}\,\alpha_{2i+1} \sim A^{(0)}_{2i,2i+1}$, $\beta^{\dagger}_{2i}\,\beta_{2i+1} \sim B^{(0)}_{2i,2i+1}$, and so on.
Furthermore, the transformation of the parafermion operators under the global $\mathbb{Z}_{n} \times \mathbb{Z}_{n}$ symmetry
reads:
\begin{equation}
\label{eq: transformation of alpha and beta -sigma x}
\begin{split}
&\,
S^{(n)}_{x,L}\,\alpha_{j}\,\Big( S^{(n)}_{x,L} \Big)^{-1} = e^{-i\frac{2\pi}{n}}\,\alpha_{j}
\,,~~
S^{(n)}_{x,R}\,\alpha_{j}\,\Big( S^{(n)}_{x,R} \Big)^{-1} =\,\alpha_{j}
\\
&\,
S^{(n)}_{x,L}\,\beta_{j}\,\Big( S^{(n)}_{x,L} \Big)^{-1} = \beta_{j}
\,,~~
S^{(n)}_{x,R}\,\beta_{j}\,\Big( S^{(n)}_{x,R} \Big)^{-1} =e^{i\frac{2\pi}{n}}\,\beta_{j}
\,.
	\end{split}	
\end{equation}

It is important to notice that the parafermions $\alpha_{i}$ in Eq.~\eqref{eq: alpha parafermions - sigma x interface}
contain the vertex operator 
$e^
{
i\,\left( \frac{1}{n}\phi_{2R} - \phi_{1L} \right)
}	
$,
which is a bound state between the right $\mathbb{Z}_{n}$ domain wall creation operator, 
$e^
{
\frac{i}{n}\phi_{2R}
}	
$,
and the left $\mathbb{Z}_{n}$ order operator,
$e^
{
i\phi_{1L}
}	
$.
Similarly, the parafermions $\beta_{i}$ in Eq.~\eqref{eq: beta parafermions - sigma x interface} encode the bound state between the left $\mathbb{Z}_{n}$ domain wall creation operator and the right $\mathbb{Z}_{n}$ order operator,
thus explicitly realizing the notion of domain wall decoration associated with the $\mathbb{Z}_{n} \times \mathbb{Z}_{n}$ SPT state.\cite{Chen-2014,Geraedts,Santos-2015-a}
It is noteworthy that, while in the lattice models of Refs.~\onlinecite{Geraedts,Santos-2015-a}, one can attach a single $\mathbb{Z}_{n}$ domain wall to $p \in \{ 1, ..., n-1\}$ charges of the other $\mathbb{Z}_{n}$ group, which accounts for $n-1$ non-trivial classes of SPT states, the low-energy field theory construction presented in this section only accounts for the $p=1$ class associated with the binding of a domain wall with the unit $\mathbb{Z}_{n}$ charge. In Appendix~\ref{sec: generalized Zn x Zn spt states} we provide the generalization that accounts
for the other classes of SPT states.

Let us now discuss the effect of small local perturbations on the ground state manifold.
If the perturbation preserves the $\mathbb{Z}_{n} \times \mathbb{Z}_{n}$ symmetry, 
then its effect on the zero-mode subspace is to induce symmetry-preserving couplings of the form
$\alpha^{\dagger}_{i}\,\alpha_{j}$,
$\beta^{\dagger}_{i}\,\beta_{j}$, $\alpha^{\dagger}_{i}\,\alpha_{j}\,\beta^{\dagger}_{\ell}\,\beta_{k}$ ($i \neq j$,
$\ell \neq k$) and so on, which, due to the presence of the gapped spectrum, have an exponentially small effect on the ground state degeneracy.
If, however, the $\mathbb{Z}_{n}\times\mathbb{Z}_{n}$ symmetry is broken by a local perturbation, then the local coupling $\alpha^{\dagger}_{i}\,\beta_{i}$ between parafermions at the same domain wall is generically induced, which lifts the ground state degeneracy.

\subsection{$K = \sigma_z$}
\label{sec: sigma z interface}

At the interface with $\mathcal{K}_{z} = \sigma_{z} \oplus (-\sigma_{z})$, 
non-trivial gapped interfaces parametrized by Eq.~\eqref{eq: ML and MR homogeneous interface - general}
are found to give rise to two classes of primitive gapped interfaces.
\\
\\
\subsubsection{First type:}
The first type of gapped interface with $\mathcal{K}_{z} = \sigma_{z}\oplus (-\sigma_{z})$ 
is parametrized by
\begin{equation}
\label{eq: form of ML and MR - sigma z - type 1}
\begin{split}
&\,
\mathcal{M}^{(2m+1)}_{L} = 
\begin{pmatrix}	
m+1 & m
\\
m & m+1
\end{pmatrix}
\,,
\\
&\,
\mathcal{M}^{(2m+1)}_{R} = W_{2}\,\mathcal{M}^{(2m+1)}_{L} 
\,,\quad
W_{2} = \sigma_{z}
\,,\quad
\end{split}
\end{equation}
for any non-zero integer $m$. 
Eq.~\eqref{eq: form of ML and MR - sigma z - type 1} implies, using 
Eqs.~\eqref{eq: relation between det of M and symmetry order} and~\eqref{eq: TEE correction}, 
that the BTES associated with this interface is
\begin{equation}
\label{eq: BTES sigma z type 1}
\{ \gamma^{(2m+1)}_{s} = \log{|2m + 1|}~\,,~ m \in \mathbb{Z}^{*} \}	
\,.
\end{equation}

Eq.~\eqref{eq: form of ML and MR - sigma z - type 1} corresponds to the null vectors
\begin{equation}
\begin{split}
&\,
L^{(2m+1)}_{1} = \left( m+1, m, m+1,m   \right),
\\
&\,
L^{(2m+1)}_{2} = \left( m, m+1, -m, -(m+1)   \right)
\,,
\end{split} 	
\end{equation}
and the local gapping interactions
\begin{equation}
\label{eq: gapping interactions sigma z - type 1}
\begin{split}
&\,
\mathcal{H}_{1,z} = \cos{\Big[ (m+1)\phi_{1L} - m \phi_{2L} - (m+1) \phi_{1R} + m \phi_{2R}   \Big]}	
\\
&\,
\mathcal{H}_{2,z} = \cos{\Big[ m \phi_{1L} - (m+1) \phi_{2L} + m \phi_{1R} - (m+1) \phi_{2R}   \Big]}	
\,.
\end{split}	
\end{equation}
As discussed in Sec.~\ref{sec: General Approach}, the interactions \eqref{eq: gapping interactions sigma z - type 1} are invariant under a $U(1)\times U(1)$ symmetry whose generators are given in Eq.~\eqref{eq: generator U(1) symmetry}. 
However, using the fact that $\mathcal{M}^{(2m+1)}_{L/R}$ are invertible for any
integer $m$ (since $\textrm{det}\Big(\mathcal{M}^{(2m+1)}_{L}\Big) = -\textrm{det}\Big(\mathcal{M}^{(2m+1)}_{R}\Big) = 2m+1$),
we can determine a subgroup of $U(1)\times U(1)$ associated with independent symmetries of the L and R fields.
For that consider the SNF of $\mathcal{M}^{(2m+1)}_{L/R}$, 
$\mathcal{D}^{(2m+1)}_{L/R} = \mathcal{U}^{(2m+1)}_{L/R}\,\mathcal{M}^{(2m+1)}_{L/R}\,\mathcal{V}^{(2m+1)}_{L/R}$,
where
\begin{equation}
\label{eq: SNF of ML for sigmaz}
\begin{split}
&\,
\mathcal{D}^{(2m+1)}_{L}  = \mathcal{D}^{(2m+1)}_{R} 
=
\begin{pmatrix}
2m+1 & 0
\\
0 & 1
\end{pmatrix}
\\
&\,
\mathcal{U}^{(2m+1)}_{L} 
=
\begin{pmatrix}
1 & 1
\\
0 & 1
\end{pmatrix}
\,,
\quad
\mathcal{U}^{(2m+1)}_{R} = \mathcal{U}^{(2m+1)}_{L}\,\sigma_{z}
\\
&\,
\mathcal{V}^{(2m+1)}_{L}  = \mathcal{V}^{(2m+1)}_{R}
=
\begin{pmatrix}
m+1 & -1
\\
-m & 1
\end{pmatrix}
\,.
	\end{split}	
\end{equation}
The first line of Eq.~\eqref{eq: SNF of ML for sigmaz} implies that the 
gapping interactions Eq.~\eqref{eq: gapping interactions sigma z - type 1}
are invariant under a $\mathbb{Z}_{2m+1}\times\,\mathbb{Z}_{2m+1}$ symmetry, whose generators are 
\begin{equation}
\label{eq: left symmetry generator sigma_z}
\begin{split}
S^{(2m+1)}_{z,L}
&\,=
\exp{\left\{\frac{i}{2m+1}\,\int_{\mathcal{I}_{z}}\,dx\, \left[ (m+1) \partial_{x}\phi_{1L} - m \partial_{x}\phi_{2L} \right] \right\} }	
\\
&\,
\simeq
\exp{\left\{-\frac{i\,m}{2m+1}\,\int_{\mathcal{I}_{z}}\,dx\, \left[ \partial_{x}\phi_{1L} + \partial_{x}\phi_{2L} \right] \right\} }
\,
\end{split}
\end{equation}
and
\begin{equation}
\label{eq: right symmetry generator sigma_z}
\begin{split}
S^{(2m+1)}_{z,R}
&\,=
\exp{\left\{\frac{i}{2m+1}\,\int_{\mathcal{I}_{z}}\,dx\, \left[ (m+1) \partial_{x}\phi_{1R} - m \partial_{x}\phi_{2R} \right] \right\} }	
\\
&\,
\simeq
\exp{\left\{-\frac{i\,m}{2m+1}\,\int_{\mathcal{I}_{z}}\,dx\, \left[ \partial_{x}\phi_{1R} + \partial_{x}\phi_{2R} \right] \right\} }
\,,
\end{split}
\end{equation}
where the domain of integration $\mathcal{I}_{z}$ contains the region of the interface that is gapped by the local interactions in Eq.~\eqref{eq: gapping interactions sigma z - type 1}.
In going from the first line to the second of Eqs.~\eqref{eq: left symmetry generator sigma_z} and 
\eqref{eq: right symmetry generator sigma_z}, we have dropped, respectively, the terms 
$\exp{\left( i \int_{\mathcal{I}_{z}}\,dx\, \partial_{x}\,\phi_{1L} \right)}$
and
$\exp{\left( i \int_{\mathcal{I}_{z}}\,dx\, \partial_{x}\,\phi_{1R} \right)}$,
that act trivially on the local fields, and thus can be removed from the definition of the symmetry generators.

In line with the discussion in Sec.~\ref{sec: Emergent Interface Symmetry}, 
we give an interpretation for the null vectors associated with Eq.~\eqref{eq: form of ML and MR - sigma z - type 1} in terms of a gauging process.
Notice that 
\begin{equation}
\label{eq: gauged K-matrix - sigma z}
\begin{split}
K_{2m+1,z} 
&\,\equiv
\mathcal{M}^{(2m+1)}_{L/R}\,\sigma_{z}\,\left( \mathcal{M}^{(2m+1)}_{L/R} \right)^{T}
= 
(2m+1)\,\sigma_{z}
\,,
\end{split}
\end{equation}
where $K_{2m+1,z}$ is identified as the K-matrix of a topologically ordered theory. 
Also notice that the only non-trivial anyonic symmetry of this theory is implemented by 
$W_{2} = \sigma_z$, which transforms the quasiparticle of the $\mathbb{Z}_{2m+1}$ gauge theory
described by the integer vector $\ell = (a,b)^{T}$ into $\ell = (a,-b)^{T}$.
So the relation appearing in Eq.~\eqref{eq: form of ML and MR - sigma z - type 1}, 
between $\mathcal{M}^{(2m+1)}_{L}$ and $\mathcal{M}^{(2m+1)}_{R}$ in terms of the anyonic symmetry 
$W_{2}$, exemplifies the general case (see Eq.~\eqref{eq: ML and MR homogeneous interface - general})
discussed in Sec.~\ref{sec: Emergent Interface Symmetry}.

The $\mathbb{Z}_{2m+1}$ topological phase associated to the K-matrix in Eq.~\eqref{eq: gauged K-matrix - sigma z} 
describes a topological order with deconfined fluxes of the global $\mathbb{Z}_{2m+1}$ 
symmetries given by Eqs.~\eqref{eq: left symmetry generator sigma_z} and \eqref{eq: right symmetry generator sigma_z},
where the symmetry flux associated to the $\mathbb{Z}_{2m+1}$ transformation is 
given by the vertex operator $e^{\frac{i}{2m+1}(\phi_{1,L/R} + \phi_{2,L/R})}$. 
Upon deconfinement of such fluxes, this global $\mathbb{Z}_{2m+1}$ symmetry is promoted to a local gauge symmetry,
and the resulting gauge theory is described by the K-matrix Eq.~\eqref{eq: gauged K-matrix - sigma z}.

Now let us move on to look for topological bound states on domain walls between regions with different gapping interactions. We have the non-trivial gapping terms with null vectors $L_{1}^{2m+1}, L_{2}^{2m+1}$ that will appear on one side of the domain wall. On the other side we will consider trivial symmetric (and primitive) gapping interactions given by:
\begin{equation}
\label{eq: trivial gapping interactions - sigma z}
\begin{split}
&\,
\mathcal{H}^{(0)}_{1,z} = \cos{\left(  \phi_{1 L} + \phi_{2 L} \right)}
\\
&\,
\mathcal{H}^{(0)}_{2,z} = \cos{\left(  \phi_{1 R} + \phi_{2 R} \right)}
\,.
\end{split}
\end{equation}
These interactions are associated with the null vectors,
\begin{equation}
\begin{split}
&\,
L^{(0)}_{1} = (1,-1,0,0)
\,,
\\
&\,
L^{(0)}_{2} = (0,0,1,-1)
\,.
	\end{split}	
\end{equation}

Now consider a series of domain walls separating segments gapped by the local
interactions in Eq.~\eqref{eq: trivial gapping interactions - sigma z}, from segments gapped by 
the interactions in Eq.~\eqref{eq: gapping interactions sigma z - type 1}. Let 
$
R^{z}_{0} = \cup_{i}\,(  x_{2i}+\varepsilon, x_{2i+1}-\varepsilon)
$
be the region gapped solely by the interactions in Eq.~\eqref{eq: trivial gapping interactions - sigma z},
and
$
R^{z}_{2m+1} = \cup_{i}\,(  x_{2i-1}+\varepsilon, x_{2i}-\varepsilon)
$
be the region gapped solely by the interactions in Eq.~\eqref{eq: gapping interactions sigma z - type 1}.
We assume also that the length of each of these segments, $|x_{k+1} - x_{k}| >> \xi$, where $\xi$ is the correlation length
set by the energy gap due to the interactions.
Consider the set of \textit{mutually commuting} operators:
\begin{subequations}
\label{eq: def C 2m+1 and D 2m+1 operators sigma z}
\begin{equation}
C^{(2m+1)}_{2i-1,2i} = 
e^
{
\frac{i}{2(2m+1)}\,\int^{x_{2i}+\varepsilon}_{x_{2i-1}-\varepsilon} dx \left(L^{(0)}_{1} + L^{(0)}_{2} \right)^{T}\mathcal{K}\,\partial_{x}\phi  
},
\end{equation}
\begin{equation}
D^{(2m+1)}_{2i-1,2i} = 
e^
{
\frac{i}{2(2m+1)}\,\int^{x_{2i}+\varepsilon}_{x_{2i-1}-\varepsilon} dx \left(L^{(0)}_{1} - L^{(0)}_{2} \right)^{T}\mathcal{K}\,\partial_{x}\phi  
}.
\end{equation}
\end{subequations}
Moreover, consider another set of \textit{mutually commuting} operators:
\begin{subequations}
\label{eq: def C0 and D0 operators sigma z}
\begin{equation}
C^{(0)}_{2i,2i+1} = 
e^
{
\frac{i}{2m+1}\,\int^{x_{2i+1}+\varepsilon}_{x_{2i}-\varepsilon} dx {L^{(2m+1)}_{2}}^{T}\mathcal{K}\,\partial_{x}\phi  
},
\end{equation}
\begin{equation}
D^{(0)}_{2i,2i+1} = 
e^
{
\frac{i}{2m+1}\,\int^{x_{2i+1}+\varepsilon}_{x_{2i}-\varepsilon} dx {L^{(2m+1)}_{1}}^{T} \mathcal{K}\,\partial_{x}\phi  
}.
\end{equation}
\end{subequations}
Importantly, the non-local operators defined in Eqs.~\eqref{eq: def C 2m+1 and D 2m+1 operators sigma z} 
and \eqref{eq: def C0 and D0 operators sigma z} commute with all terms appearing in the interaction part of the Hamiltonian in
Eqs.~\eqref{eq: trivial gapping interactions - sigma z} and \eqref{eq: gapping interactions sigma z - type 1}.

Making use of the result derived in Appendix~\ref{sec: Commutation relation between vertex operators},
we find that the nontrivial commutation relations among these operators read:
\begin{equation}
\label{eq: commutation relations non-local ops - sigma z}
\begin{split}
&\,
C^{(2m+1)}_{2i-1,2i}\,C^{(0)}_{2j,2j+1}
=
e^{-i\,\frac{2\pi}{2m+1}\left( \delta_{j,i} - \delta_{j,i-1}  \right)}
\,
C^{(0)}_{2j,2j+1}\,C^{(2m+1)}_{2i-1,2i}	
\,,
\\
&\,
D^{(2m+1)}_{2i-1,2i}\,D^{(0)}_{2j,2j+1}
=
e^{-i\,\frac{2\pi}{2m+1}\left( \delta_{j,i} - \delta_{j,i-1}  \right)}
\,
D^{(0)}_{2j,2j+1}\,D^{(2m+1)}_{2i-1,2i}	
\,.
\end{split}
\end{equation}
Eq.\eqref{eq: commutation relations non-local ops - sigma z}
implies two independent sets of $\mathbb{Z}_{2m+1}$ clock operators from which one can identify
the ground state degeneracy $(2m+1)^{g-1} \times (2m+1)^{g-1}$ for a configuration with $2g$ domain walls.
A pair of parafermion operators with quantum dimension $\sqrt{2m+1}$ can be shown to be supported at each
of these domain walls.
We omit the demonstration because it mirrors the one discussed in Sec.~\ref{sec: sigma x interface}.

\subsubsection{Second type:}
The second type of gapped interface with $\mathcal{K}_{z} = \sigma_{z}\oplus (-\sigma_{z})$ 
can be parametrized by
\begin{equation}
\label{eq: form of ML and MR - sigma z - type 2}
\begin{split}
&\,
\mathcal{M}^{(4m)}_{L} = 
\begin{pmatrix}	
2m & 2m
\\
1 & -1
\end{pmatrix}
\,,
\\
&\,
\mathcal{M}^{(4m)}_{R} = W_{2}\,\mathcal{M}^{(4m)}_{L} 
\,,\quad
W_{2} = \sigma_{x}
\,,\quad
\end{split}
\end{equation}
for any non-zero integer $m$. 
Eq.~\eqref{eq: form of ML and MR - sigma z - type 2} implies, using 
Eqs.~\eqref{eq: relation between det of M and symmetry order} and~\eqref{eq: TEE correction}, 
that the BTES associated with this interface is
\begin{equation}
\label{eq: BTES sigma z type 2}
\{ \gamma^{(4m)}_{s} = \log{|4m|}~\,,~ m \in \mathbb{Z}^{*} \}	
\,.
\end{equation}

Eq.~\eqref{eq: form of ML and MR - sigma z - type 2} is associated with the null vectors
\begin{equation}
\begin{split}
&\,
L^{(4m)}_{1} = \left( 2m, 2m, 1, -1  \right),
\\
&\,
L^{(4m)}_{2} = \left( 1, -1, 2m, 2m   \right)
\,,
\end{split} 	
\end{equation}
and the local gapping interactions:
\begin{equation}
\label{eq: gapping interactions sigma z - type 2}
\begin{split}
&\,
\tilde{\mathcal{H}}_{1,z} = \cos{\Big[ 2m \phi_{1L} - 2m \phi_{2L} - \phi_{1R} - \phi_{2R}   \Big]},
\\
&\,
\tilde{\mathcal{H}}_{2,z} = \cos{\Big[ \phi_{1L} + \phi_{2L} - 2m \phi_{1R}  + 2m \phi_{2R}   \Big]}	
\,.
\end{split}	
\end{equation}
The SNF of $\mathcal{M}^{(4m)}_{L/R}$, 
$\mathcal{D}^{(4m)}_{L/R} = \mathcal{U}^{(4m)}_{L/R}\,\mathcal{M}^{(4m)}_{L/R}\,\mathcal{V}^{(4m)}_{L/R}$,
\begin{equation}
\label{eq: SNF of ML for sigmaz - type 2}
\begin{split}
&\,
\mathcal{D}^{(4m)}_{L}  = \mathcal{D}^{(4m)}_{R} 
=
\begin{pmatrix}
4m & 0
\\
0 & 1
\end{pmatrix}
\\
&\,
\mathcal{U}^{(4m)}_{L} 
=
\begin{pmatrix}
1 & 2m
\\
0 & -1
\end{pmatrix}
\,,
\quad
\mathcal{U}^{(4m)}_{R} = \mathcal{U}^{(4m)}_{L}\,\sigma_{x}
\\
&\,
\mathcal{V}^{(4m)}_{L}  = \mathcal{V}^{(4m)}_{R}
=
\begin{pmatrix}
1 & 0
\\
1 & 1
\end{pmatrix}
\,,
	\end{split}	
\end{equation}
implies that the gapping interactions in Eq.~\eqref{eq: gapping interactions sigma z - type 2}
are invariant under a $\mathbb{Z}_{4m}\times\,\mathbb{Z}_{4m}$ symmetry, whose left and right generators are,
respectively,
\begin{equation}
\label{eq: left symmetry generator sigma_z - type 2}
\begin{split}
S^{(4m)}_{z,L}
&\,=
\exp{\left\{\frac{i}{4m}\,\int_{\mathcal{I}^{'}_{z}}\,dx\, \left( \partial_{x}\phi_{1L} + \partial_{x}\phi_{2L} \right) \right\} }	
\,,
\end{split}
\end{equation}
and
\begin{equation}
\label{eq: right symmetry generator sigma_z - type 2}
\begin{split}
S^{(4m)}_{z,R}
&\,=
\exp{\left\{\frac{i}{4m}\,\int_{\mathcal{I}^{'}_{z}}\,dx\, \left( \partial_{x}\phi_{1R} + \partial_{x}\phi_{2R} \right) \right\} }	
\,,
\end{split}
\end{equation}
where the domain of integration $\mathcal{I}^{'}_{z}$ contains the region of the interface that is gapped by the local interactions in Eq.~\eqref{eq: gapping interactions sigma z - type 2}.

In line with the discussion in Sec.~\ref{sec: Emergent Interface Symmetry}, we give an interpretation for the null vectors associated with Eq.~\eqref{eq: form of ML and MR - sigma z - type 2} in terms of a gauging process. 
Notice that 
\begin{equation}
\label{eq: gauged K-matrix - sigma z - type 2}
\begin{split}
K_{4m,x} 
&\,\equiv 
\mathcal{M}^{(4m)}_{L/R}\,\sigma_{z}\,\left( \mathcal{M}^{(4m)}_{L/R} \right)^{T}
= 
4m\,\sigma_{x}
\,,
\end{split}
\end{equation} 
where $K_{4m,x}$ is identified as the K-matrix of the deconfined $\mathbb{Z}_{4m}$ gauge theory.
As such, the $\mathbb{Z}_{4m}$ gauge theory obtained in Eq.~\eqref{eq: gauged K-matrix - sigma z - type 2}
describes a theory where the symmetry fluxes associated with the global $\mathbb{Z}_{4m}$ 
symmetries given by Eqs.~\eqref{eq: left symmetry generator sigma_z - type 2} and~\eqref{eq: right symmetry generator sigma_z - type 2} are deconfined.
The symmetry flux associated to the $\mathbb{Z}_{4m}$ transformation is 
given by the vertex operator $e^{\frac{i}{4m}(\phi_{1,L/R} + \phi_{2,L/R})}$. 
The interactions \eqref{eq: gapping interactions sigma z - type 2} describe a gapped interface 
which can be obtained in the zero thickness limit of the slab of $\mathbb{Z}_{4m}$ gauge theory \eqref{eq: gauged K-matrix - sigma z - type 2},
as depicted in Fig.~\ref{fig: interface-2}(b).

Now consider a series of domain walls that separate segments gapped by the local
interactions in Eq.~\eqref{eq: trivial gapping interactions - sigma z} from segments gapped by 
the interactions in Eq.~\eqref{eq: gapping interactions sigma z - type 2}. Let 
$
R^{z}_{0} = \cup_{i}\,(  x_{2i}+\varepsilon, x_{2i+1}-\varepsilon)
$
be the region gapped solely by the interactions in Eq.~\eqref{eq: trivial gapping interactions - sigma z},
and
$
R^{z}_{4m} = \cup_{i}\,(  x_{2i-1}+\varepsilon, x_{2i}-\varepsilon)
$
be the region gapped solely by the interactions in Eq.~\eqref{eq: gapping interactions sigma z - type 2}.
We assume also that the length of each of these segments, $|x_{k+1} - x_{k}| >> \xi$, where $\xi$ is the correlation length
set by the energy gap due to the interactions.
Consider the set of \textit{mutually commuting} operators:
\begin{subequations}
\label{eq: def C 4m and D 4m operators sigma z - type 2}
\begin{equation}
E^{(4m)}_{2i-1,2i} = 
e^
{
\frac{i}{4m}\,\int^{x_{2i}+\varepsilon}_{x_{2i-1}-\varepsilon} dx {L^{(0)}_{1}}^{T}\mathcal{K}\,\partial_{x}\phi  
},
\end{equation}
\begin{equation}
F^{(4m)}_{2i-1,2i} = 
e^
{
\frac{i}{4m}\,\int^{x_{2i}+\varepsilon}_{x_{2i-1}-\varepsilon} dx {L^{(0)}_{2}}^{T}\mathcal{K}\,\partial_{x}\phi  
}.
\end{equation}
\end{subequations}
In addition, consider another set of \textit{mutually commuting} operators:
\begin{subequations}
\label{eq: def C0 and D0 operators sigma z - type 2}
\begin{equation}
E^{(0)}_{2i,2i+1} = 
e^
{
\frac{i}{4m}\,\int^{x_{2i+1}+\varepsilon}_{x_{2i}-\varepsilon} dx {L^{(4m)}_{2}}^{T}\mathcal{K}\,\partial_{x}\phi  
},
\end{equation}
\begin{equation}
F^{(0)}_{2i,2i+1} = 
e^
{
\frac{i}{4m}\,\int^{x_{2i+1}+\varepsilon}_{x_{2i}-\varepsilon} dx {L^{(4m)}_{1}}^{T} \mathcal{K}\,\partial_{x}\phi  
}.
\end{equation}
\end{subequations}
The non-local operators defined in Eqs.~\eqref{eq: def C 4m and D 4m operators sigma z - type 2} 
and \eqref{eq: def C0 and D0 operators sigma z - type 2} commute with all the local terms of the interacting part of the Hamiltonian in
Eqs.~\eqref{eq: trivial gapping interactions - sigma z} and \eqref{eq: gapping interactions sigma z - type 2}.

Making use of the result derived in Appendix~\ref{sec: Commutation relation between vertex operators},
we find that the only nontrivial commutation relations among these operators reads:
\begin{equation}
\label{eq: commutation relations non-local ops - sigma z - type 2}
\begin{split}
&\,
E^{(4m)}_{2i-1,2i}\,E^{(0)}_{2j,2j+1}
=
e^{-i\,\frac{2\pi}{4m}\left( \delta_{j,i} - \delta_{j,i-1}  \right)}
\,
E^{(0)}_{2j,2j+1}\,E^{(4m)}_{2i-1,2i}	
\,,
\\
&\,
F^{(4m)}_{2i-1,2i}\,F^{(0)}_{2j,2j+1}
=
e^{-i\,\frac{2\pi}{4m}\left( \delta_{j,i} - \delta_{j,i-1}  \right)}
\,
F^{(0)}_{2j,2j+1}\,F^{(4m)}_{2i-1,2i}	
\,.
\end{split}
\end{equation}
Eq.~\eqref{eq: commutation relations non-local ops - sigma z - type 2}
implies two independent sets of $\mathbb{Z}_{4m}$ clock operators from which one can identify
the ground state degeneracy $(4m)^{g-1} \times (4m)^{g-1}$ for a configuration with $2g$ domain walls.
A pair of parafermion operators with quantum dimension $\sqrt{4m}$ can be shown to be supported at each
of these domain walls.
Again, we omit the explicit demonstration that parallels the one provided in Sec.\ref{sec: sigma x interface}.

In summary, the BTES associated with the homogeneous interface with $K = \sigma_z$,
given by Eqs.~\eqref{eq: BTES sigma z type 1} and~\eqref{eq: BTES sigma z type 2}, is identified with the sequence of integers
not congruent with $2$ mod $4$.~\cite{sigmaz-sequence}

\subsection{$1$D SPTs as the boundary of $2$D phases}
\label{sec: folding mechanism}
We now discuss an important implication of the existence of homogeneous self-gappable SPT interfaces. 
Imagine folding the left and right sides of the interface such that it now constitutes the gapped boundary between the bilayer 2D system and the topologically trivial 2D vacuum.
The existence of the gap-generating interactions in Eqs.~\eqref{eq: Zn x Zn interactions - sigma x} or \eqref{eq: trivial gapping interactions - sigma x} implies that the bilayer bosonic 2D system belongs to the trivial $\mathbb{Z}_{n} \times \mathbb{Z}_{n}$ \emph{$2$D SPT class}, since these interactions gap the boundary modes without breaking the $\mathbb{Z}_{n} \times \mathbb{Z}_{n}$ symmetry.
Thus, given a self-gappable 2D topological phase, the bilayer system (obtained from the aforementioned folding mechanism) represents a trivial $2$D SPT phase of matter that supports 1D non-trivial SPT boundary states 
formed by the interactions~\eqref{eq: Zn x Zn interactions - sigma x}. 
Furthermore, as seen in Sec.~\ref{sec: sigma x interface}, zero-dimensional domain walls along the $1$D boundary, 
separating regions gapped by the interactions in Eq.~\eqref{eq: trivial gapping interactions - sigma x}
from those gapped by the interactions in Eq.~\eqref{eq: Zn x Zn interactions - sigma x}, trap 
symmetry-protected ``non-Abelian" zero modes accounting for a non-trivial ground state degeneracy of the system.

Remarkably, we prove in 
Sec.~\ref{sec: gauged phases share TEE correction}
and
Sec.~\ref{sec: proportional K matrices share TEE correction} 
that $2$D Abelian topological phases described by K-matrices that are either gauge-related, or proportional to one another,
can generate the same set of SPT interfaces. As an example, from our analysis of $K=\sigma_x$ above,
we conclude that all topologically ordered bulk phases characterized by
the K-matrix $K_{\textrm{folded}} = p\,\sigma_{x} \oplus (-p\,\sigma_{x})$ 
(obtained from $K_{\textrm{bulk}} = p\,\sigma_{x}$ by the folding procedure) 
can support $\mathbb{Z}_{n} \times \mathbb{Z}_{n}$ 1D SPT boundary phases for all integer $n$.
The discussion above also applies to bulk fermionic phases as discussed in Sec.~\ref{sec: sigma z interface}.
Under the folding mechanism, we conclude that the $2$D phases described by the K-matrix
$K_{\textrm{folded}} = p\,\sigma_{z} \oplus (-p\,\sigma_{z})$ support SPT boundary chains protected by symmetries
$\mathbb{Z}_{2m+1} \times \mathbb{Z}_{2m+1}$ or $\mathbb{Z}_{4m} \times \mathbb{Z}_{4m}$.

\section{Heterogeneous Stable Interfaces}
\label{sec: Heterogeneous Stable Interfaces}

In this Section we more thoroughly study the heterogeneous interface introduced in Sec.~\ref{sec: Emergent Interface Symmetry},
\begin{equation}
\label{eq: heterogeneuos KL and KR - second expression}
K_{L} = K
\,,\quad
K_{R} = G\,K\,G^{T}
\,,
\end{equation}
where $G \in \textrm{GL}(r,\mathbb{Z})$, and specialize to the case where the edge states associated with $K_{L}$ and $K_{R}$ 
are stable in the sense that the L and R edge modes can not be gapped independently.
(Note that if the left bulk phase with K-matrix $K_{L}=K$ has a stable edge, 
the right bulk phase with K-matrix $K_{R} = G\,K\,G^{T}$ also has a stable edge.
To show this, assume the contrapositive, i.e., that the right edge can be gapped by local interactions.
Then there exists a set of integer null vectors $\ell_{i}$, $i \in  \{ 1, ..., r/2 \}$ such that 
$\ell_{i}^{T}\, G\,K\,G^{T} \,\ell_{j} = 0$. 
But then $G^{T}\,\ell_{i}$, $i \in  \{ 1, ..., r/2 \}$ is a set of integer null vectors associated with local interactions which gap the left edge.)
While an edge theory with non-zero chiral central charge always satisfies this property, the edge states of certain non-chiral bulk phases can also be stable, if the phase supports fractionalized excitations in the bulk and at the edge, as discussed in Ref.~\onlinecite{Levin-2013}.

The stable property of the left and right phases implies that the interface can only be gapped by correlated backscattering processes that mix the left and right modes across the interface. 
These are characterized by the null vectors in Eq~(\ref{eq: null vectors heterogeneous interface - anyonic symmetry}), which we repeat here for convenience:
\begin{equation}
\label{eq: null vectors - stable heterogeneous interface}
\Lambda^{T}_i = \left( e^{T}_{i}\,W\,G,~~  e^{T}_{i} \right)^{T}
\,,
\end{equation}
where $W \in \textrm{SL}(r,\mathbb{Z})$ is any anyonic symmetry of $K_{R} = G\,K\,G^{T}$, i.e., $W\,K_{R}\,W^{T} = K_{R}$.
The null vectors in Eq.~(\ref{eq: null vectors - stable heterogeneous interface}) correspond to the local gapping interactions,
\begin{equation}
\label{eq: interactions - stable heterogeneous interface}
\mathcal{H}_i = 
\cos{\left( e^{T}_i\,W\,G\,K\,\phi_{L}  - e^{T}_i G\,K\,G^{T} \phi_{R}   \right)}
\,,
\end{equation}
for $i = 1, ..., r$.
The interactions in Eq.~\eqref{eq: interactions - stable heterogeneous interface} are invariant
under a discrete global symmetry $\mathcal{G}_{L}$ acting on the left modes of the interface, such that 
$|\mathcal{G}_{L}| = |\textrm{det}(W\,G)|= |\textrm{det}(G)|$.

\begin{figure}[h!]
\includegraphics[width=0.5\textwidth]{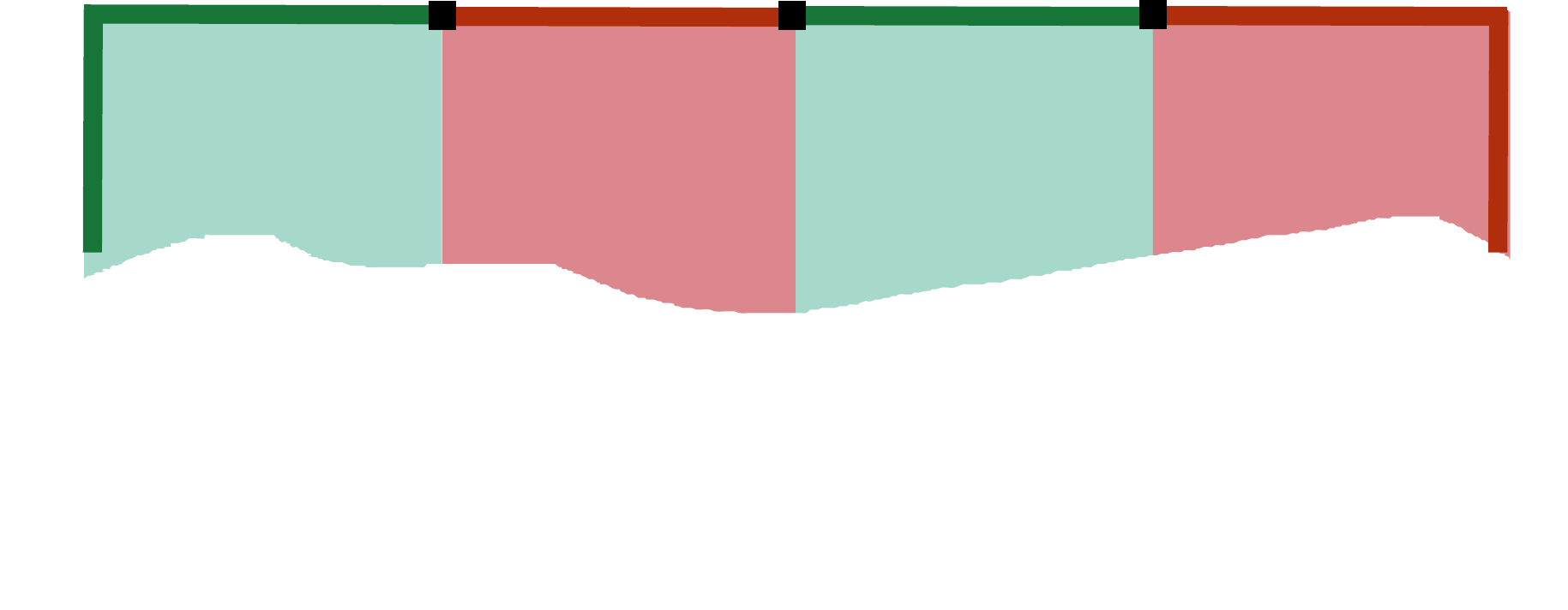}
\caption
{
(Color online)
An array of $2$D topological phases associated to the matrices, $K$ (green) and $G\,K\,G^{T}$ (red), as defined in 
Eq.~\eqref{eq: heterogeneuos KL and KR - second expression}.
Each bulk phase supports  stable edges states represented by thick green and red lines.
Each interface running along the vertical direction is gapped by the interaction in Eq.\eqref{eq: interactions - stable heterogeneous interface}, which terminates at a point of transition (black dot) 
between the  stable edge states (running along the horizontal direction) supporting distinct sets of low-energy quasiparticle excitations. 
}
\label{fig: heterogeneous-dws}
\end{figure}

When $|\textrm{det}(G)| > 1$, the two phases associated to the K-matrices in Eq.~\eqref{eq: heterogeneuos KL and KR - second expression} have distinct quasiparticle spectra. 
If the theory is defined on the lower-half plane (with the topologically trivial vacuum in the upper-half plane) as shown in Fig.~\ref{fig: heterogeneous-dws}, any gapped interface formed by the interactions in Eq.~\eqref{eq: interactions - stable heterogeneous interface} 
terminates at a domain wall between one dimensional gapless edge states that support different low-energy quasiparticle excitations. 
The transition between these two distinct types of stable edge states, running along the horizontal axis (with coordinate $y$) depicted in Fig.~(\ref{fig: heterogeneous-dws}), can be accessed with a theory controlled by the effective $3r \times 3r$ block diagonal K-matrix:
\begin{equation}
\label{eq: effective K-matrix chiral edge transitions}
\mathcal{K}_{\textrm{t}} = 
\textrm{diag}(K_1, K_2, K_3)	
=
\textrm{diag}(K,~G\,K\,G^{T}, -K)	
\,,
\end{equation}
where $K_{1}$, $K_{2},$ and $K_{3}$ are $r \times r$ K-matrices associated with 
the $r$-component fields $\phi_{1}$, $\phi_{2},$ and $\phi_{3}$, respectively. We 
denote by $\Phi(t,y) = (\phi_{1}(t,y), \phi_{2}(t,y), \phi_{3}(t,y))$ the total $3r$-component field content
of the edge states running along the $y$ axis.
The rationale behind the effective transition K-matrix in Eq.~\eqref{eq: effective K-matrix chiral edge transitions}
is the following: if local interactions gap the $\phi_{2}$ and $\phi_{3}$ modes,
then the remaining gapless chiral modes $\phi_{1}$ describe the chiral edge states of the topological phase
to the left of the interface.
Alternatively, if local interactions gap the $\phi_{1}$ and $\phi_{3}$ modes
then the remaining gapless chiral modes $\phi_{2}$ describe the chiral edge states of the topological phase
to the right of the interface. 
By ``combining" the two aforementioned types of gapping processes, 
we can probe the universal properties of the transition between the chiral theories described by $\phi_1$ and $\phi_2$.
In the following we show that the two aforementioned interactions that partially gap a sector of the 
low energy modes of $\mathcal{K}_{\textrm{t}}$ - associated with locking the modes $(\phi_1, \phi_3)$ and $(\phi_2, \phi_3)$ - respect a discrete symmetry and can be interpreted, respectively, as trivial and non-trivial gapped SPT chains
supporting domain wall parafermions.

We now discuss in detail the two distinct gapless phases that descend from the theory with effective K-matrix given in Eq.~\eqref{eq: effective K-matrix chiral edge transitions}.
The interactions responsible for gapping the low-energy modes of the second and third blocks
of the K-matrix Eq.~\eqref{eq: effective K-matrix chiral edge transitions} are
\begin{equation}
\label{eq: gapping interactions - stable edge - type 1}
\mathbb{H}_i = 
\cos{\left( e^{T}_i G\,K\,G^{T} \phi_{2} - e^{T}_i\,W\,G\,K\,\phi_{3}   \right)}
\,,
\end{equation}
which are parametrized by the null vectors
\begin{equation}
\mathbb{L}^{T}_{i} = \left( 0,~~  e^{T}_{i}, ~~ e^{T}_{i}\,W\,G \right)
\,,
\end{equation}
for $i = 1, ..., r$.
To instead gap the first and third blocks of the K-matrix in Eq.~\eqref{eq: effective K-matrix chiral edge transitions}, we can add the interactions,
\begin{equation}
\label{eq: gapping interactions - stable edge - type 2}
\tilde{\mathbb{H}}_i = 
\cos{\left[ e^{T}_{i}\,K\,\left( \phi_{1} - \phi_{3} \right)  \right]}
\,,
\end{equation}
which are parametrized by the null vectors,
\begin{equation}
\tilde{\mathbb{L}}_{i} = \left( e^{T}_{i},~~  0,~~ e^{T}_{i} \right)
\,,
\end{equation}
for $i = 1, ..., r$.

We can parametrize the SNF of $WG$ as 
\begin{equation}
\begin{split}
&\,
\mathbb{D} = \mathbb{U}\,(W\,G)\,\mathbb{V} = \textrm{diag}(d_{1}, ..., d_{r})
\,,
\end{split}
\end{equation} 
with  $\mathbb{U}, \mathbb{V} \in \textrm{SL}(r,\mathbb{Z})$.
The interactions in Eq.~\eqref{eq: gapping interactions - stable edge - type 1} and \eqref{eq: gapping interactions - stable edge - type 2} are then invariant under the global symmetry group
$\mathbb{G} = \prod^{r}_{i = 1}\mathbb{Z}_{d_{i}}$ with $\mathbb{Z}_{d_{p}}$ symmetry generators given by
\begin{equation}
\begin{split}
&\,
\mathbb{S}^{(p)} 
= 
\exp{\left[ i\,\int_{\mathbb{I}}\,dy\, \left( \alpha^{(p)} \right)^{T} \partial_{x} (\phi_{1} - \phi_{3})(y) \right]}
\,,
\\
&\,
\alpha^{(p)} 
=
\frac{1}{d_{p}}\,\mathbb{V}\,e_{p}
\,,\quad
p = 1, ..., r
\,.
\end{split}
\end{equation}
The domain of integration $\mathbb{I}$ contains the regions of the boundary that are gapped by the local interactions in
Eqs.~\eqref{eq: gapping interactions - stable edge - type 1} and \eqref{eq: gapping interactions - stable edge - type 2}.

We now consider domain walls separating the edge states of $K_{L}$ and $K_{R}$.
Let 
$
\mathbb{R} = \cup_{i}\,(  y_{2i}+\varepsilon, y_{2i+1}-\varepsilon)
$
be the region describing the edge state of $K_{L}$ where the interactions in
Eq.~\eqref{eq: gapping interactions - stable edge - type 1} gap the modes $\phi_{2}$ and $\phi_{3}$,
and let
$
\tilde{\mathbb{R}} = \cup_{i}\,(  y_{2i-1}+\varepsilon, y_{2i}-\varepsilon)
$
be the region describing the edge state of $K_{R}$ where the interactions in Eq.~\eqref{eq: gapping interactions - stable edge - type 2} gap the modes $\phi_{1}$ and $\phi_{3}$.
We can define two sets of non-local operators acting on finite segments of the edge of the system,
which commute with all the interactions in Eqs.~\eqref{eq: gapping interactions - stable edge - type 1} and
\eqref{eq: gapping interactions - stable edge - type 2}.
The first set is given by
\begin{subequations}
\begin{equation}
\label{eq: S operators - heterrogeneous interface}
\begin{split}
&\,
\mathbb{S}^{(p)}_{2i,2i+1} = 
e^
{
\frac{i}{d_{p}} \int^{y_{2i+1}+\varepsilon}_{y_{2i}-\varepsilon} dy\, e^{T}_{p}\,V^{T}\,\partial_{y} (\phi_{1} - \phi_{3}) 
}
\,,
\\
&\,
p = 1, ..., r
\,,
\end{split}
\end{equation}
\begin{equation}
\begin{split}
&\,
[ \mathbb{S}^{(p)}_{2i,2i+1}, \mathbb{S}^{(p')}_{2j,2j+1} ] = 0 
\,,
\quad
\forall~ (p,p')~ \textrm{and}~(i,j)
\,,
\end{split}
\end{equation}
\end{subequations}
which correspond to operators that measure the $\mathbb{Z}_{d_{p}}$ symmetry quantum numbers associated with 
the segment $(y_{2i}, y_{2i+1})$. 
The trivial commutation relations between the operators $\mathbb{S}^{(p)}_{2i,2i+1}$ 
defined by Eq.~\eqref{eq: S operators - heterrogeneous interface} and the interactions in 
Eqs.~\eqref{eq: gapping interactions - stable edge - type 1} and \eqref{eq: gapping interactions - stable edge - type 2}
are a consequence of the fact that these interactions respect the global $\prod^{r}_{p = 1}\,\mathbb{Z}_{d_{p}}$ symmetry.

Now consider another set of operators:
\begin{subequations}
\begin{equation}
\label{eq: O operators - heterrogeneous interface}
\begin{split}
&\,
\mathbb{O}^{(p)}_{2i-1,2i} = 
e^
{
\frac{i}{d_{p}} \sum^{r}_{m=1} \mathbb{U}_{p m} 
\int^{y_{2i}+\varepsilon}_{y_{2i-1}-\varepsilon} dy\, \mathbb{L}^{T}_{m}\,\mathcal{K}_{\textrm{t}}\,\partial_{x}\Phi
}
\,,
\\
&\,
p = 1, ..., r
\,,
\end{split}
\end{equation}
\begin{equation}
\begin{split}
&\,
[ \mathbb{O}^{(p)}_{2i-1,2i}, \mathbb{O}^{(p')}_{2j-1,2j} ] = 0 
\,,
\quad
\forall~ (p,p')~ \textrm{and}~(i,j)
\,.
\end{split}
\end{equation}
\end{subequations}
The only possible non-trivial action of the operators $\mathbb{O}^{(p)}_{2i-1,2i}$  on the interactions in Eqs.~\eqref{eq: gapping interactions - stable edge - type 1} and \eqref{eq: gapping interactions - stable edge - type 2} is to act as a shift operator for the terms $\cos{\left(\tilde{\mathbb{L}}_{a}^{T}\,\mathcal{K}_{\textrm{t}}\,\Phi(y)\right)}$
for $y \in [2i-1, 2i]$, in which case, it can be checked that 
$
\mathbb{O}^{(p)}_{2i-1,2i}
\,
\cos{\left(\tilde{\mathbb{L}}_{a}^{T}\,\mathcal{K}_{\textrm{t}}\,\Phi(y)\right)}
\,
\Big(\mathbb{O}^{(p)}_{2i-1,2i}\Big)^{-1}
=
\cos{\left(\tilde{\mathbb{L}}_{a}^{T}\,\mathcal{K}_{\textrm{t}}\,\Phi(y) - 2\pi\,(\mathbb{V}^{-1} K)_{p a}\right)}
$.
However, since $\mathbb{V}^{-1} K \in \textrm{GL}(r,\mathbb{Z})$, this shift is an integer multiple of $2\pi$.
This implies trivial commutation relations of the operators $\mathbb{O}^{(p)}_{2i-1,2i}$ with the interaction terms.

Making use of the result derived in Appendix \ref{sec: Commutation relation between vertex operators},
the commutation relations between the two sets of operators in Eqs.~\eqref{eq: S operators - heterrogeneous interface}
and \eqref{eq: O operators - heterrogeneous interface} read:
\begin{equation}
\label{eq: commutation O and S -heterogeneous interface}
\begin{split}
&\,
\mathbb{O}^{(q)}_{2i-1,2i}
\,
\mathbb{S}^{(p)}_{2j,2j+1}
=
e^
{
\frac{2\pi i}{d_{p}} \delta_{p,q}	
\left(
\delta_{j,i} -\delta_{j,i-1} 
\right)
}
\,
\mathbb{S}^{(p)}_{2j,2j+1}
\,
\mathbb{O}^{(q)}_{2i-1,2i}
\,,
\\
&\,
p, q = 1, ..., r
\,.
\end{split}	
\end{equation}
Eq.~\eqref{eq: commutation O and S -heterogeneous interface} encodes $r$ independent algebras
since $\mathbb{O}^{(q)}$ and $\mathbb{S}^{(p)}$ commute whenever $p \neq q$. 
For each value of $p$, $\mathbb{O}^{(p)}_{2i-1,2i}$ ($\mathbb{O}^{(p)}_{2i+1,2i+2}$) acts as a raising (lowering) operator for the $d_{p} > 1$ possible eigenvalues of the symmetry operator $\mathbb{S}^{(p)}_{2i,2i+1}$ defined on the neighboring segment.
In the presence of $2g$ domain walls separating the edge states of $K_{L}$ and $K_{R}$,
the ground state degeneracy equals $\prod^{r}_{p = 1}\,(d_{p})^{g-1}$, which is associated with the presence of
$r$ independent parafermion zero modes with quantum dimension $\sqrt{d_{p}}$ for $p=1,2,\ldots r.$

We note that the algebra \eqref{eq: commutation O and S -heterogeneous interface} and, the ground state degeneracy 
that follows from it, are closely related to the results of Cano {\it et al.} in Ref.~\onlinecite{Cano-2015-a}, who studied
transitions along topologically distinct chiral edge states of the same bulk phase. 
The physical scenario we consider here does differ from that of Ref.~\onlinecite{Cano-2015-a} in that we
consider the formation of a gapped interface via the local interactions in Eq.~\eqref{eq: interactions - stable heterogeneous interface}, which stabilize an SPT interface that terminates at each chiral edge transition.
The global symmetry respected by these local gapping interactions is a consequence of the gauging relation between the  topological phases forming the interface (see Eq.\eqref{eq: heterogeneuos KL and KR - second expression}). Due to the distinct topological phases, the gapped interface forms an kind of anyon Andreev reflector, since only certain quasiparticles can be fully transmitted across the interface, while other quasiparticles are partially transmitted and reflected by the condensate at the interface. The Andreev reflecting SPT interfaces discussed in this section generalize those gapped interfaces separating one-component chiral Abelian topological phases discussed in Ref.~\onlinecite{Santos-2017}.

\section{Homogeneous interface between two chiral phases}
\label{sec: Homogeneous chiral interfaces}
Let us now move on to discuss the last interface type, a homogeneous interface between two chiral topological phases. The simplest type of interface one might imagine is an interface between two one-component Laughlin quantum Hall theories with a $K$-matrix given by a single integer. However, this type of interface does not give rise to a robust left-right emergent symmetry because the allowed tunneling terms between the two sides of the interface are strongly constrained by the primitivity condition. Instead, to uncover a possible emergent interface symmetry we need to consider multiple components. For our purposes we will consider an interface between two $\nu=2$ integer quantum Hall states, and our methods can be straightforwardly adapted to other theories.
\subsection{$K = \textrm{I}_{2\times 2}$ interface}
\label{sec: nu = 2 interface}
At the interface defined by the K-matrix, $\mathcal{K}_{\nu=2} = \textrm{I}_{2 \times 2} \oplus (-\textrm{I}_{2 \times 2})$, 
non-trivial gapped interfaces of the form given in Eq.~\eqref{eq: ML and MR homogeneous interface - general}
are found to be parametrized by
\begin{equation}
\label{eq: form of ML and MR nu 2}
\begin{split}
&\,
\mathcal{M}^{(m)}_{\nu=2,L} = 
\begin{pmatrix}	
a & b
\\
b & -a
\end{pmatrix}
\,,
\\
&\,
\mathcal{M}^{(m)}_{\nu=2,R} = W_{2}\,\mathcal{M}^{(m)}_{\nu=2,L} 
\,,\quad
W_{2} = \sigma_{z}
\,,\quad
\end{split}
\end{equation}
for $a, b \in \mathbb{Z}$ and $\textrm{gcd}(a,b) = 1$ such that 
\begin{equation}
\label{eq: condition on (a,b) nu 2}
\begin{split}
&\,
a^2 + b^2 = m
\,,
\end{split}	
\end{equation}
where $m = 5, 13, 17, 25, ...$ are numbers that are divisible only by primes congruent to $1$ mod $4$.~\cite{nu2-sequence}
The primitive null vectors associated with Eq.\eqref{eq: form of ML and MR nu 2} are
\begin{equation}
\begin{split}
&\,
\Lambda^{(m)}_{\nu=2,1} = \left( a, b, a, b  \right)
\,,
\\
&\,
\Lambda^{(m)}_{\nu=2,2} = \left( b, -a, -b, a   \right)
\,,
\end{split} 	
\end{equation}
such that the local gapping interactions read
\begin{equation}
\label{eq: gapping interactions nu 2}
\begin{split}
&\,
\mathcal{H}_{\nu=2,1} = \cos{\Big[ a \phi_{1L} + b \phi_{2L} -a \phi_{1R} -b \phi_{2R}   \Big]}	
\,,
\\
&\,
\mathcal{H}_{\nu=2,2} = \cos{\Big[ b \phi_{1L} -a \phi_{2L} +b \phi_{1R}  -a \phi_{2R}   \Big]}	
\,.
\end{split}	
\end{equation}

The primitive property of the interactions in Eq.~\eqref{eq: gapping interactions nu 2} can be checked by computing the minors
of the rectangular matrix, $\Big( \mathcal{M}^{(m)}_{\nu=2,l}, \mathcal{M}^{(m)}_{\nu=2,R}  \Big)$, which are 
$\{ \pm (a^2+b^2), \pm 2 a\,b, \pm (a^2 - b^2)\}$. 
Eq.~\eqref{eq: condition on (a,b) nu 2} ensures that the greatest
common divisor of this set is equal to one, hence the primitivity condition is satisfied.

The matrices in Eq.~\eqref{eq: form of ML and MR nu 2} satisfy
$\textrm{det}\Big(\mathcal{M}^{(m)}_{\nu=2,R}\Big) = -\textrm{det}\Big(\mathcal{M}^{(m)}_{\nu=2,L}\Big) = a^2+b^2 = m$,
which encodes the fact that the TEE correction associated with this gapped interface equals $\log(m)= \log(a^{2} + b^{2}),$
and that the interactions in Eq.~\eqref{eq: gapping interactions nu 2} are invariant under a discrete 
$\mathbb{Z}_{a^2 + b^2} \times \mathbb{Z}_{a^2 + b^2}$ symmetry. 

In order to simplify the discussion, we shall take hereafter $a =1$ and $b = 2n$ ($n \in \mathbb{Z}$) such that $m = 1+4n^{2}$.
Then, the SNF 
$\mathcal{D}^{}_{\nu=2,L/R} = \mathcal{U}^{}_{\nu=2,L/R}\,\mathcal{M}^{}_{\nu=2,L/R}\,\mathcal{V}^{}_{\nu=2,L/R}$ 
is given by
\begin{equation}
\label{eq: SNF of ML for nu 2}
\begin{split}
&\,
\mathcal{D}_{\nu=2,L}  = \mathcal{D}_{\nu=2,R} 
=
\begin{pmatrix}
1+4\,n^2 & 0
\\
0 & 1
\end{pmatrix},
\\
&\,
\mathcal{U}_{\nu=2,L} 
=
\begin{pmatrix}
1 & 2n
\\
0 & -1
\end{pmatrix}
\,,
\quad
\mathcal{U}_{\nu=2,R} = \mathcal{U}_{\nu=2,L}\,\sigma_{z},
\\
&\,
\mathcal{V}_{\nu=2,L}  = \mathcal{V}_{\nu=2,R}
=
\begin{pmatrix}
1 & 0
\\
2n & 1
\end{pmatrix}
\,.
	\end{split}	
\end{equation}
Eq.~\eqref{eq: SNF of ML for nu 2} implies that the gapping interactions in Eq.~\eqref{eq: gapping interactions nu 2}
are invariant under a $\mathbb{Z}_{1+4n^2}\times\,\mathbb{Z}_{1+4n^2}$ symmetry, whose left and right generators are, respectively, 
\begin{equation}
\label{eq: left symmetry generator nu 2}
\begin{split}
S^{(1+4n^2)}_{\nu=2,L}
&\,=
\exp{\left\{\frac{i}{1+4n^2}\,\int_{\mathcal{I}_{\nu=2}}\,dx\, \left( \partial_{x}\phi_{1L} + 2n\,\partial_{x}\phi_{2L} \right) \right\} }	
\,,
\end{split}
\end{equation}
and
\begin{equation}
\label{eq: right symmetry generator nu 2}
\begin{split}
S^{(1+4n^2)}_{\nu=2,R}
&\,=
\exp{\left\{\frac{i}{1+4n^2}\,\int_{\mathcal{I}_{\nu=2}}\,dx\, \left( \partial_{x}\phi_{1R} + 2n\,\partial_{x}\phi_{2R} \right) \right\} }	
\,,
\end{split}
\end{equation}
where the domain of integration $\mathcal{I}_{\nu=2}$ contains the region of the interface that is gapped by the local interactions in
Eq.~\eqref{eq: gapping interactions nu 2} (with $a=1$ and $b = 2n$). 

In line with the discussion in Sec.~\ref{sec: Emergent Interface Symmetry}, 
we give an interpretation for the null vectors associated with
Eq.~\eqref{eq: gapping interactions nu 2} in terms of a gauging process. Notice that 
\begin{equation}
\label{eq: gauged K-matrix - nu 2}
\begin{split}
K_{(m,m,0)} 
&\,\equiv 
\mathcal{M}^{(m)}_{\nu=2,L/R} 
\,
\textrm{I}_{2 \times 2}
\,
\left( \mathcal{M}^{(m)}_{\nu=2,L/R}  \right)^{T}
= 
m\,\textrm{I}_{2 \times 2}
\,,
\end{split}
\end{equation}
where $K_{(m,m,0)}$ represents the K-matrix of a chiral bilayer topologically
ordered state, which we denote by $(m,m,0)$ in connection with the Halperin bilayer FQH states.
The gauge theory associated to the K-matrix in Eq.~\eqref{eq: gauged K-matrix - nu 2} describes a phase where the 
symmetry fluxes associated with the global $\mathbb{Z}_{m}$ symmetries given by 
Eqs.~\eqref{eq: left symmetry generator nu 2} and \eqref{eq: right symmetry generator nu 2} are deconfined.
The symmetry flux associated to the $\mathbb{Z}_{m}$ transformation is given by the vertex operator 
$e^{\frac{i}{1+4n^2}\,\left( \partial_{x}\phi_{1,L/R} + 2n\,\partial_{x}\phi_{2,L/R} \right)}$. 

Similar to the non-trivial gapped interfaces discussed in Sec.~\ref{sec: Homogeneous Self-Gappable interfaces}, 
the gapped interface formed by the interactions in Eq.~\eqref{eq: gapping interactions nu 2} can be realized as the 
zero thickness limit of a topological slab with K-matrix Eq.~\eqref{eq: gauged K-matrix - nu 2} that is placed between 
$\nu=2$ states, as shown in Fig.~\ref{fig: interface-2}.
Since each interface between the $\nu=2$ and the $(m,m,0)$ states is gapped by interactions that are 
$\mathbb{Z}_{m}$ symmetric [see Fig.~(\ref{fig: interface-2}a)], as the thickness of the $(m,m,0)$ slab goes to zero, one recovers local gapping interactions in Eq.\eqref{eq: gapping interactions nu 2} between the left and right $\nu=2$ degrees of freedom,
which possess a $\mathbb{Z}_{m}\times \mathbb{Z}_{m}$ symmetry [see Fig.~(\ref{fig: interface-2}b)].
However, in contrast to the situation investigated in Sec.~\ref{sec: Homogeneous Self-Gappable interfaces}, where
we identified trivial, symmetry-preserving gapped interfaces that were associated with gapping processes that did not mix left and 
right modes across the interface, in chiral phases such as the $\nu=2$ phase, there is no such gapped interface, since one cannot gap each side of the interface independently. The same situation holds for non-chiral phases with stable edge states.
Therefore, in chiral and in stable bulk phases, a non-trivial gapped homogeneous interface 
either terminates at a point along the gapless boundary of the system 
or it can terminate at another gapped region in the bulk which is associated with a different symmetry. 

Applying the discussion of Sec.~\ref{sec: Heterogeneous Stable Interfaces} to the present case, we conclude that
an array of $\nu = 2$ states intercalated by $(m,m,0)$ states (as shown in Fig.~(\ref{fig: heterogeneous-dws})) 
hosts a $\mathbb{Z}_{m}$ parafermion mode at the end points of the gapped interface. 
Thus, as the thickness of the $(m,m,0)$ slab goes to zero, the situation described in Fig.~(\ref{fig: heterogeneous-dws}) reduces 
to a sequence of $\mathbb{Z}_{m} \times \mathbb{Z}_{m}$ gapped interfaces between $\nu=2$ bulk phases, each of which hosts a pair of $\mathbb{Z}_{m}$ parafermions along the chiral edge. If the tunneling interactions between the two neighboring $\nu=2$ phases are primitive then the $\mathbb{Z}_{m} \times \mathbb{Z}_{m}$ symmetry is not broken (explicitly or spontaneously) and the pair of parafermions are forbidden to couple to each other.

\section{General Properties of Boundary Topological Entanglement Sequences}
\label{sec:general}
One of the primary goals of this article is to clarify the origin of the constant, subleading contributions to the entanglement entropy. 
It was predicted in Ref. \onlinecite{cano-2015} using the wire-construction, and later confirmed using Chern-Simons calculations,\cite{Fliss2017} that the subleading correction to the area law is not unique for an Abelian topological phase. At first this may seem problematic, especially since the topological entanglement entropy is a typical diagnostic for identifying topological phases in numerical simulations. Indeed, what we have discovered is that the precise value of the subleading correction to the area law depends on how the topological phase is created microscopically, i.e., it depends on the nature of the interactions that give rise to the topological phase. Thus, different microscopic (e.g., lattice) realizations of a given topological phase may yield different values for the TEE. However, despite this detail dependence, the contribution to the subleading correction still only depends on universal, topological data in the $K$-matrix as we will illustrate below.

In addition, one of the shortcomings of the conventional TEE is that it is a single number and, by itself, it cannot precisely distinguish and classify topological phases. Our results show that a given Abelian topological phase can support an entire sequence of possible universal contributions to the entanglement entropy.
This sequence might be a useful distinguishing characteristic for a given topological phase. 
We have given some examples above where we can calculate the full entanglement sequence, e.g., for interfaces with $K=\sigma^x$ we showed that the sequence is the full set of integers, but in general it is a very difficult exercise in number theory.

In this section we will provide some additional intuition about the nature of the entanglement correction in the context of the possible use of the entanglement sequences to classify topological phases. First, to show that the correction only depends on topological data, we will provide a geometric construction that illustrates the mathematical origin of the correction to the entanglement entropy. Then we will prove two general results about the sequence of entanglement corrections: (i) that two topological phases that are related by a gauging procedure, or (ii) have proportional $K$-matrices share the same sequence of corrections to the entropy. Finally, we will make comments about the utility and limitations of the entanglement sequences as a classification tool.

\subsection{Emergent symmetry and entanglement correction from lattice maps}
Each $K$-matrix specifying an Abelian topological phase also has a geometric interpretation. The rows of $K$ can be treated as integer basis vectors for a Bravais lattice, where the lattice points represent the local-particle excitations of the theory. We can think of the gapping terms at an interface as maps between the lattice generated by $K_L$ and that generated by $K_R.$ The null vectors that specify the gapping terms provide such a map between both sides as follows. 
Consider the set of integer vectors $\left(v^{L/R}_{i}\right)^{T} = e^{T}_{i}\cdot K_{L/R}$, $i = 1, ..., r$, 
formed by the rows of $K_{L/R}$. Since $K_{L/R}$ are invertible, $\{ v^{L/R}_{i}, i = 1, ..., r \}$ are two sets of linearly independent vectors such that $\psi^{L/R}_{i} \equiv e^{i \left(v^{L/R}_{i}\right)^{T} \phi_{L/R}}$ can be thought
as a basis of local quasiparticle operators. More precisely, $v^{L/R}_{i}$ generate independent translations 
vectors on the integer lattice, which connect points which are equivalent up to local particles.
Then $\mathsf{V}_{l/R} = |\textrm{det}(K_{L/R})|$ correspond to the volume of the unit cell generated by these
vectors, such that points within the unit cell represent topologically distinguishable quasiparticles.

Whenever $\mathcal{M}_{L/R}$ are non-singular, we can recast the 
null equation \eqref{eq: null condition matrix form} as
\begin{equation}
K_{L} = \mathcal{O}\,K_{R}\,\mathcal{O}^{T}	
\,,
\quad 
\mathcal{O} = \mathcal{M}_{L}^{-1} \mathcal{M}_{R}
\,,
\end{equation}
from which follows the relationship
\begin{equation}
v^{L}_{i} = \sum^{r}_{j=1} \mathcal{O}_{ij} \left( \mathcal{O} v^{R}_{j} \right)	
\end{equation}
between the generators of local operators on the L and R topological phases. This equation indicates that gapping interactions are maps from one lattice to another such that the ratio of volume of the left and right unit cells (as described in the previous paragraph) is $|\textrm{det}(\mathcal{O})|^{2}$, which is the square of a rational number. This relation then establishes a commensurability condition that allows for the formation of a bosonic condensate at the gapped interface.

The lattice points cannot be topologically distinguished from one another, they are all local and hence topologically equivalent to the vacuum state. However, when a given set of gapping interactions generates an emergent symmetry it allows for a subset of the local particles to be distinguished by symmetry labels of the emergent symmetry. To unveil this property we proceed to determine the symmetry charges of the local particle. Recalling from Sec.~\ref{sec: Left/right discrete symmetries} that $\mu = L, R$ local particles 
$\psi_{\mu, \ell} = \exp{(i \ell_{\mu}^{T} K_{\mu} \phi_{\mu})}$, $\ell_{\mu} \in \mathbb{Z}^{r}$, transform as
\begin{equation}
\psi_{\mu, \ell} \rightarrow 
e^{i 2\pi \left( \mathcal{V}_{\mu}^{T} \ell_{\mu} \right)^{T} \,\mathcal{D}_{\mu}^{-1}\, \left(\mathcal{U}_{\mu}\,t_{\mu}\right)}
\psi_{\mu, \ell}
\,,
\end{equation}
where $\mathcal{D}_{\mu} = \textrm{diag}(d_{1, \mu}, ..., d_{r, \mu})$,
we identify the generator of the $\mathbb{Z}_{d_{j, \mu}}$ symmetry with the vector
$t_{\mu, j} = \mathcal{U}_{\mu}^{-1} e_{j}$ from which the $\mathbb{Z}_{d_{j, \mu}}$ symmetry transformation follows
\begin{equation}
\mathbb{Z}_{d_{j, \mu}}:
\quad
\psi_{\mu, \ell} \rightarrow 
e^{i \frac{2\pi}{d_{j, \mu}} \left( e^{T}_{j} \cdot \mathcal{V}_{\mu}^{T} \ell_{\mu} \right)}\,
\psi_{\mu, \ell}
\,.
\end{equation}
This equation implies that the local particles carrying charge $e^{\frac{i 2\pi p}{d_{j, \mu}}}$
are parametrized by the integer vectors
\begin{equation}
\label{eq: integer vector Zd symmetry}
\begin{split}
&\,	
\ell^{(p)}_{j, \mu}	
=
\Big(\mathcal{V}_{\mu}^{T}\Big)^{-1}\,\Big( p\,e_{j} + u_{j}\Big)
\,,
\quad
e^{T}_{j} \cdot u_{j} = 0
\,,
\\
&\,
p = 0, ..., d_{j, \mu}-1
\,.
\end{split}
\end{equation}
Any two local particles whose integer vectors differ by $\ell^{(p)}_{j, \mu}$ differ by
$e^{\frac{i 2\pi p}{d_{j, \mu}}}$ in their global $\mathbb{Z}_{d_{j, \mu}}$ charge.
According to \eqref{eq: integer vector Zd symmetry}, 
\begin{equation}
\label{eq: translation vectors Zd symmetry}
\mathcal{T}_{j,\mu} = d_{j, \mu} \Big( \mathcal{V}_{\mu}^{T} \Big)^{-1}  e_{j}	
\,
\quad
\mathcal{T}_{i \neq j, \mu} = \Big( \mathcal{V}_{\mu}^{T} \Big)^{-1}  e_{i \neq j}
\end{equation}
define translation vectors on the integer lattice connecting local particles with the same $\mathbb{Z}_{d_{j, \mu}}$ charge, which then defines a unit cell of volume $d_{j, \mu}$, corresponding to the number of distinguishable local particle under the
symmetry. This then defines a direct relationship between the size of the unit cell and the entanglement correction.

Let us now illustrate the previous discussion with some examples. First, consider the case $K_L = 1$ and $K_R = 4$
corresponding to an interface between fermionic and bosonic Laughlin states with filling $\nu=1$ and $\nu=1/4$,
respectively. In this case, the null vector reads $\Lambda = (2,1)$, which corresponds to an interaction
that binds two local fermions of the $\nu=1$ phase with one local boson of the $\nu=1/4$ state. As such,
this interaction possess a $\mathbb{Z}_2$ symmetry associated with transformations of the local fermion by
$\psi_{L} \rightarrow e^{i\,\pi}\psi_{L}$. Since local excitations of the $\nu=1$ are described
by vertex operators $\exp{(i\ell \phi_{L})}$, $\ell \in \mathbb{Z}$, the $\mathbb{Z}_2$ symmetry identifies
local operators $\ell \sim \ell+2$ as having the same charge, which implies the translation operator $\mathcal{T} = 2$
and a unit cell of size $2$. The $\log{2}$ value of the the entanglement entropy across this heterogeneous interface,
is then a measure of the $\mathbb{Z}_2$ symmetry which, in turn, originates from the nature of the bosonic condensate at the interface that binds two local particles on the left to one local particle on the right.
In Figs.~\eqref{fig: lattice-nu2} and~{\eqref{fig: lattice-sigmax} we illustrate these concepts for the gapped interfaces
of $K = I_{2 \times 2}$ and $K=\sigma_x$. 

\begin{figure}[h!]
\includegraphics[width=0.4\textwidth]{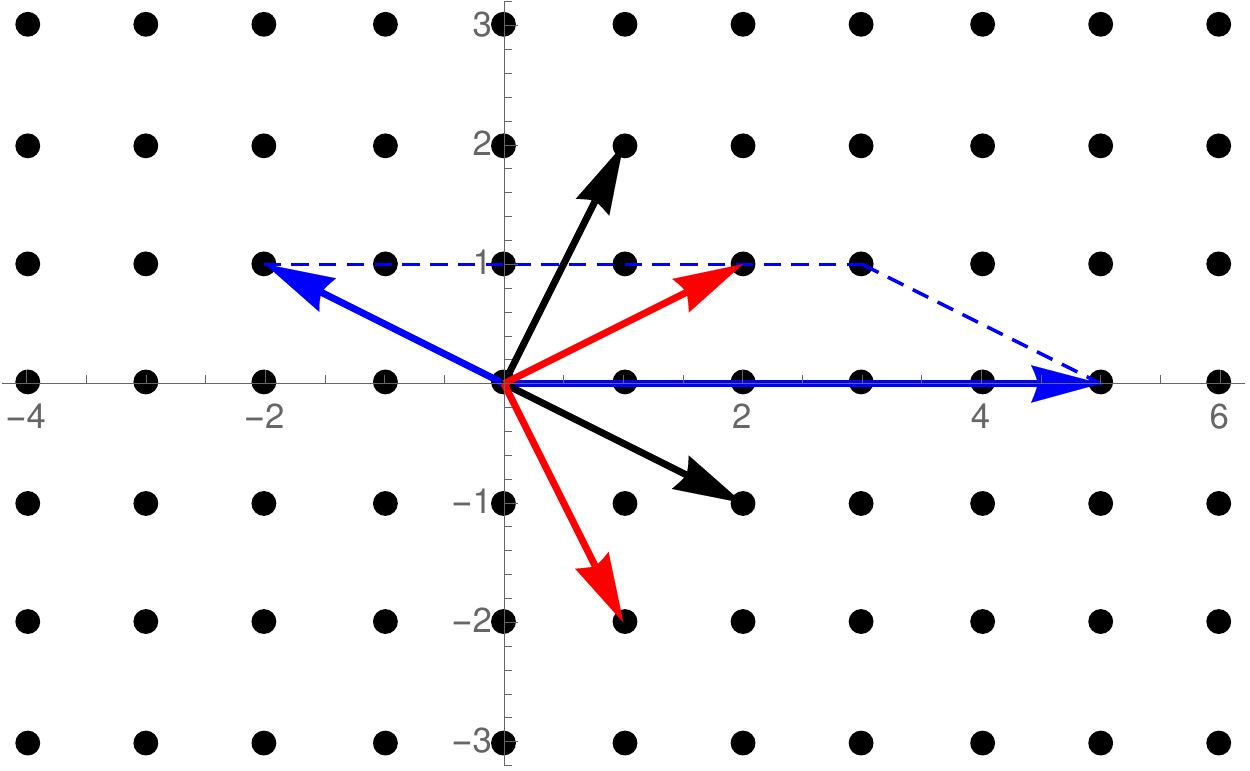}
\caption
{
(Color online)
Points $\ell \in \mathbb{Z}^{2}$ on the integer lattice parametrize
local particles $\psi_{\mu, \ell} = \exp{(i \ell_{\mu}^{T} K_{\mu} \phi_{\mu})}$,
$\mu= L, R$, of the $\nu=2$ state described by $K_{\nu=2} = I_{2\times 2}$.
When the interface is gapped by $\mathbb{Z}_{5} \times \mathbb{Z}_{5}$ interactions,
we identify, using Eq.\eqref{eq: SNF of ML for nu 2} and Eq.\eqref{eq: translation vectors Zd symmetry},
the blue vectors $\mathcal{T}_{1}=(5,0)$ and $\mathcal{T}_{2} =(-2,1)$ as the generators of translations 
connecting local operators carrying the same $\mathbb{Z}_{5}$ charge, with the corresponding
unit cell marked by the blue parallelogram. This parallelogram has area $|\mathcal{T}_1 \times \mathcal{T}_{2}|= 5$,
corresponding to five distinguishable local particles carrying different $\mathbb{Z}_{5}$ charges.
The black vectors $m_{1L}=(1,2)$ and $m_{2L}=(2,-1)$ are defined by the columns of 
$
\mathcal{M}_{L} 
= 
\begin{pmatrix}
1 & 2
\\
2 & -1
\end{pmatrix}
$
and the red vectors $m_{1R}=(1,-2)$ and $m_{2R}=(2,1)$ are defined by the columns of 
$
\mathcal{M}_{R} 
= 
\sigma_{z}\,\mathcal{M}_{L}
=
\begin{pmatrix}
1 & 2
\\
-2 & 1
\end{pmatrix}
$.
The minors of the $2\times 4$ matrix $\Big( \mathcal{M}_{L} ~ \mathcal{M}_{R} \Big)$, 
$\{-5, -4, -3,  -3, 4, 5 \}$, are given by the determinant of each of the six $2 \times 2$ sub-matrices, which is
interpreted as the (oriented) area of the parallelogram spanned by pairs of distinguished vectors $m_{1L}$, $m_{2L}$, $m_{1R}$ and $m_{2R}$.
The primitive condition of the gapping interactions reflects the condition
that the greatest common divisor the set of minors is equal to $1$.
Importantly, the relation between $\mathcal{M}_{R} = \sigma_{z}\,\mathcal{M}_{L}$
implies that $m_{1R}$ and $m_{2R}$ are obtained from $m_{1L}$ and $m_{2L}$ upon reflection
on the horizontal axis. This change in orientation is crucially tied to the primitive condition:
had we chosen the case $\mathcal{M}_{L} = \pm \mathcal{M}_{R}$, which is a solution of the null equation,
the g.c.d. of the minors would have been $|\textrm{det}(\mathcal{M}_{L})| = |\textrm{det}(\mathcal{M}_{R})| = 5 > 1$.
}
\label{fig: lattice-nu2}
\end{figure}

\begin{figure}[h!]
\includegraphics[width=0.3\textwidth]{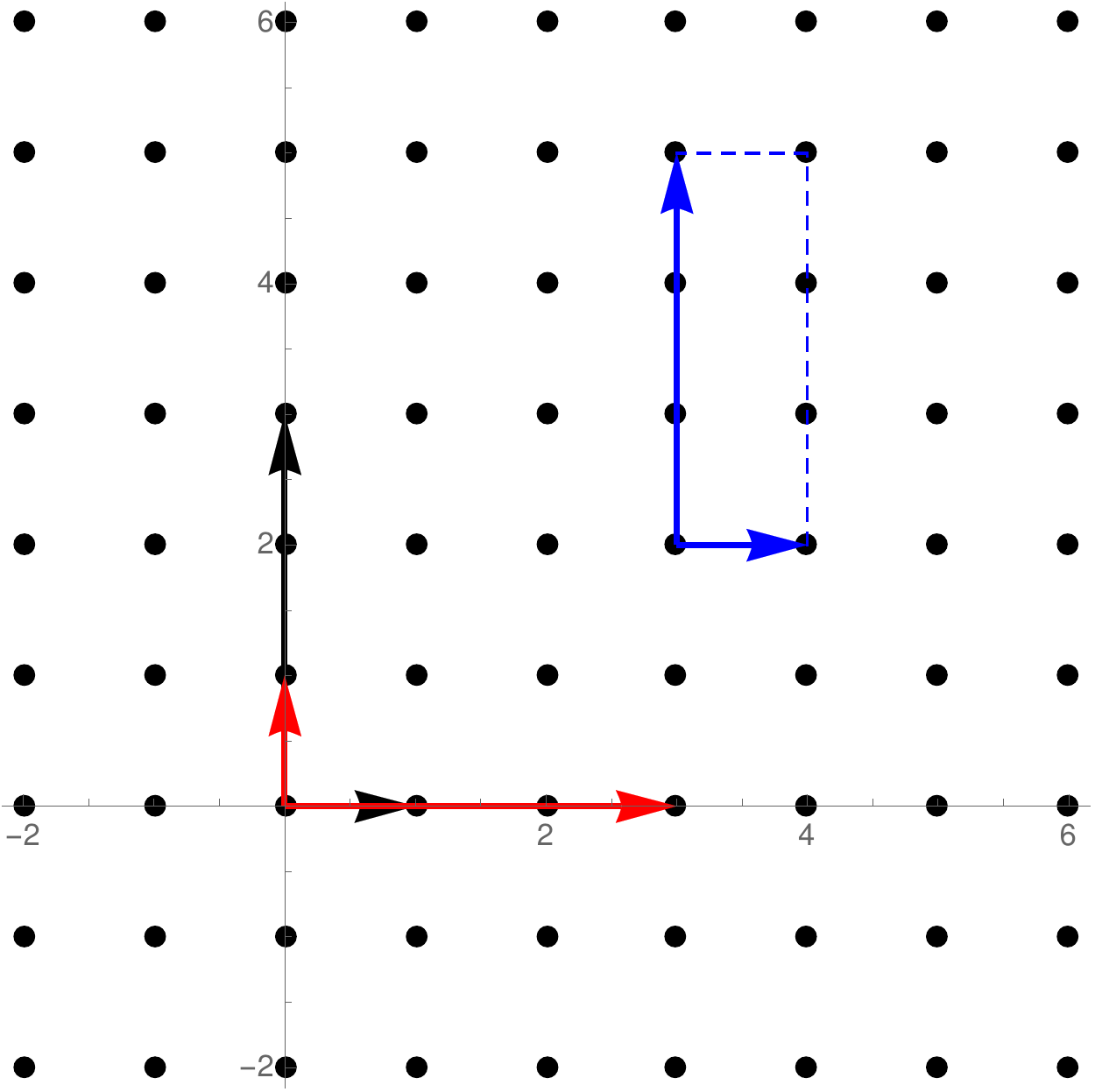}
\caption
{
(Color online)
Points $\ell \in \mathbb{Z}^{2}$ on the integer lattice parametrize
local particles $\psi_{\mu, \ell} = \exp{(i \ell_{\mu}^{T} K_{\mu} \phi_{\mu})}$,
$\mu= L, R$, of bosonic state described by $K = \sigma_{x}$.
When the interface is gapped by $\mathbb{Z}_{3} \times \mathbb{Z}_{3}$ interactions,
we identify, using 
Eq.\eqref{eq: SNF of ML and MR for sigma x}
and 
Eq.\eqref{eq: translation vectors Zd symmetry},
the blue vectors $\mathcal{T}_{1}=(0,3)$ and $\mathcal{T}_{2} =(1,0)$ as the generators of translations 
connecting local operators carrying the same $\mathbb{Z}_{3}$ charge, with the corresponding
unit cell marked by the blue parallelogram. This parallelogram has area $|\mathcal{T}_1 \times \mathcal{T}_{2}|= 3$,
corresponding to three distinguishable local particles carrying different $\mathbb{Z}_{3}$ charges.
The black vectors $m_{1L}=(1,0)$ and $m_{2L}=(0,3)$ are defined by the columns of 
$
\mathcal{M}_{L} 
= 
\begin{pmatrix}
1 & 0
\\
0 & 3
\end{pmatrix}
$
and the red vectors $m_{1R}=(0,1)$ and $m_{2R}=(3,0)$ are defined by the columns of 
$
\mathcal{M}_{R} 
= 
\sigma_{x}\,\mathcal{M}_{L}
=
\begin{pmatrix}
0 & 3
\\
1 & 0
\end{pmatrix}
$.
The minors of the $2\times 4$ matrix $\Big( \mathcal{M}_{L} ~ \mathcal{M}_{R} \Big)$, 
$\{3, 1, 0,  0, -9, -3 \}$, are given by the determinant of each of the six $2 \times 2$ sub-matrices, which is
interpreted as the (oriented) area of the parallelogram spanned by pairs of distinguished vectors $m_{1L}$, $m_{2L}$, $m_{1R}$ and $m_{2R}$.
The primitive condition of the gapping interactions reflects the condition
that the greatest common divisor the set of minors is equal to $1$.
Importantly, the relation between $\mathcal{M}_{R} = \sigma_{x}\,\mathcal{M}_{L}$
implies that $m_{1R}$ and $m_{2R}$ are obtained from $m_{1L}$ and $m_{2L}$ upon a $90$ degree
rotation. This change in orientation is crucially tied to the primitive condition:
had we chosen the case $\mathcal{M}_{L} = \pm \mathcal{M}_{R}$, which is a solution of the null equation,
the g.c.d of the minors would
have been $|\textrm{det}(\mathcal{M}_{L})| = |\textrm{det}(\mathcal{M}_{R})|  = 3  > 1$.
}
\label{fig: lattice-sigmax}
\end{figure}

\subsection{Gauge-related topological phases share the same BTES}
\label{sec: gauged phases share TEE correction}

In this subsection, we prove that the BTES associated with the gapped interface between Abelian topological bulk phases described by the matrices $K_{R/L}$ is the same as the BTES associated with the gapped interface of the gauged matrices, $\tilde{K}_{R/L} \equiv \tilde{G} K_{R/L} \tilde{G}^T$,
where $\tilde{G}$ is an integer matrix.

\subsubsection{Proof that each TEE correction obtainable at the $\tilde{K}_{R/L}$ interface is obtainable at the $K_{R/L}$ interface}
\label{sec:proof gauge 1}

Suppose that an interface within the phase described by $\tilde{K}_{R/L}$ is gapped by terms parameterized by $\mathcal{M}_{R/L}$, so that
\begin{equation}
\mathcal{M}_R \tilde{K}_R \mathcal{M}_R^T - \mathcal{M}_L \tilde{K}_L \mathcal{M}_L^T = 0.
\label{eq: gap gauged theories}
\end{equation}
We show in Appendix~\ref{sec:TEE proof} that if the rows of the matrix $\begin{pmatrix} \mathcal{M}_R & \mathcal{M}_L\end{pmatrix}$ are primitive, then the TEE correction is equal to $\log |\det \mathcal{M}_{R/L}|$. (Note that for a heterogeneous interface, $ |\det \mathcal{M}_{R}| \neq  |\det \mathcal{M}_{L}|$ because which part of the TEE is considered the ``correction'' is relative to the $K$ matrix of either side. The total TEE does not depend on the $R/L$ label.)

We will now construct a primitive gapping term for the interface across $K_{R/L}$ that has the same TEE correction.
We will need the SNF of the matrix $\begin{pmatrix} \mathcal{M}_R \tilde{G} & \mathcal{M}_L \tilde{G} \end{pmatrix}$:
\begin{equation}
\mathcal{U}_{\tilde{G}} \begin{pmatrix} \mathcal{M}_R \tilde{G} & \mathcal{M}_L \tilde{G} \end{pmatrix} \mathcal{V}_{\tilde{G}} = \begin{pmatrix} \mathcal{S}_{\tilde{G}} & 0\end{pmatrix},
\label{eq: SNF KG}
\end{equation}
where $\mathcal{U}_{\tilde{G}},\mathcal{V}_{\tilde{G}}$ are unimodular matrices, and $\mathcal{S}_{\tilde{G}}$ is diagonal.
It is useful to write the $2r \times 2r$ matrix $\mathcal{V}_{\tilde{G}}^{-1}$ in terms of its $r\times r$ blocks:
\begin{equation}
\label{eq: def VG inverse}
\mathcal{V}_{\tilde{G}}^{-1} = \begin{pmatrix} V_1 & V_2\\ V_3 & V_4\end{pmatrix},
\end{equation}
so that Eq.~(\ref{eq: SNF KG}) yields
\begin{equation} 
\mathcal{U}_{\tilde{G}} \mathcal{M}_{R/L} \tilde{G} = \mathcal{S}_{\tilde{G}} V_{1,2}.
\label{eq: def V12}
\end{equation}
The matrices $V_{1,2}$ furnish a gapping term for an interface across $K_{R/L}$: Eqs.~(\ref{eq: gap gauged theories}) and (\ref{eq: def V12}) yield
\begin{equation}
V_1 K_R V_1^T - V_2 K_L V_2^T
=0.
\end{equation}
Furthermore, from Eq.~\eqref{eq: def VG inverse}, $\begin{pmatrix} V_1 & V_2 \end{pmatrix}\mathcal{V}_G = \begin{pmatrix} \mathbb{I} & 0 \end{pmatrix}$, which shows that the rows of $\begin{pmatrix} V_1 & V_2 \end{pmatrix}$ are a set of primitive vectors.
From the result of Appendix~\ref{sec:TEE proof}, the TEE correction is given by $\log |\det V_{1,2}|= \log |\det \mathcal{M}_{R/L} \det \tilde{G}/\det \mathcal{S}_{\tilde{G}}|$, where the last equality follows from Eq.~(\ref{eq: def V12}).
Following the steps in Appendix~\ref{sec:proof det S}, we find that when the gapping terms are primitive, $|\det \mathcal{S}_{\tilde{G}}| = |\det \tilde{G}|$, from which it follows that the TEE correction associated with the $K_{R/L}$ interface is equal to $\log |\det \mathcal{M}_{R/L}| $, as desired.

\subsubsection{Proof that each TEE correction obtainable at the $K_{R/L}$ interface is obtainable at the $\tilde{K}_{R/L}$ interface}
\label{sec:proof gauge 2}
We now reverse the analysis. Consider an interface between the phases with K-matrices $K_{R/L}$ and a set of primitive gapping terms encoded in the matrices $V_{1,2}$, which satisfy
\begin{equation}
V_1 K_R V_1^T - V_2 K_L V_2^T = 0.
\label{eq: gap K}
\end{equation}
As proved in Appendix~\ref{sec:TEE proof}, the TEE correction corresponding to these gapping terms is $\log | \det V_{1,2} |$, where, again, the TEE correction does not have to be the same on either side of the interface because it depends on which part of the TEE is the ``correction.'' (The total TEE, of course, does not depend on which side of the interface is considered.)

We now construct gapping terms across an interface between $\tilde{K}_{R/L}$.
We will need the SNF for the $r\times 2r$ integer matrix $(\det \tilde{G}) \begin{pmatrix}  V_1 \tilde{G}^{-1} & V_2 \tilde{G}^{-1} \end{pmatrix}$, where the overall factor of $\det \tilde{G}$ ensures that the matrix has integer entries:
\begin{equation}
\mathcal{U}_0 ( \det \tilde{G} ) \begin{pmatrix}V_1 \tilde{G}^{-1} & V_2 \tilde{G}^{-1} \end{pmatrix} \mathcal{V}_0 = \begin{pmatrix} \mathcal{S}_0 & 0 \end{pmatrix},
\label{eq: SNF VG}
\end{equation}
where $\mathcal{U}_0,\mathcal{V}_0$ are unimodular matrices, and $\mathcal{S}_0$ is an integer diagonal matrix.
From Appendix~\ref{sec:proof det S}, 
\begin{equation}
|\det \mathcal{S}_0 | = | \det \left( (\det \tilde{G})\tilde{G}^{-1} \right)|.
\label{eq: det s}
\end{equation}
We write $\mathcal{V}_0^{-1}$ in terms of its $r\times r$ blocks,
\begin{equation}
\mathcal{V}_0^{-1} = \begin{pmatrix} V_1' & V_2' \\ V_3' & V_4' \end{pmatrix}.
\end{equation}
Eq.~(\ref{eq: SNF VG}) then yields:
\begin{equation}
\mathcal{U}_0(\det \tilde{G}) V_{1,2}\tilde{G}^{-1} = \mathcal{S}_0 V_{1,2}'.
\label{eq: udetGVGinverse}
\end{equation}
It is clear that the matrix $\begin{pmatrix} V_1' & V_2'\end{pmatrix}$ is primitive because $\begin{pmatrix} V_1' & V_2' \end{pmatrix} \mathcal{V}_0 = \begin{pmatrix} \mathbb{I} & 0 \end{pmatrix}$. Furthermore, $V_{1,2}'$ serve as gapping terms across an interface in the phase with K-matrices $\tilde{K}_{R/L}$:
\begin{equation}
V_1' \tilde{K}_R (V_1')^T - (V_2') \tilde{K}_L (V_2')^T = 0,
\label{eq: gap KG}
\end{equation}
which can be found by plugging Eqs.~(\ref{eq: gap K}) and (\ref{eq: udetGVGinverse}) into Eq.~(\ref{eq: gap KG}).
Using the result in Appendix~\ref{sec:TEE proof}, the TEE correction corresponding to these terms is given by $\log |\det V_{1,2}'|$. From Eqs~(\ref{eq: det s}) and (\ref{eq: udetGVGinverse}), $\log |\det V_{1,2}' | = \log |\det V_{1,2}|$, exactly the TEE correction across the $K_{R/L}$ interface.

\subsection{Proof that $K$ and $p\,K$ ($p \in \mathbb{Z}^{*}$) share the same BTES}
\label{sec: proportional K matrices share TEE correction}

To see that bulk phases whose K-matrices are proportional to each other share the same BTES, note that 
$K$ and $p\,K$ admit the same set of solutions of the null equation \eqref{eq: null condition matrix form}.
This implies the same null vectors $\mathcal{M}_{R/L}$, hence the same symmetry and BTES.

\subsection{Comments on using entanglement sequences for classification}
We can define an equivalence class of topological phases as those which are related by a gauging operation, i.e., $K_1 \sim K_2$ if there exist integer matrices $G_{1,2}$ such that $G_1 K_1 G_1^T = G_2 K_2 G_2^T$.
Two topological phases can share a gapped interface if and only if they are in the same equivalence class.
Furthermore, the proofs in Secs.~\ref{sec:proof gauge 1} and \ref{sec:proof gauge 2} show that all phases within the same equivalence class share the same BTES. Thus, the BTES can be used to identify the equivalence class.

However, the identification is not unique.
For example, $K_1 = \begin{pmatrix} 2 & 0 \\ 0 & 3 \end{pmatrix}$ and $K_2 = \begin{pmatrix} 1 & 0\\ 0 & 6\end{pmatrix}$ cannot share a gapped interface. (For suppose they could: then there exists four integers $(a,b,c,d)$ such that 
\begin{equation}
\frac{a^2}{2} + \frac{b^2}{3}  = c^2 + \frac{d^2}{6} 
\label{eq:2316 not equivalent}
\end{equation}
Multiplying by $6$ yields $2b^2 = d^2 \mod 3$, which implies $b=d=0\mod 3$. Thus, $b=3\tilde{b}, d=3\tilde{d}$, where $\tilde{b},\tilde{d}\in \mathbb{Z}$. Plugging into Eq~(\ref{eq:2316 not equivalent}), $a^2/2 + 3\tilde{b}^2 = c^2 + 3/2\tilde{d}^2$. Multiplying by $2$ yields $a^2 = 2c^2 \mod 3$, which implies $a=c=0\mod 3$. Thus, $a=3\tilde{a}$, $c=3\tilde{c}$, where $\tilde{a},\tilde{c}\in \mathbb{Z}$. Plugging in to Eq~(\ref{eq:2316 not equivalent}) yields $\tilde{a}^2/2 + \tilde{b}^2/3 = \tilde{c}^2 + \tilde{d}^2/6$. Hence, if $(a,b,c,d)$ is a solution to Eq~(\ref{eq:2316 not equivalent}), then so is $(a/3,b/3,c/3,d/3)$. This yields an infinite sequence of smaller integers, which presents a contradiction. Hence, there is no integer solution to Eq~(\ref{eq:2316 not equivalent}).)
Yet, although the theories described by $K_1$ and $K_2$ cannot share a gapped edge,
we checked through brute force construction of possible gapping terms that the first 100 entries of their BTES are identical, and we fully expect the theories to yield identical sequences.~\cite{1623-sequence}
The same is true for $K_1 = \begin{pmatrix} 3 & 0 \\ 0 & 5 \end{pmatrix}$ and $K_2 = \begin{pmatrix} 1 & 0\\ 0 & 15\end{pmatrix}$.

\section{Conclusions}
\label{sec:conclusions}

In this paper, we established a direct connection between the corrections to the topological entanglement entropy of  $2$D Abelian topological phases, found in Ref.~\onlinecite{cano-2015}, and the existence of $1$D SPT phases supported along the entanglement cut.
The set of non-negative corrections $\gamma_s$, which we named the Boundary Topological Entanglement 
Sequence (BTES),
was then shown to be associated with the order of the symmetry group of the SPT state.
Our work shows that the entanglement entropy of the ground state of gapped Abelian $2$D phases 
admits a symmetry-protected refinement beyond the standard bulk topological value.~\cite{kitaevpreskill,levinwen}

The relationship between the BTES and $1$D SPT states was established as follows.
First, we proved (Sec.~\ref{sec: General Approach}) that whenever $\gamma_s \neq 0$, 
the gapping interactions along the boundary (or interface) are invariant under a discrete symmetry group 
$\mathcal{G}_{L} \times \mathcal{G}_{R},$ which transforms local particles on the left and right sides of the interface and, furthermore, that the entanglement correction provides a measure of the rank of the symmetry group
via $\gamma_{s,L/R} = \log(|{\cal G}_{L,R}|)$, depending upon whether the left or right interface is traced over.
The SPT nature of the gapped interface is mathematically contained
in the non-singular matrices $\mathcal{M}_{L}$ and $\mathcal{M}_{R}.$ These matrices parameterize the gapping interactions,
specify how local particles transform under the global symmetry, and determine whether the interactions are primitive.
Local quasiparticles with the same global symmetry charge are related by translation vectors on the integer
lattice, whose directions are determined from the Smith normal form of $\mathcal{M}_{L/R}$. 
As such,
local particles, that are not distinguishable by their statistics, but instead by the action of the symmetry transformation, are represented by 
points on the integer lattice that reside within a unit cell whose volume is $|\textrm{det}(\mathcal{M}_{L/R})| = e^{\gamma_{s, L/R}}$.
Furthermore, the primitivity condition of the local interactions was shown to have
a geometrical interpretation in terms of a condition on the area of the parallelograms 
spanned by the column vectors of $\left(\mathcal{M}_{L}~ \mathcal{M}_{R} \right)$.
These properties were discussed in Sec.~\ref{sec:general} and illustrated in Figs.~\ref{fig: lattice-nu2} 
and~\ref{fig: lattice-sigmax}.

Second, we have confirmed the SPT nature of the gapped interface by directly studying the low energy properties of zero-dimensional edge states of the $1$D SPT interface. Our analysis, has shown a ground state degeneracy associated with symmetry-protected parafermion-like edge modes of the SPT system. Although we have primarily focused upon representative examples of bosonic and fermionic, chiral and non-chiral Abelian phases with or without topological order that are associated with $2 \times 2$ K-matrices, our approach can be straightforwardly applied to Abelian phases described by K-matrices of higher rank.
As such, our approach and results are more general than Ref.~\onlinecite{Zou-2016}, which concentrated on $2$D topologically trivial bosonic states.
%
Interestingly, Ref.~\onlinecite{Zou-2016} identified a necessary condition for when TEE corrections can occur 
using a so-called replica correlation function: states with vanishingly small replica correlation length are not expected to exhibit TEE corrections, while those with large replica correlation length can.
For instance, the authors explicitly show that the double semion and $\mathbb{Z}_{n}$ gauge theories have vanishingly small replica correlation length, in contrast to the topologically trivial examples that were shown to allow a TEE correction. 
Our work shows that gapped homogeneous interfaces between $\mathbb{Z}_{n}$ gauge theories ($K = n\,\sigma_x$) admit the BTES given
in Eqs.~\ref{eq: BTES sigma x}, while interfaces 
between double semion phases ($K = 2\,\sigma_z$) admit the BTES given in 
Eqs.~\ref{eq: BTES sigma z type 1} and~\ref{eq: BTES sigma z type 2}.
Thus, it is of interest to consider a generalization of the replica correlation function that is sensitive to the TEE corrections that can occur in topologically ordered systems; alternatively, it may only be necessary to generalize the topologically non-trivial lattice models so as to induce 1D SPT order along a potential entanglement cut which might, in turn, be detected by the replica correlation function. 

We note that the ground state degeneracy associated with the $1$D SPT is a lower bound on the
total possible ground state degeneracy for a 2D system with gapped edges on a cylinder. As shown in Refs.~\onlinecite{hung-wan-2015,lan-wang-wen-2015}, a topologically ordered system placed on a cylinder with gapped edge states may possess a ground state degeneracy -- of topological origin -- associated with creation and annihilation of anyons on distinct gapped edges. 
Different from the physics discussed in these references, in this paper, we focused on the ground state degeneracy intrinsically associated with each 1D SPT boundary (or interface). For instance, we discussed examples where there was a ground state
degeneracy that originated from the boundary/interface modes even when the bulk system possessed no deconfined anyon excitations.

An important conceptual framework underlying this work, discussed in Sec.~\ref{sec: Inhomogeneous gauging related interfaces},
is the fact that $2$D Abelian topological phases related by a gauging mechanism can always share a gapped interface, as depicted in Fig.~\ref{fig: interface-1}. 
In the language of Chern-Simons theories, this gauging mechanism relates two Abelian phases characterized by rank-$r$ K-matrices $K$ and $G\,K\,G^{T}$, where $G \in \textrm{GL}(r,\mathbb{Z})$ encodes the deconfinement of symmetry fluxes of a global symmetry $\mathcal{G}$ whose rank $|\mathcal{G}| = |\textrm{det}(G)|$. 
We have shown in Sec.~\ref{sec: Heterogeneous Stable Interfaces} 
that heterogeneous interfaces separating Abelian phases related by such a gauging mechanism
behave as $1$D SPT interfaces where the gapping interactions are invariant under the global symmetry $\mathcal{G}$,
the correction $\gamma_s = \frac{1}{2}\log{|\textrm{det}(G)|},$ and the end points of the gapped interface contain parafermion-like excitations.
Moreover, due to the different topological properties of the bulk phases,
the interface behaves as a generalization of the anyon Andreev reflectors discussed in Ref.~\onlinecite{Santos-2017},
since only certain quasiparticles can be fully transmitted across the interface, while other quasiparticles are partially transmitted and reflected by the condensate at the interface. 

In Sec.~\ref{sec: Homogeneous interface with emergent symmmetry}
we studied how $1$D SPT states protected by global discrete Abelian symmetry $\mathcal{G} \times \mathcal{G}$
can be formed at homogeneous interfaces.
The one-dimensional symmetry-protected topological state can then be understood to arise via the limit depicted in Figs.~\ref{fig: interface-2} (a) and (b) in which a finite slab of the $G\,K\,G^{T}$ gauge theory, inserted between two regions of the $K$ phase, is reduced to zero width. This limit provides an intuitive way by which to produce a one-dimensional symmetry-protected phase.
We have exemplified this approach by studying the cases $K = \sigma_{x}$ (Sec.~\ref{sec: sigma x interface}), 
$K = \sigma_{z}$ (Sec.~\ref{sec: sigma z interface}), and $K = \textrm{I}_{2 \times 2}$ (Sec.~\ref{sec: nu = 2 interface})
and discussed the nature of their symmetry-protected edge states and determined their full BTES.

In summary, our work unveils an important connection between the entanglement entropy of $2$D Abelian phases of matter 
and $1$D SPT states protected by Abelian discrete symmetries. 
There are a variety of directions for future work, including the study of gapped states occurring at 
interfaces within non-Abelian topological phases and the behavior of the BTES near phase transitions.
It would also be appealing to find signatures of the symmetry-protected parafermion zero modes 
in numerical simulations such as those employed in Ref.~\onlinecite{vaezi-2017}.
Furthermore, it would be interesting to understand the relation, if any, between our results and the studies of the entanglement entropy of gauge theories in Refs.~\onlinecite{casinihuertarosabal, Ghosh2015},
as well to understand any possible connections between the SPT interfaces discussed here and gapped interfaces 
obtained via symmetry ``group extension" discussed in Ref.~\onlinecite{wangwenwitten-2017}.

\acknowledgements
We acknowledge useful discussions with Aris Alexandradinata, Moshe Goldstein, Eugeniu Plamadeala, Cenke Xu, and Michael Zaletel. 
We thank Kevin Slagle for bringing Ref.~\onlinecite{Zou-2016} to our attention.
L.H.S.  is supported  by  a  fellowship  from  the  Gordon  and  Betty Moore Foundation’s 
EPiQS Initiative through Grant No. GBMF4305  at  the  University  of  Illinois.   
M.M. is supported in part by the UCR Academic Senate.
T.L.H.  is  supported  by  the  US  National  Science  Foundation  under
grant DMR 1351895-CAR.
J.C. and T.L.H. acknowledge the hospitality of the KITP, where this work was initiated.
This research was supported in part by the National Science Foundation under Grant No. NSF PHY17-48958.
\appendix

\section{Derivation of TEE correction}
\label{sec:TEE correction}

Given the gapping terms defined by the matrices $\mathcal{M}_{R/L}$ in Eq.~(\ref{eq: definition of M Left and Right}), we showed in Ref.~\onlinecite{cano-2015} that the correction to the TEE is given by
\begin{equation}
\gamma_{s,L/R} = \log{ | \det v_{L/R}|},
\end{equation}
where the matrices $v_{L/R}$ are determined by the SNF of the $r\times 2r$ matrix:
\begin{equation}
\begin{pmatrix} \mathcal{M}_R K_R &  \mathcal{M}_L K_L \end{pmatrix} = U^{-1} \begin{pmatrix} S & 0 \end{pmatrix} V^{-1},
\label{eq:SNF MK}
\end{equation}
where $U$ and $V$ are unimodular matrices, $\mathcal{S}$ is diagonal, and the blocks of $V$ are given by
\begin{equation}
V = \begin{pmatrix} v_1 & v_R \\ v_3 & v_L \end{pmatrix}.
\label{eq: V blocks}
\end{equation}
(Note that the $\mathcal{M}_{R/L}$ of this paper were called $m_{R/L}$ in Ref.~\onlinecite{cano-2015}).
We prove in Appendix~(\ref{sec:TEE proof}) that when the $r \times 2r$ matrix, $\begin{pmatrix} \mathcal{M}_R & \mathcal{M}_L \end{pmatrix}$, is primitive, as defined in Sec.~\ref{sec: Primitive Condition of the Null Vectors}, that $|\det v_{L/R}| = |\det \mathcal{M}_{R/L}|$.
This is useful because, for primitive gapping terms, it allows the TEE correction to be computed without computing the SNF.
Eq.~(\ref{eq: TEE correction}) then follows immediately from Eq.~(\ref{eq: relation between det of M and symmetry order}).

A consequence that we prove in Appendix~(\ref{sec:proof det S}) is that $| \det S| = \sqrt{ |\det K_L \det K_R |}$ when the gapping terms are primitive, which we use in Sec.~\ref{sec: Emergent Interface Symmetry}.

\subsection{Proof that when $\begin{pmatrix} \mathcal{M}_R & \mathcal{M}_L \end{pmatrix}$ is primitive, $|\det v_{R/L}| = |\det \mathcal{M}_{R/L}|$}
\label{sec:TEE proof}

Let the SNF of $\begin{pmatrix} \mathcal{M}_R & \mathcal{M}_L \end{pmatrix}$ be given by
\begin{equation}
\begin{pmatrix} \mathcal{M}_R & \mathcal{M}_L \end{pmatrix} = t^{-1} \begin{pmatrix} s & 0 \end{pmatrix} w^{-1},
\label{eq:SNF MR ML}
\end{equation}
where $t, w$ are unimodular matrices, and $s$ is an integer diagonal matrix.
When the SNF is canonically ordered, the diagonal entries of $s$ are $s_1, s_2, \dots s_r$, where $s_i = d_i /d_{i-1}$, $d_i$ is the greatest common divisor of the $i\times i$ minors of $\begin{pmatrix} \mathcal{M}_R & \mathcal{M}_L \end{pmatrix}$ and $d_0 = 1$.
If the rows of $\begin{pmatrix} \mathcal{M}_R & \mathcal{M}_L \end{pmatrix}$ are primitive, then by definition (Sec.~\ref{sec: Primitive Condition of the Null Vectors}), $d_r=1$.
Since $d_{i-1}$ divides  $d_i$, it must be that $d_1 = d_2 = \cdots = d_r = 1$ and, consequently, 
\begin{equation}
s = \mathbb{I}_{r\times r}.
\label{eq:s-identity}
\end{equation}
Let us write the $2r\times 2r$ matrix $w^{-1}$ in terms of its $r\times r$ blocks,
\begin{equation}
w^{-1} = \begin{pmatrix} w_1 & w_2 \\ w_3 & w_4 \end{pmatrix}.
\label{eq:w inverse}
\end{equation}
Eqs.~(\ref{eq:SNF MR ML}) and (\ref{eq:w inverse}) together yield: 
\begin{equation}
t\mathcal{M}_{R/L} = w_{1,2}.
\label{eq: tMw}
\end{equation}

We will also need the SNF,
\begin{equation}
w_1 K_R w_3^T - w_2 K_L w_4^T = (t')^{-1} s' (w')^{-1},
\label{eq:SNF Kw}
\end{equation}
where $t', w'$ are unimodular matrices, and $s'$ is a diagonal integer matrix.
Then, using Eqs~(\ref{eq:SNF MR ML}), (\ref{eq:s-identity}) and (\ref{eq:SNF Kw}), we can rewrite Eq~(\ref{eq:SNF MK}) as
\begin{widetext}
\begin{align}
\begin{pmatrix} \mathcal{M}_R K_R &  \mathcal{M}_L K_L \end{pmatrix} &= t^{-1} t \begin{pmatrix} \mathcal{M}_R &  \mathcal{M}_L \end{pmatrix} ww^{-1} \begin{pmatrix} K_R & 0 \\ 0& K_L \end{pmatrix} \nonumber\\
&= t^{-1} \begin{pmatrix} \mathbb{I} & 0 \end{pmatrix} w^{-1} \begin{pmatrix} K_R & 0 \\ 0& K_L \end{pmatrix} \nonumber\\
&= t^{-1} \begin{pmatrix} w_1 K_R & -w_2 K_L \end{pmatrix}\begin{pmatrix} \mathbb{I} & 0 \\ 0 & -\mathbb{I} \end{pmatrix} \nonumber\\
&= t^{-1} \begin{pmatrix} w_1 K_R & -w_2 K_L \end{pmatrix}(w^{-1})^T w^T\begin{pmatrix} \mathbb{I} & 0 \\ 0 & -\mathbb{I} \end{pmatrix}  \nonumber\\
&= t^{-1}   \begin{pmatrix} w_1 K_Rw_1^T - w_2 K_L w_2^T & w_1 K_R w_3^T - w_2 K_L w_4^T \end{pmatrix} w^T\begin{pmatrix} \mathbb{I} & 0 \\ 0 & -\mathbb{I} \end{pmatrix}  \nonumber\\
&= t^{-1}   \begin{pmatrix} 0 & w_1 K_R w_3^T - w_2 K_L w_4^T \end{pmatrix} w^T\begin{pmatrix} \mathbb{I} & 0 \\ 0 & -\mathbb{I} \end{pmatrix}\nonumber\\
&= t^{-1} \begin{pmatrix} 0 & (t')^{-1} s' (w')^{-1} \end{pmatrix} w^T \begin{pmatrix} \mathbb{I} & 0 \\ 0 & -\mathbb{I} \end{pmatrix} \nonumber\\
&= (t' t)^{-1} \begin{pmatrix} 0 & s' \end{pmatrix} \begin{pmatrix} \mathbb{I} & 0 \\ 0 & (w')^{-1} \end{pmatrix}w^T \begin{pmatrix} \mathbb{I} & 0 \\ 0 & -\mathbb{I} \end{pmatrix}\nonumber\\
&= (t' t)^{-1} \begin{pmatrix} s' & 0 \end{pmatrix} \begin{pmatrix} 0 & \mathbb{I} \\ \mathbb{I} & 0 \end{pmatrix} \begin{pmatrix} \mathbb{I} & 0 \\ 0 & (w')^{-1} \end{pmatrix}w^T \begin{pmatrix} \mathbb{I} & 0 \\ 0 & -\mathbb{I} \end{pmatrix} \nonumber\\
&=   (t' t)^{-1} \begin{pmatrix} s' & 0 \end{pmatrix}  \begin{pmatrix} w_3^T w' & w_1^T \\ -w_4^T w' & -w_2^T \end{pmatrix}^{-1} ,
\label{eq:SNF MK2}
\end{align} 
where we have used Eq.~(\ref{eq: tMw}), and the null matrix condition in Eq.~(\ref{eq: null condition matrix form}), to prove:
\begin{equation}
w_1 K_Rw_1^T - w_2 K_L w_2^T  = t \left( \mathcal{M}_R K_R \mathcal{M}_R^T  - \mathcal{M}_L K_L \mathcal{M}_L^T \right) t^T = 0.
\end{equation}
\end{widetext}
Equating Eqs~(\ref{eq:SNF MK}) and (\ref{eq:SNF MK2}) shows that $S=s'$ and that one could choose:
\begin{equation}
U=t't, V=\begin{pmatrix} w_3^T w' & w_1^T \\ -w_4^T w' & -w_2^T \end{pmatrix}.
\label{eq: find UV}
\end{equation}
Comparing $V$ in Eq.~(\ref{eq: find UV}) to its block decomposition in Eq.~(\ref{eq: V blocks}) shows that
\begin{equation}
v_R = w_1^T, v_L = -w_2^T.
\end{equation}
Plugging in Eq.~(\ref{eq: tMw}) yields 
\begin{equation}
v_R = \mathcal{M}_R^T t^T, v_L = -\mathcal{M}_L^T t^T.
\label{eq:equate v and M}
\end{equation}
Since $t$ is unimodular (it is defined in Eq~(\ref{eq:SNF MR ML})), this completes the proof.

\subsection{Proof that $| \det S| = \sqrt{ |\det K_L \det K_R |}$ when $\begin{pmatrix} \mathcal{M}_R & \mathcal{M}_L\end{pmatrix}$ is primitive}
\label{sec:proof det S}

Eqs.~(\ref{eq:SNF MK}) and (\ref{eq: V blocks}) can be rewritten as two equations:
\begin{subequations}
\begin{equation}
U(\mathcal{M}_R K_R v_1 + \mathcal{M}_L K_L v_3) = S, 
\label{eq:MRKRv1}
\end{equation}
\begin{equation}
U(\mathcal{M}_R K_R v_R + \mathcal{M}_L K_L v_L) = 0.
\label{eq:MRKRvR}
\end{equation}
\end{subequations}
Further, when $v_L$ is invertible (which is always the case when $\mathcal{M}_L$ is invertible, via Eq.~(\ref{eq:equate v and M})), the determinant of $V$ can be written as:
\begin{equation}
\label{eq:det V}
\det V = \det v_L \det (v_1 - v_R v_L^{-1} v_3)= \pm 1,
\end{equation}
where the last equality follows because $V$ is unimodular.

Eqs.~(\ref{eq:MRKRv1}), (\ref{eq:MRKRvR}) and (\ref{eq:det V}) together yield:
\begin{align} 
| \det S| &= |\det (\mathcal{M}_R K_R v_1 + \mathcal{M}_L K_L v_3)| \nonumber\\
&= | \det (\mathcal{M}_R K_R )\det (v_1 + K_R^{-1} \mathcal{M}_R^{-1} \mathcal{M}_L K_L v_3) |\nonumber\\
&= |\det (\mathcal{M}_R K_R )\det (v_1 - v_R v_L^{-1} v_3)| \nonumber\\
&= |\det (\mathcal{M}_R K_R )/\det v_L |.
\label{eq:compute det S}
\end{align}
Eq.~(\ref{eq:equate v and M}) shows that $|\det v_L |= |\det \mathcal{M}_L|$. 
Furthermore, the null condition in (\ref{eq: null condition matrix form}) requires $|\det \mathcal{M}_R /\det \mathcal{M}_L | = \sqrt{ \det K_R /\det K_L}$.
Thus, Eq.~(\ref{eq:compute det S}) yields:
\begin{equation}
\label{eq: det S}
|\det S| = \sqrt{ \det K_R \det K_L }.
\end{equation}

\section{Commutation relations between vertex operators}
\label{sec: Commutation relation between vertex operators}
Consider
\begin{equation}
\begin{split}
&\,		
\Gamma_{k} 
=
e^
{
i\,c_{k}\,\int^{z_{k}}_{y_{k}}\,L^{T}_{k}\partial_{x}\phi(x)	
}
\equiv
e^
{
A_{k}
}
\,,
	\end{split}	
\end{equation}
where $[y_{k}, z_{k}]$ is a segment on the real line. We want to obtain a general expression for
the commutation relations of two operators, $\Gamma_{k}$ and $\Gamma_{p}$. Since $[A_{k}, A_{p}] \in \mathbb{C}$,
we have $\Gamma_{k}\,\Gamma_{p} = e^{[A_{k},A_{p}]}\,\Gamma_{p}\,\Gamma_{k}$, where
\begin{equation}
\begin{split}
[A_{k},A_{p}]
=
i\,\pi\,c_{k}\,c_{p}\,L^{T}_{k}\,\mathcal{K}^{-1}\,L_{p}\,F(y_{k},z_{k}\,|\,y_{p},z_{p})		
\,
	\end{split}	
\end{equation}
and 
\begin{equation}
\begin{split}
F(y_{k},z_{k}\,|\,y_{p},z_{p})
&\,=
\textrm{sgn}(z_{k}-z_{p})	
-
\textrm{sgn}(z_{k}-y_{p})	
\\
&\,
-
\textrm{sgn}(y_{k}-z_{p})	
+
\textrm{sgn}(y_{k}-y_{p})	
\,.
\end{split}
\end{equation}

One readily verifies that $F(y_{k},z_{k}\,|\,y_{p},z_{p})  = 0$  
when the intervals $(y_{k},z_{k})$ and $(y_{p},z_{p})$ are either non-overlapping
or if one interval is fully within the other, which implies the commutation relation 
$\Gamma_{k}\,\Gamma_{p} = \Gamma_{p}\,\Gamma_{k}$. Non-trivial commutation relation may occur with 
\begin{equation}
\begin{split}
&\,	
\Gamma_{k}\,\Gamma_{p} = e^{-2 i\,\pi\,c_{k}\,c_{p}\,L^{T}_{k}\,\mathcal{K}^{-1}\,L_{p}}\,\Gamma_{p}\,\Gamma_{k}	
\,,
\\
&\,
\textrm{for}~
y_{k} < y_{p} < z_{k} < z_{p}
\,.
\end{split}
\end{equation}

\section
{Generalization of the $\mathbb{Z}_{n} \times \mathbb{Z}_{n}$ SPT states discussed in Sec.~\ref{sec: sigma x interface}
} 
\label{sec: generalized Zn x Zn spt states}
In this Appendix, we present a generalization of the $\mathbb{Z}_{n} \times \mathbb{Z}_{n}$ SPT states
discussed in Sec.~\ref{sec: sigma x interface} formed at the interface with $\mathcal{K}_{x} = \sigma_{x} \oplus (-\sigma_{x})$.
Specifically, we will focus on the bound states that arise at domains between different topological classes of these SPT states.

Consider a series of domain walls separating segments gapped by the
interactions parametrized by the null vectors:
\begin{equation}
\label{eq: null vectors generalized - 1}
\Lambda^{(p)}_1 = (1,0,0,p\,n)
\,,\quad
\Lambda^{(p)}_2 = (0,p\,n,1,0)
\,,
\end{equation}
from segments gapped by 
the interactions parametrized by the null vectors:
\begin{equation}
\label{eq: null vectors generalized - 2}
\Lambda^{(q)}_1 = (1,0,0,q\,n)
\,,\quad
\Lambda^{(q)}_2 = (0,q\,n,1,0)
\,,
\end{equation}
where $p,q \in \mathbb{Z}_{n}$ label the distinct classes of SPT states. The interactions defined by the null vectors in Eqs.~\eqref{eq: null vectors generalized - 1} and~\eqref{eq: null vectors generalized - 2} 
are invariant under the action of the symmetry generators in Eqs.~\eqref{eq: left symmetry generator sigma_x}
and~\eqref{eq: right symmetry generator sigma_x}.

Let 
$
R^{x}_{q} = \cup_{i}\,(  x_{2i}+\varepsilon, x_{2i+1}-\varepsilon)
$
be the region gapped solely by the interactions in Eq.~\eqref{eq: null vectors generalized - 2}
and
$
R^{x}_{p} = \cup_{i}\,(  x_{2i-1}+\varepsilon, x_{2i}-\varepsilon)
$
be the region gapped solely by the interactions in Eq.~\eqref{eq: null vectors generalized - 1}, where $\varepsilon = 0^{+}$ is an infinitesimally small regulator distance.
We assume also that the length of each of these segments, $|x_{k+1} - x_{k}| \gg \xi$, where $\xi$ is the correlation length
set by the energy gap due to the interactions.

Following similar steps as in Sec.~\ref{sec: sigma x interface}, 
consider the two sets of \textit{mutually commuting} operators:
\begin{subequations}
\label{eq: def Aq and Bq operators sigma x}
\begin{equation}
A^{(q)}_{2i-1,2i} = \exp{\Big( \frac{i}{n}\,\int^{x_{2i}+\varepsilon}_{x_{2i-1}-\varepsilon}\,dx\,
[\Lambda^{(q)}_{1}]^{T}\,\mathcal{K}_{x}\,\partial_{x}\Phi  \Big)}
\,,
\end{equation}
\begin{equation}
B^{(q)}_{2i-1,2i} = \exp{\Big( \frac{i}{n}\,\int^{x_{2i}+\varepsilon}_{x_{2i-1}-\varepsilon}\,dx\,
[\Lambda^{(q)}_{2}]^{T}\,\mathcal{K}_{x}\,\partial_{x}\Phi  \Big)}
\,,
\end{equation}
\end{subequations}
and
\begin{subequations}
\label{eq: def Ap and Bp operators sigma x}
\begin{equation}
A^{(p)}_{2i,2i+1} = \exp{\Big( \frac{i}{n}\,\int^{x_{2i+1}+\varepsilon}_{x_{2i}-\varepsilon}\,dx\,
[\Lambda^{(p)}_{2}]^{T}\,\mathcal{K}_{x}\,\partial_{x}\Phi  \Big)}
\,,
\end{equation}
\begin{equation}
B^{(p)}_{2i,2i+1} = \exp{\Big( \frac{i}{n}\,\int^{x_{2i+1}+\varepsilon}_{x_{2i}-\varepsilon}\,dx\,
[\Lambda^{(p)}_{1}]^{T}\,\mathcal{K}_{x}\,\partial_{x}\Phi  \Big)}
\,.
\end{equation}
\end{subequations}
Eqs.~\eqref{eq: def Aq and Bq operators sigma x} and~\eqref{eq: def Ap and Bp operators sigma x} define two sets of non-local operators that commute with all the gapping interactions associated with
the null vectors in Eqs.~\eqref{eq: null vectors generalized - 1} and \eqref{eq: null vectors generalized - 2}.
Hence, the commutation relations between these operators offer useful information about the ground state manifold, 
since the states belonging to this manifold form a representation of the operator algebra whose dimension is the ground state degeneracy.

Making use of the result derived in Appendix~\ref{sec: Commutation relation between vertex operators},
we find that the only non-trivial commutation relations read:
\begin{equation}
\label{eq: commutation relations non-local ops - generalized sigma x}
\begin{split}
&\,
A^{(q)}_{2i-1,2i}\,A^{(p)}_{2j,2j+1}
=
e^{-i\,\frac{2\pi(p-q)}{n}\left( \delta_{j,i} - \delta_{j,i-1}  \right)}
\,
A^{(p)}_{2j,2j+1}\,A^{(q)}_{2i-1,2i}	
\,,
\\
&\,
B^{(q)}_{2i-1,2i}\,B^{(p)}_{2j,2j+1}
=
e^{i\,\frac{2\pi(p-q)}{n}\left( \delta_{j,i} - \delta_{j,i-1}  \right)}
\,
B^{(p)}_{2j,2j+1}\,B^{(q)}_{2i-1,2i}	
\,.
\end{split}
\end{equation}
The commutation relations in \eqref{eq: commutation relations non-local ops - generalized sigma x},
which are a generalization of those in \eqref{eq: commutation relations non-local ops - sigma x},
imply that the commutation relations
between these non-local operators break into two independent sets. 
For a configuration with $2g$ domain walls, Eq.~\eqref{eq: commutation relations non-local ops - generalized sigma x}
implies a ground state degeneracy of $n^{g-1} \times n^{g-1}$ if $p-q \neq 0$ (mod $n$).

This ground state degeneracy can be attributed to the presence of a pair of zero modes, $(\alpha_{i},\beta_{i})$, at the domain walls:
\begin{equation}
\label{eq: generalized alpha parafermions - sigma x interface}
\begin{split}
&\,
\alpha_{2i} 
=
e^
{
i\,\left( \phi_{2R}/n - p\,\phi_{1L} \right)(x_{2i} - \varepsilon)
+
i\,\left( \phi_{2L}/n - q\,\phi_{1R} \right) (x_{2i} + \varepsilon) 
}	
\,,
\\
&\,
\alpha_{2i+1} 
=
e^
{
i\,\left( \phi_{2L}/n - q\,\phi_{1R} \right)   (x_{2i+1} - \varepsilon)
+
i\,\left( \phi_{2R}/n - p\,\phi_{1L} \right)(x_{2i+1} + \varepsilon)
}	
\,,	
	\end{split}	
\end{equation}
and
\begin{equation}
\label{eq: generalized beta parafermions - sigma x interface}
\begin{split}
&\,
\beta_{2i} 
=
e^
{
i\,\left( \phi_{2L}/n - p\,\phi_{1R} \right)(x_{2i} - \varepsilon)
+
i\,\left( \phi_{2R}/n - q\,\phi_{1L} \right) (x_{2i} + \varepsilon) 
}	
\,,
\\
&\,
\beta_{2i+1} 
=
e^
{
i\,\left( \phi_{2R}/n - q\,\phi_{1L} \right)   (x_{2i+1} - \varepsilon)
+
i\,\left( \phi_{2L}/n - p\,\phi_{1R} \right)(x_{2i+1} + \varepsilon)
}	
\,.	
\end{split}	
\end{equation}
These zero modes satisfy the algebra $\alpha_{2i+1}\,\alpha_{2i} = e^{-i\frac{2\pi (p-q)}{n}}\,\alpha_{2i}\,\alpha_{2i+1}$,
$\beta_{2i+1}\,\beta_{2i} = e^{i\frac{2\pi(p-q)}{n}}\,\beta_{2i}\,\beta_{2i+1}$ and $\alpha_{i}\,\beta_{j} = \beta_{j}\,\alpha_{i}$,
generalizing the parafermions in 
Eqs.~\eqref{eq: alpha parafermions - sigma x interface} and~\eqref{eq: beta parafermions - sigma x interface}.
The result is consistent the additive property of SPT classes: 
If we glue two SPT states with labels $p_1$ and $p_2$, the effective SPT state has label $p_1 +p_2$ mod ($n$). Thus on the $n=12$ with $p=9$ and $q=3$, we can think about the $p=9$ class
as a $p_1 =3$ class glued with a $p_2 =3$ class. The problem then maps to the edge of the $p_1=6$ with vacuum ($p=0$), where the 
parafermions transform projectively with phases $e^{\frac{i 2\pi \times 6}{12} }$.

Importantly, as anticipated at the end of Sec.~\ref{sec: sigma x interface} and promised at the beginning of
this section, the generalized parafermions in \eqref{eq: generalized alpha parafermions - sigma x interface}
are formed by combinations of the vertex operators
$
e^
{
i\,\left( \phi_{2L}/n - q\,\phi_{1R} \right)
}
$
and
$
e^
{
i\,\left( \phi_{2R}/n - p\,\phi_{1L} \right)
}
$,
which represent, respectively, a bound state between a domain operator of the left(right) $\mathbb{Z}_{n}$ symmetry
and $q$($p$) charges of the right(left) $\mathbb{Z}_{n}$ symmetry, with similar properties satisfied by the generalized 
parafermions in \eqref{eq: generalized beta parafermions - sigma x interface}.
Thus, the $\mathbb{Z}_{n}\times\mathbb{Z}_{n}$ SPT states obtained here realize the concept of decorated
domain walls encoded in the lattice models of Refs.~\onlinecite{Geraedts,Santos-2015-a}.


\begin{thebibliography}{99}

\bibitem{Wen-book}
X.-G. Wen, \textit{Quantum Field Theory of Many-body Systems,}
Oxford University Press, 2004.

\bibitem{TQC-Review}
C. Nayak, S. H. Simon, A. Stern, M. Freedman, and S. Das Sarma, 
Rev. Mod. Phys. \textbf{80}, 1083 (2008).

\bibitem{kitaevpreskill}
A. Kitaev and J. Preskill,
Phys.Rev.Lett. \textbf{96}, 110404 (2006).

\bibitem{levinwen}
M. Levin and X.-G. Wen,
Phys. Rev. Lett., \textbf{96}, 110405 (2006).

\bibitem{lihaldane}
H. Li and F. D. M. Haldane,
Phys. Rev. Lett. \textbf{101}, 010504 (2008).

\bibitem{pollmann-2010}
Frank Pollmann, Erez Berg, Ari M. Turner, Masaki Oshikawa,
Phys. Rev. B \textbf{81}, 064439 (2010).

\bibitem{cirac-2011}
J. Ignacio Cirac, Didier Poilblanc, Norbert Schuch, and Frank Verstraete
Phys. Rev. B \textbf{83}, 245134 (2011).


\bibitem{regnault-2011}
N. Regnault, B. Andrei Bernevig
Phys. Rev. X \textbf{1}, 021014 (2011).

\bibitem{qikatsuraludwig}
X.-L.~Qi and H.~Katsura and A.~W.~W.~Ludwig, Phys. Rev.~Lett.~{\bf 108}, 196402 (2012).

\bibitem{lundgrenentanglement}
R.~Lundgren, Y.~Fuji, S.~Furukawa, and M.~Oshikawa, Phys.~Rev.~B {\bf 88}, 245137 (2013).

\bibitem{chen-fradkin-2013}
X. Chen and E. Fradkin,
J. Stat. Mech. Theor. Exp. \textbf{2013}, P08013 (2013).

\bibitem{furukawa-kim-2013}
S. Furukawa and Y. B. Kim,
Phys. Rev. B \textbf{83} 085112 (2011); Phys. Rev. B \textbf{87} 119901(E) (2013); 


\bibitem{kapustin-2011}
A. Kapustin and N. Saulina,
Nucl. Phys. B \textbf{845}, 393 (2011).

\bibitem{Levin-2013}
M. Levin,
Phys. Rev. X \textbf{3}, 021009 (2013).

\bibitem{Barkeshli-2013-a}
M. Barkeshli, C.-M. Jian, and X.-L. Wen,
Phys. Rev. B \textbf{88}, 241103(R) (2013).


\bibitem{Lindner-2012}
N. H. Lindner, E. Berg, G. Refael, and A. Stern, 
Phys. Rev. X \textbf{2}, 041002 (2012).

\bibitem{Clarke-2013}
D. J. Clarke, J. Alicea, and K. Shtengel, 
Nature Commun. \textbf{4}, 1348 (2013).

\bibitem{Cheng-2012}
M. Cheng, 
Phys. Rev. B \textbf{86}, 195126 (2012).

\bibitem{Vaezi-2013}
A. Vaezi, 
Phys. Rev. B \textbf{87}, 035132 (2013).

\bibitem{Barkeshli-2013-b}
M. Barkeshli, C.-M. Jian, and X.-L. Qi,
Phys. Rev. B \textbf{87}, 045130 (2013).

\bibitem{Plamadeala2013}
E.~Plamadeala, M. Mulligan, and C. Nayak, Phys.~Rev.~B {\bf 88}, 045131 (2013).

\bibitem{Mong-2014}
R. S. K. Mong, D. J. Clarke, J. Alicea, N. H. Lindner, P. Fendley, C. Nayak, Y. Oreg, A. Stern, E. Berg, K. Shtengel, M. P. A. Fisher,
Phys. Rev. X \textbf{4}, 011036 (2014).

\bibitem{Khan-2014}
M. N. Khan, J. C. Y. Teo, and T. L. Hughes,
Phys. Rev. B \textbf{90}, 235149 (2014).

\bibitem{Lu-2014}
Y.-M. Lu and D.-H. Lee,
Phys. Rev. B \textbf{89}, 205117 (2014).

\bibitem{Cano-2014}
J. Cano, M. Cheng, M. Mulligan, C. Nayak, E. Plamadeala, and J. Yard,
Phys. Rev. B \textbf{89}, 115116 (2014).

\bibitem{Cano-2015-a}
J. Cano, M. Cheng, M. Barkeshli, D. J. Clarke, and C. Nayak,
Phys. Rev. B \textbf{92}, 195152 (2015).

\bibitem{Khan-2016}
M. N. Khan, J. C. Y. Teo, T. L. Hughes, and S. Vishveshwara,
Phys. Rev. B \textbf{95}, 205112 (2017).

\bibitem{Santos-2017}
L. H. Santos and T. L. Hughes,
Phys. Rev. Lett. \textbf{118}, 136801 (2017).  


\bibitem{cano-2015}
J. Cano, T. L. Hughes, and M. Mulligan,
Phys. Rev. B \textbf{92}, 075104 (2015). 


\bibitem{Senthil-SPT-review}
T. Senthil,
Ann. Rev. Cond. Matt. Phys. \textbf{6}, 299 (2015).

\bibitem{Haldane-1983-a}
F. D. M. Haldane,
Phys. Lett. \textbf{93A}, 464 (1983).


\bibitem{Haldane-1983-b}
F. D. M. Haldane,
Phys. Rev. Lett. \textbf{50}, 1153 (1983).


\bibitem{Affleck-1987}
I. Affleck, T. Kennedy, E. H. Lieb, and H. Tasaki,
Phys. Rev. Lett. \textbf{59}, 799 (1987).


\bibitem{Affleck-1988}
I. Affleck, T. Kennedy, E. H. Lieb, and H. Tasaki,
Comm. Math. Phys. \textbf{115}, 477 (1988).


\bibitem{Schuch-2011}
N. Schuch, D. Perez-Garcia, and I. Cirac,
Phys. Rev. B \textbf{84}, 165139 (2011).


\bibitem{Chen-2013}
X. Chen, Z. -C. Gu, Z. -X. Liu and X. -G. Wen, 
Phys. Rev. B \textbf{87}, 155114 (2013).

\bibitem{Gu-Wen-2014}
Z.-C. Gu and X.-G. Wen,
Phys. Rev. B \textbf{90}, 115141 (2014).


\bibitem{Bi-2013}
Zhen Bi, Alex Rasmussen, and Cenke Xu,
Phys. Rev. B \textbf{91}, 134404 (2015).


\bibitem{Freedman2004}
M. Freedman, C. Nayak, K. Shtengel, K. Walker, and Z. Wang, 
Ann. Phys. \textbf{310}, 428 (2004).

\bibitem{LevinStern2013}
M. Levin and A. Stern, 
Phys. Rev. Lett. \textbf{103}, 196803 (2009).


\bibitem{Neupert-2011}
T. Neupert, L. Santos, S. Ryu, C. Chamon, and C. Mudry,
Phys. Rev. B \textbf{84}, 165107 (2011).

\bibitem{Santos-2011}
L. Santos, T. Neupert, S. Ryu, C. Chamon, and C. Mudry,
Phys. Rev. B \textbf{84}, 165138 (2011).

\bibitem{Levin2011}
M. Levin, F. J. Burnell, M. Koch-Janusz, and A. Stern,
Phys. Rev. B \textbf{84}, 235145 (2011).

\bibitem{LevinStern-2012}
M. Levin and A. Stern,
Phys. Rev. B \textbf{86}, 115131 (2012).

\bibitem{Ryu-2012}
S. Ryu and S.-C. Zhang,
Phys. Rev. B \textbf{85}, 245132 (2012).

\bibitem{Levin-2012}
M. Levin and Z.-C. Gu,
Phys. Rev. B \textbf{86}, 115109 (2012).

\bibitem{Lu-2012}
Y.-M. Lu and A. Vishwanath, 
Phys. Rev. B \textbf{86}, 125119 (2012).

\bibitem{Vishwanath-2013}
A. Vishwanath and T. Senthil, 
Phys. Rev. X \textbf{3}, 011016 (2013).

\bibitem{Metliski-2013-a}
M. A. Metlitski, C. L. Kane, and M. P. A. Fisher,
Phys. Rev. B \textbf{88}, 035131 (2013).

\bibitem{Kapustin-2014}
A. Kapustin, 
arXiv:1403.1467; arXiv:1404.6659 (unpublished).


\bibitem{J-Wang-2015}
J. Wang, L. H. Santos, and X.-G. Wen,
Phys. Rev. B \textbf{91}, 195134 (2015).


\bibitem{Ye-2016}
P. Ye and Z.-C. Gu,
Phys. Rev. B \textbf{93}, 205157 (2016).

\bibitem{Fidkowski-2011}
L. Fidkowski and A. Kitaev,
Phys. Rev. B \textbf{83}, 075103 (2011).

\bibitem{Fidkowski-2010}
L. Fidkowski and A. Kitaev,
Phys. Rev. B \textbf{81}, 134509 (2010).

\bibitem{C-Wang-2014}
C. Wang, A. Potter, and T. Senthil, 
Science \textbf{343}, 629 (2014).

\bibitem{Zou-2016}
L. Zou and J. Haah,
Phys. Rev. B \textbf{94}, 075151 (2016).

\bibitem{casinihuertarosabal}
H.~Casini, M.~Huerta, and J.~A.~Rosabal, Phys.~Rev.~D {\bf 89}, 085012 (2014).

\bibitem{Ghosh2015}
S.~Ghosh, R.~M.~Soni, S.~P.~Trivedi, JHEP {\bf 9}, 69 (2015).

\bibitem{Fliss2017}
J.~R.~Fliss, X.~Wen, O.~Parrikar, C.~T.~Hsieh, T.~L.~Hughes, and R.~G.~Leigh, JHEP {\bf 9}, 56 (2017).


\bibitem{Geraedts}
S. D. Geraedts and O. I. Motrunich,
arXiv:1410.1580.

\bibitem{Santos-2015-a}
L. H. Santos,
Phys. Rev. B \textbf{91}, 155150 (2015).

\bibitem{Fradkin-1980}
E. Fradkin and L. P. Kadanoff,
Nucl. Phys. B\textbf{170}, (1980) 1-15.


\bibitem{Haldane-1995}
F. D. M. Haldane,
Phys. Rev. Lett. \textbf{74}, 2090 (1995).


\bibitem{read-1990}
N. Read,
Phys. Rev. Lett. \textbf{65}, 1502 (1990).


\bibitem{wen-1992}
X. G. Wen and A. Zee,
Phys. Rev. B \textbf{46}, 2290 (1992).


\bibitem{Wen-1995}
X.-G. Wen,
Advances in Physics, \textbf{44}, 405 (1995).


\bibitem{Lu-Fidkowski-2014}
Y.-M. Lu and L. Fidkowski,
Phys. Rev. B \textbf{89}, 115321 (2014).

\bibitem{Hung-2013}
L.-Y. Hung and Y. Wan,
Phys. Rev. B \textbf{87}, 195103 (2013).

\bibitem{Lu-2016}
Y.-M. Lu and A. Vishwanath,
Phys. Rev. B \textbf{93}, 155121 (2016).


\bibitem{comment-1}
The process of obatining the null vectors \eqref{eq: null vectors homogeneous interface - general}
from \eqref{eq: null vectors interfaces 1} and \eqref{eq: null vectors interfaces 2} is a special 
case of the general statement. Let $\{ \Lambda^{(1)}_{i} = (\alpha_{i},\beta_{i}),~ i = 1, ..., r \}$
and $\{ \Lambda^{(2)}_{i} = (\gamma_{i},\delta_{i}),~ i = 1, ..., r \}$ be, respectively, 
two sets of null vectors of the interfaces with K-matrices $\mathcal{K}_{1} = \textrm{diag}(K, -G\,K\,G^{T})$
and $\mathcal{K}_{2} = \textrm{diag}(G\,K\,G^{T},-K)$. Then $\{(\alpha_{i},\delta_{i}),~ i = 1, ..., r \}$
is a set of null vectors of $\textrm{diag}(K, -K)$ if and only if $\{(\beta_{i},\gamma_{i}),~ i = 1, ..., r \}$ 
is a set of null vectors of $\textrm{diag}(-G\,K\,G^{T}, G\,K\,G^{T})$.


\bibitem{anyonicsymmetry-K}
This can be verified by inspecting the form of the local gapping interactions
under a change of basis:
\begin{equation}
\begin{split}
\mathcal{H}_{i} 
&\,=
\cos{\left( \Lambda^{T}_{L, i}\,K_{L}\,\phi_{L}  - \Lambda^{T}_{R, i}\,K_{R}\,\phi_{R}\right)}
\\
&\,		
=
\cos{\left( \Lambda^{T}_{L, i}\,\tilde{W}_{L}\,K_{L}\,\tilde{W}^{T}_{L}\,\phi_{L}  - 
\Lambda^{T}_{R, i}\,\tilde{W}_{R}\,K_{R}\,\tilde{W}^{T}_{R}\,\phi_{R}\right)}
\\
&\,
=
\cos{\left( \Lambda^{T}_{L, i}\,\tilde{W}_{L}\,K_{L}\,\tilde{\phi}_{L}  - 
\Lambda^{T}_{R, i}\,\tilde{W}_{R}\,K_{R}\,\tilde{\phi}_{R}\right)}		
\,,
	\end{split}	
\end{equation}
where $\tilde{\phi}_{L/R} = \tilde{W}^{T}_{L/R}\,\phi_{L/R}$ and the anyonic symmetries
of $K_{L/R}$ were invoked in passing from the first to the second line. 
A change of basis then amounts to
$\mathcal{M}_{L/R} \rightarrow \mathcal{M}_{L/R}\,\tilde{W}_{L/R}$.


\bibitem{wanglevin}
Chenjie Wang and Michael Levin,
Phys. Rev. B \textbf{88}, 245136 (2013).


\bibitem{Chen-2014}
X. Chen, Y.-M. Lu, and A. Vishwanath,
Nat. Comm. \textbf{5}, 3507 (2014).


\bibitem{sigmaz-sequence}
For further details, see http://oeis.org/A042965.



\bibitem{nu2-sequence}
For further details, see http://oeis.org/A004613.



\bibitem{1623-sequence}
For further details, see http://oeis.org/A007310.

\bibitem{hung-wan-2015}
L.-Y. Hung and Y. Wan,
Phys. Rev. Lett. \textbf{114}, 076401 (2015)


\bibitem{lan-wang-wen-2015}
T. Lan, J. C. Wang and X.-G. Wen,
Phys. Rev. Lett. \textbf{114}, 076402 (2015)


\bibitem{vaezi-2017}
M.-S. Vaezi and A. Vaezi,
arXiv: 1706.01192 (unpublished).

\bibitem{wangwenwitten-2017}
J. Wang, X.-G. Wen and E. Witten,
arXiv: 1705.06728 (unpublished).







\end{thebibliography}
\end{document}